\documentclass[times]{nagauth}
\usepackage{amsfonts}
\usepackage{amssymb}
\usepackage{mathtools}
\usepackage{microtype}
\usepackage{booktabs}
\usepackage{bbm}
\usepackage{todonotes}
\usepackage{arydshln}
\usepackage{verbatim} %\includecomment{comment}
\usepackage{multirow}
\usepackage{lineno}
\newcommand{\sign}{\operatorname{sign}}
\newcommand{\detr}{\operatorname{det}}

\newcommand{\LB}{\left[\begin{MAT}(r){l}}
\newcommand{\RB}{\\ \end{MAT}\right]}
\begin{document}
%\linenumbers

%
\title{Stiffness pathologies in discrete granular systems:
       bifurcation, neutral equilibrium, and instability
       in the presence of kinematic constraints} 
\author{Matthew R. Kuhn\affil{1}\corrauth,
        Florent Prunier\affil{2}, and Ali Daouadji\affil{2}}
\address{\affilnum{1}Br. Godfrey Vassallo Prof. of Engrg.,
             Donald P. Shiley School of Engrg., Univ. of Portland,
             5000 N. Willamette Blvd., Portland, OR, USA 97231
         \break
         \affilnum{2}University of Lyon, INSA-Lyon, GEOMAS, F-69621, France}
\corraddr{Donald P. Shiley School of Engineering,
          University of Portland,
          5000 N. Willamette Blvd.,
          Portland, OR, 97203, USA.
          Email: \texttt{kuhn@up.edu}}
\runningheads{Kuhn, Prunier, Daouadji}{Stiffness pathologies}
\begin{abstract}
The paper develops the stiffness relationship between the
movements and forces among a system of discrete interacting
grains.
The approach is similar to that used in structural analysis,
but the stiffness matrix of granular material is inherently
non-symmetric because of the geometrics of particle interactions
and of the frictional behavior of the contacts.
Internal geometric constraints are imposed by the particles'
shapes, in particular, by the surface curvatures of the particles
at their points of contact.
Moreover, the stiffness relationship is incrementally
non-linear, and even small assemblies require the analysis
of multiple stiffness branches, with each branch region being
a pointed convex cone in displacement-space.
These aspects of the particle-level stiffness relationship
gives rise to three
types of micro-scale failure:
neutral equilibrium,
bifurcation and path instability, and instability of equilibrium.
These three pathologies are defined in the context of four types
of displacement constraints,
which can be readily analyzed with certain generalized inverses.
That is, instability and non-uniqueness
are investigated in the presence of kinematic constraints.
Bifurcation paths can be either stable or unstable,
as determined with the Hill--Ba\v{z}ant--Petryk criterion.
Examples of simple granular systems of three, sixteen,
and sixty four disks are analyzed.
With each system, multiple contacts were assumed to
be at the friction limit.
Even with these small systems,
micro-scale failure is expressed in many different forms, with some
systems having hundreds of micro-scale failure modes.
The examples suggest that micro-scale failure
is pervasive within granular materials,
with particle arrangements being in a nearly
continual state of instability.
\end{abstract}
\maketitle
\section{\large Introduction}
%
%\todo[inline]{Expand this introduction}
%
In the mid-2000s, several independent works were published
on the nature of internal rigidity, uniqueness,
and stability of discrete granular materials
\cite{Bagi:2007a,Kuhn:2005b,Nicot:2007b}.
When violated, these favorable
conditions
%, to be given precise meaning the paper,
give rise to failure, weakening, and localized deformation,
conditions that we broadly designate as
\emph{stiffness pathologies}.
The incremental macro-scale behavior of granular materials
is known to be exceedingly complex, and the strength, stiffness,
and various
forms of failure (diffuse, localized, static, dynamic, etc.)
at the continuum, macro-scale have received extensive
investigation in the past decades.
At the risk of making the nearly impenetrable and
confounding behavior of these materials
even more so, we return to a study of discrete
failure~--- in its many forms~--- by
addressing the internal particle-scale
stiffness and rigidity of granular
materials.
\par
The works of Bagi \cite{Bagi:2007a},
Nicot and Darve \cite{Nicot:2007b},
and Kuhn and Chang \cite{Kuhn:2005b},
viewed granular materials at the micro-level,
treating a material region, which might appear continuous
at a larger scale, as a collection of
discrete and notionally rigid
granules that interact when touching each other
at idealized contact points.
The paper takes a similar primitive approach,
departing from a continuum framework and
treating a granular medium as a discrete system
of interacting grains.
%while avoiding any extension to a continuous medium.
As one difference between discrete and and continuous media,
the discrete topology of a granular medium can be expressed as
a multi-graph \cite{Satake:1993b,Bagi:1996a}
with a finite (or at most, countable) number
of vertices (grains) and edges
(contacts between grains), and we adopt this elemental
view of a granular material.
Because a discrete graph expresses the topology of a finite open set
(e.g., a planar graph is homeomorphic with a sphere),
a graph has no interior, and without an interior, it has
no boundary, no unit normal on the boundary, no volume,
and no surface area.
Our vocabulary is instead of movement and force
and of the relationship between
them~--- stiffness.
Movements and forces are associated with particles and contacts:
the particle forces are external forces, and the contact forces are
internal forces, but no distinction is made between boundary and interior
forces, as there is no boundary or interior.
%Stress and strain are occasionally computed but are based upon
%a crude averaging of the forces and movements.
%
\par
Another difference between discrete and continuous systems
arises in the choice of a reference configuration
of stable behavior.
With continuous systems having simple boundaries,
one can usually distinguish a ``fundamental deformation''
with which a buckled or bifurcated deformation can be compared
(for example, the fundamental
deformation of a persistently straight column
or of a region that
deforms in a fundamental, uniform mode without shear bands).
Sliding between discrete particles at their contacts,
a primary mechanism of deformation and failure,
usually occurs in only a subset of the contacts,
and predetermination of this sliding subset is usually quite difficult.
%A single contact within a granular material can remain elastic
%or slide, but for an entire assembly,
%sliding at some contacts can begin a very small
%strains, and the pre-determination of this sliding subset
%is quite difficult.
As such, the fundamental deformation of a discrete granular system
can rarely be presaged.
\par
The paper extends the current understanding of discrete systems,
by presenting a thorough accounting of four different types of geometric
effects in granular media, for which three of the effects
behave as internal follower forces.
These geometric effects are a results of the
internal geometric constraints that are imposed by the particles'
shapes, in particular, by the surface curvatures of the particles
at their points of contact.
The paper also describes four types of external displacement constraint,
using the theory of generalized inverses to address three of
the types.
We also extend an accepted definition of stability by incorporating
the internal geometric effects within granular systems and
give a comprehensive accounting of three types of pathologic
behaviors in discrete systems:  neutral equilibrium,
bifurcation and path instability, and instability of equilibrium,
with Ba\v{z}ant--Petryk theory applied
to the problem of path instability.
This accounting is detailed for each of the four types of
displacement constraints.
An analysis of potential pathologies is particularly vexing
with granular materials, as behavior is incrementally
non-linear, and we present a systematic means of
determining the consistency of a possible pathology with
the assumed contact-level stiffness conditions.
We also present examples of simple granular assemblies and
show how the various pathologies arise in these systems.
\par
The plan of the paper is as follows.
In the next section, we develop essential elements for
characterizing the stiffness of discrete systems.
These elements include the derivation of the stiffness
matrix of a granular assembly, including those stiffness components
that depend upon the current forces and their geometric alteration
(Section~\ref{sec:equilibrium}).
A thermodynamic framework for analyzing the stability of a granular system
is then developed in Section~\ref{sec:stability}.
In Section~\ref{sec:constraints},
we consider possible constraints placed upon a granular
system by its surroundings, as these limitations will
restrict the available modes of deformation and instability.
Four categories of displacement constraints are developed in this
section.
To provide a specific framework for these principles,
Section~\ref{sec:contacts} describes a standard two-branch frictional
model for the contact interaction between particles,
and Section~\ref{sec:branches} places this model in the context of an
entire assembly's stiffness matrix.
\par
After establishing these principles,
we define various stiffness pathologies
in Section~\ref{sec:Pathologies}
(controllability, bifurcation, and instability).
Examples of several simple granular systems
are then analyzed in Section~\ref{sec:Examples}.
These examples are engaged using the methods
and language of linear algebra, which we hope will clarify
distinctions among the different pathologies.
\par
With occasional exceptions,
we use vector and matrix, rather than index,
notation for most objects and operations.
Vectors and matrices are enclosed in brackets when they
contain information for an entire assembly; but
brackets are usually excluded when the vector or matrix
is referenced to a single contact.
Vectors are written with bold lower case letters;
matrices are with bold upper case letters;
and scalars are normal lower case letters.
Inner, outer, and dyad products are denoted as follows:
inner products $\mathbf{x}\cdot\mathbf{y}=x_{i}y_{i}$ and
$\mathbf{A}\cdot\mathbf{x}=A_{ij}x_{j}$;
outer product $\mathbf{x}\times\mathbf{y}=e_{ijk}x_{j}y_{k}$;
and dyad product $\mathbf{x}\otimes\mathbf{y}=x_{i}y_{j}$.
%
%
%\par
%Many constitutive relations have been proposed for contact
%force and displacement, including the quite complex,
%history-dependent models of Cattaneo \cite{Cattaneo:1938a}
%and Mindlin \cite{Mindlin:1953a}
%and of Kalker \cite{Kalker:1967a}
%and models that include resistive moments \cite{Iwashita:1998a}.
%Among the simplest models is the zero-tension
%zero-moment contact with
%linear stiffnesses in both normal and tangential directions
%but with a frictional limit on the tangential force.
%This prototype linear-frictional contact can produce a
%negative contribution
%$\mathfrak{d}\mathbf{f}^{pq}\cdot\delta\mathbf{u}^{pq,\text{def}}$
%when frictional sliding is accompanied by unloading
%of the normal component of force
%\cite{Mandel,Bazant:1991a}.
%This type of sliding among a significant number
%of contacts can lead to a negative $W_{2}$
%and loss of sustainability \cite{NicotsWorkHere}.
%
\section{\large Stiffness framework for discrete systems}%
\label{sec:Framework}
Bagi \cite{Bagi:2007a}
and Kuhn and Chang \cite{Kuhn:2005b},
approached the discrete mechanics of granular
materials from a structural mechanics perspective.
These concurrent works developed a stiffness relationship between
movement and force in matrix form, so that established
concepts of stability, bifurcation, softening
%\todo{Softening?}
and controllability, already used in structural analysis,
could also be applied to granular assemblies
(see, for example, Bazant \cite{Bazant:1991a}).
Both works were preceded by others that developed
stiffness matrices for granular systems
\cite{Serrano:1973a,Kishino:1988a,McNamara:2006a}, but these
earlier works neglected second-order geometric changes,
which are essential to the developments
that follow.
A general stiffness relationship was also developed by
Agnolin and Roux \cite{Agnolin:2007a,Agnolin:2007c},
who had considered geometric effects
in the appendices of these works. 
We will use the notation of
Kuhn and Chang  \cite{Kuhn:2005b} as it makes useful
distinctions between objective and non-objective
quantities, which become relevant when distinguishing
various stiffness pathologies.
\subsection{\normalsize Notation}\label{sec:notation}
\begin{figure}
  \centering
  \includegraphics{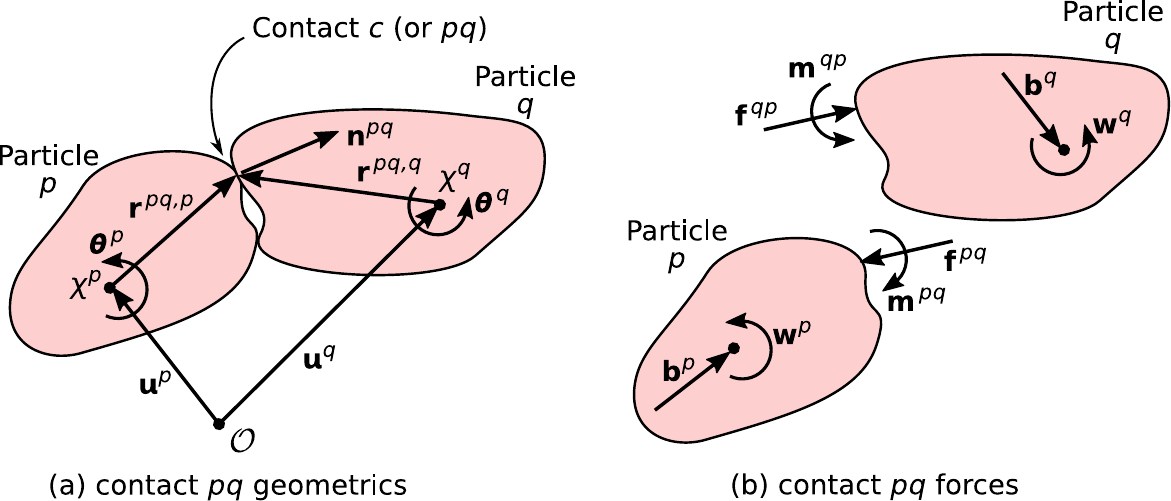}
  \caption{Contact between particles $p$ and $q$:
           (a)~geometry; and (b)~forces.
           \label{fig:ContactVectors}}
\end{figure}

The shapes, positions, orientations, contact forces,
and loading history of the $N$ particles in a granular
assembly are assumed known at
time $t$,
and we develop the conditions for equilibrium
in a deformed state at $t+dt$.
Each particle is assumed a hard body that interacts
with neighboring particles at their shared compliant contacts.
Movement and deformation are assumed slow and quasi-static,
so that we can neglect viscous or gyroscopic forces.
As shown in Fig.~\ref{fig:ContactVectors},
a particle $p$ touches particle $q$ at a contact $c$,
with $q$ being a member of the set $\mathcal{C}_{p}^{q}$
of $p$'s neighboring
particles, $q\in\mathcal{C}_{p}^{q}$, and with
$c$ being a member of the set $\mathcal{C}_{p}^{c}$
of $p$'s contacts,
$c\in\mathcal{C}_{p}^{c}$.
\par
In this work, an assembly's contact topology is represented as
a directed multi-graph (a multi-digraph).
The graph is ``multi'', because a pair of particles
can share multiple contacts, as can occur with non-convex particles.
The graph is ``directed'', because we treat the $pq$ contact
(from $p$ to $q$) as distinct from the contact $qp$ 
(from $q$ to $p$).
This approach allows a more straightforward derivation
and is consistent with the non-symmetry of the stiffness
matrix,
as we will see that
the $pq$ stiffness can differ from the $qp$ stiffness.
The system can also include isolated ``rattler'' particles
that are not in contact with other particles.
With this situation, we can adopt two different approaches.
We can include these rattlers as isolated nodes of a complete
topology, which will lead to numerous zero-stiffness modes
of neutral equilibrium,
or we can consider only the load-bearing network of
contacting (non-rattler) particles and then update the
topology whenever rattlers are freshly released from or
incorporated into the network.
For reasons given in Section~\ref{sec:equilibrium},
we take the latter approach and consider only the current
load-bearing network of contacts and particles.
\par
Vector $\mathbf{u}^{p}$ is the location of a material reference
point $\chi^{p}$ attached to $p$;
$\boldsymbol{\theta}^{p}$ are the orientation cosines of $p$;
and
$\mathbf{b}^{p}$ and $\mathbf{w}^{p}$ are the external
force and moment that act upon $p$ at the point $\chi^{p}$.
The entire particle system has $M$ contacts.
The single contact vector $\mathbf{r}^{c,p}$,
is from $\chi^{p}$ to its contact $c$;
$\mathbf{n}^{c}$ is the outward unit vector normal to the
surface of $p$ at contact $c$;
and $\mathbf{f}^{c}$ and $\mathbf{m}^{c}$
are the contact force and moment exerted upon $p$ at $c$.
\par
Although the two particles, $p$ and $q$, can share multiple contacts $c$,
we will use the more convenient notations
$\mathbf{r}^{pq,p}$, $\mathbf{n}^{pq}$,
$\mathbf{f}^{pq}$, and $\mathbf{m}^{pq}$
with the understanding that the pair $pq$ can represent one of several
contacts $c$ between $p$ and $q$.
With this notation,
force $\mathbf{f}^{pq}$ and moment $\mathbf{m}^{pq}$
act upon particle $p$,
and $\mathbf{n}^{pq}$ is directed outward from $p$.
We must distinguish, however the two ``$\mathbf{r}^{pq}$'' contact
vectors for contact $pq$:
vector $\mathbf{r}^{pq,p}$ is directed from $p$ to the contact;
whereas vector $\mathbf{r}^{pq,q}$ is directed from $q$ to the
contact.
\par
We gather the particle positions and orientations of all $N$ particles
into a stacked column vector $[\mathbf{u}/\boldsymbol{\theta}]$,
the external forces and moments
into the stacked complementary vector $[\mathbf{b}/\mathbf{w}]$,
and the internal contact forces and moments into the stacked
vector $[\mathbf{f}/\mathbf{m}]$:
\begin{equation}\label{eq:gather}
  \left. \begin{array}{r} \mathbf{u}^{p}\\
  \boldsymbol{\theta}^{p}
  \end{array}\right\}_{6\times 1}
  \;\rightsquigarrow\;
  \def\arraystretch{1.00}
  \left[
    \begin{array}{@{\extracolsep{\fill}}c@{\extracolsep{\fill}}}
      \mathbf{u} \\\hdashline[1pt/1pt]
      \boldsymbol{\theta}
    \end{array}
  \right]_{6N \times 1}
  \!,\quad
  \left.\begin{array}{r}
  \mathbf{b}^{p}\\
  \mathbf{w}^{p}\end{array}\right\}_{6\times 1}
  \;\rightsquigarrow\;
  \def\arraystretch{1.00}
  \left[
    \begin{array}{@{\extracolsep{\fill}}c@{\extracolsep{\fill}}}
      \mathbf{b} \\\hdashline[1pt/1pt]
      \mathbf{w}
    \end{array}
  \right]_{6N \times 1}
  ,\quad
  \left.\begin{array}{r}
  \mathbf{f}^{pq}\\\mathbf{m}^{pq}
  \end{array}\right\}_{6\times 1}
  \;\rightsquigarrow\;
  \def\arraystretch{1.00}
  \left[
    \begin{array}{@{\extracolsep{\fill}}c@{\extracolsep{\fill}}}
      \mathbf{f} \\\hdashline[1pt/1pt]
      \mathbf{m}
    \end{array}
  \right]_{2(6M) \times 1}
\end{equation}
where the ``$\rightsquigarrow$'' represents a matrix assembly
process that gathers the individual ``$p$'' and ``$pq$''
vectors into vectors for the entire assembly.
In these equations, $N$ is the number of particles,
with each particle involving 3-vectors for the position,
orientation, external force, and external moment; whereas,
$M$ is the number of contacts, each with a
3-vector of contact force and a 3-vector of contact moment.
Recalling that the $pq$ and $qp$ contact variants are treated
separately, the stacked vector $[\mathbf{f}/\mathbf{m}]$
has $2(6M)$ components.
Although two-dimensional (2D) systems can be represented
with smaller
vectors and matrices, these objects for a 2D system can
be readily extracted from the 3D counterparts.
%where $6N$
%is the number of degrees of freedom of the system
%($6N=6N$ for three-dimensional assemblies,
%and $6N=3N$ for two-dimensional assemblies);
%whereas, $(6M)$ is the number of available contact
%forces and moments.
%For systems with moment-resisting contacts,
%$(6M)=6M$ for three-dimensional assemblies,
%$(6M)=3M$ for two-dimensional assemblies;
%whereas, for systems with no contact moments,
%$(6M)=3M$ for three-dimensional assemblies,
%and $(6M)=2M$ for two-dimensional assemblies.
%To distinguish between the $pq$ and $qp$
%variants of the incremental contact forces,
%we collect both variants in a vector
%$[\mathbf{f}/\mathbf{m}]$ with $2D_{m}$ rows
%(see the discussion that follows Eq.~\ref{eq:dmpqall}).
%\par
We will sometimes compress the notation,
with $[\mathbf{x}]$ , $[\mathbf{p}]$,
and $[\,\boldsymbol{\mathfrak{f}}\,]$ representing
the configuration, loading, and internal force vectors:
\begin{equation}\label{eq:gather2}
  \left[\mathbf{x}\right]_{6N \times 1} =
  \left[
    \begin{array}{@{\extracolsep{\fill}}c@{\extracolsep{\fill}}}
      \mathbf{u} \\\hdashline[1pt/1pt]
      \boldsymbol{\theta}
    \end{array}
  \right]%_{6N\times 1}
  \;\text{, }\quad
  \left[\mathbf{p}\right]_{6N \times 1} =
  \left[
    \begin{array}{@{\extracolsep{\fill}}c@{\extracolsep{\fill}}}
      \mathbf{b} \\\hdashline[1pt/1pt]
      \mathbf{w}
    \end{array}
  \right]%_{6N \times 1}
  \;\text{, and}\quad
  \left[\,\boldsymbol{\mathfrak{f}}\,\right]_{2(6M) \times 1} =
  \left[
    \begin{array}{@{\extracolsep{\fill}}c@{\extracolsep{\fill}}}
      \mathbf{f} \\\hdashline[1pt/1pt]
      \mathbf{m}
    \end{array}
  \right]%_{(6M) \times 1}
\end{equation}
\par
\begin{figure}
  \centering
  \includegraphics{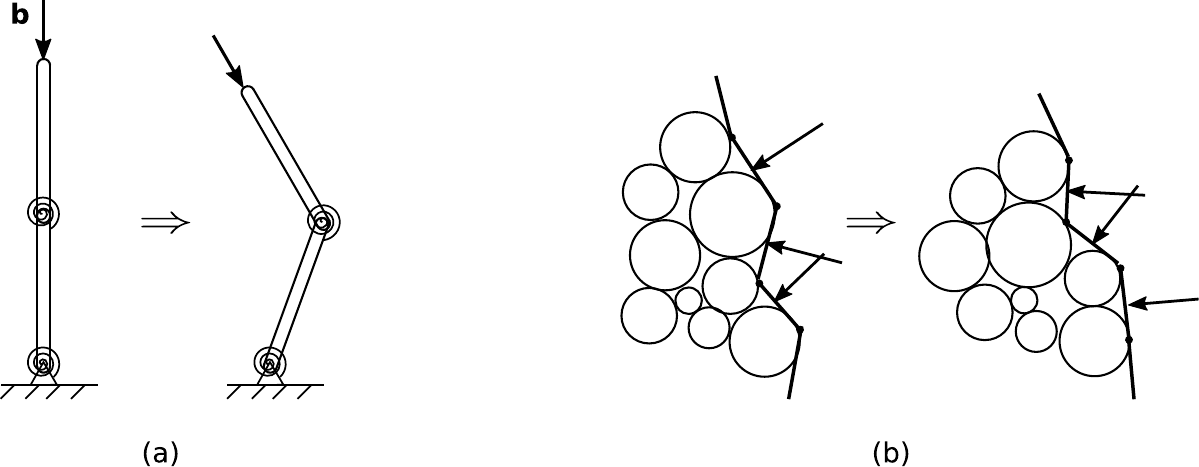}
  \caption{Position-dependent external forces:
           (a)~follower load on Ziegler column
           \cite{Nicot:2012a}; and
           (b)~membrane loads.
           \label{fig:Follower}}
\end{figure}

Nicot et al. \cite{Nicot:2012a} considered the stability of
non-conservative structural systems,
in which the external loads $[\mathbf{b}/\mathbf{w}]$
depend upon the positions and orientations,
$[\mathbf{u}]$ and $[\boldsymbol{\theta}]$, of material
points $\chi$ within the system.
In this general setting, the $6N$ elements of
the force vector $[\mathbf{p}]$ are functions %$G_{i},i=1,2,\ldots,6N$
of the positions $[\mathbf{x}]$ and an $L$-list of loading parameters
$\mathbf{q}$:
\begin{equation}
  \left[\mathbf{p}\right]_{6N \times 1} =
  \left[
    \rule{0ex}{2.5ex}
    \mathbf{p}\left(\rule{0ex}{2.2ex}
    [\mathbf{x}]_{6N\times 1},[\mathbf{q}]_{L\times 1}
    \right)
  \right]_{6N\times 1}
  \label{eq:pGx}
\end{equation}
These systems include those with external follower forces having
directions and magnitudes that change as the system is deformed,
such as the two-bar (i.e. two-particle) Ziegler column
with tangential loading,
analyzed in \cite{Ziegler:1968a,Nicot:2012a}
(Fig.~\ref{fig:Follower}a).
For this articulated column,
the list $[\mathbf{q}]$ is simply the
single magnitude of the follower force,
but the directed force $\mathbf{b}$ will depend upon
the rotations of the bars.
With compliant loading machines,
the forces exerted by the machinery
on peripheral particles can also depend
upon these particles' positions and orientations.
Another example is a granular assembly enclosed within a
rubber membrane that presses against the the assembly's peripheral
particles with an external confining
pressure $p$ (Fig.~\ref{fig:Follower}b).
Such membranes are commonly modeled as flat pieces of
``virtual membrane'' that apply the pressure as
discrete external forces to points $\chi$
within the peripheral particles.
The magnitudes and directions of
the membrane forces depend on the sizes and orientations of the
pieces, which depend, in turn,
upon %the locations of the centers $\chi$:
the positions $\mathbf{u}$
and orientations $\boldsymbol{\theta}$
of the particle points
%the forces on the peripheral are a function of their positions
and of the pressure $p$,
which serves as the single loading parameter in list $[\mathbf{q}]$.
\par
If the external forces $[\mathbf{b}]$ and $[\mathbf{w}]$
depend on the particles' positions $[\mathbf{u}]$ and $[\boldsymbol{\theta}]$,
as in Eq.~(\ref{eq:pGx}),
then the incremental forces will depend upon changes in these positions
and upon the $L$ loading parameters,
%then one must incoroporate this dependence in the incremental
%equilibrium of the particles,
%
\begin{equation}\label{eq:dbG}
  \left[d\mathbf{p}\right]%_{6N \times 1}
  =
%  \left[
%    \begin{array}{@{\extracolsep{\fill}}c@{\extracolsep{\fill}}}
%      d\mathbf{b} \\\hdashline[1pt/1pt]
%      d\mathbf{w}
%    \end{array}
%  \right]%_{6N \times 1}
%  =
  \left[
    \partial \mathbf{p}/\partial \mathbf{x}
  \right]_{6N\times 6N}
  \left[ d\mathbf{x}\right]%_{6N\times 1}
  \:+\:
  \left[
    \partial \mathbf{p}/\partial \mathbf{q}
  \right]_{6N\times L}
  \left[ d\mathbf{q}\right]_{L\times 1}
\end{equation}
%
%where the matrices
%$[\partial \mathbf{p}/\partial\mathbf{x}]$ and
%$[\partial \mathbf{p}/\partial\mathbf{q}]$ are multiplied by
%the displacements $[d\mathbf{x}]$ and by increments of
%the loading parameters $[d\mathbf{q}]$.
where the final term
$[\partial\mathbf{p}/\partial\mathbf{q}][d\mathbf{q}]$
gives the loading increments $[d\mathbf{p}]$ produced by
increments in the $L$ loading parameters
(e.g., increments in the confining pressure,
increments in the applied platen loads, etc.).
We designate the loading
matrix $[\partial\mathbf{p}/\partial\mathbf{q}]$
as $[\mathbf{Q}]$, and we designate
the last product in Eq.~(\ref{eq:dbG})
as the incremental loading vector $[d\mathfrak{p}]$:
\begin{equation}\label{eq:frakp}
  \left[\mathbf{Q}\right]_{6N\times L} =
  \left[
    \partial \mathbf{p}/\partial \mathbf{q}
  \right]%_{6N\times L}
  \quad\text{and}\quad
  \left[d\mathfrak{p}\right]_{6N\times 1} =
  \left[
    \partial \mathbf{p}/\partial \mathbf{q}
  \right]%_{6N\times L}
  \left[ d\mathbf{q}\right]%_{L\times 1}
  =
  \left[\mathbf{Q}\right]\left[d\mathbf{q}\right]
\end{equation}
where $[d\mathfrak{p}]$ serves as a lower-dimensional
subspace of the more general incremental loading vector
$[d\mathbf{p}]=[d\mathbf{b}/d\mathbf{w}]$.
The term $[\partial\mathbf{p}/\partial\mathbf{x}]$
in Eq.~(\ref{eq:dbG}) is one of several geometric
effects that we will consider.
%\par
%In the simplest case of dead loads that are independent of
%of positions $[\mathbf{x}]$, the relation in Eq.~(\ref{eq:pGx})
%is the identity $[\mathbf{p}]=[\mathbf{I}][\mathbf{q}]$, with $[\mathbf{q}]$
%serving as a surrogate for $[\mathbf{p}]$.
%
\subsection{\normalsize Equilibrium in initial and displaced states}\label{sec:equilibrium}
The nature of the external loads~--- whether dead loads or follower loads~---
is known to affect the stability
of structural systems.
More subtle, however, are the forces among the particles
within a granular system,
which also can act as \emph{internal follower forces},
since the directions of the inter-particle
forces can depend upon the positions and orientations of
the particles, $[\mathbf{u}]$ and $[\boldsymbol{\theta}]$.
To reveal this dependence, we analyze a system that is assumed
in equilibrium in both its initial and displaced states~--- at
times $t$ and $t+dt$.
The initial equilibrium of a particle $p$ requires
\begin{equation}\label{eq:Equilibrium0}
  -\sum_{\mathclap{c\in\mathcal{C}^{c}_{p}}}
    \mathbf{f}^{pq} =
  \mathbf{b}^{p}, \qquad
  -\sum_{\mathclap{c\in\mathcal{C}^{c}_{p}}}
    \left(
      \mathbf{m}^{pq} +
      \mathbf{r}^{pq,p}\times \mathbf{f}^{pq}
    \right) =
  \mathbf{w}^{p}
\end{equation}
in which both summations include all of $p$'s contacts $c$.
Recall that $\mathbf{r}^{pq,p}$ is the contact vector of
contact $pq$, from the reference point $\chi^{p}$ of $p$
to its contact with $q$.
The equilibrium equations of all $N$ particles are gathered
into the matrix relation
\begin{equation}\label{eq:Equilbrium0a}
  \left[\mathbf{A}\right]_{6N \times 2(6M)}
  \def\arraystretch{1.00}
  \left[
    \begin{array}{@{\extracolsep{\fill}}c@{\extracolsep{\fill}}}
      \mathbf{f} \\\hdashline[1pt/1pt]
      \mathbf{m}
    \end{array}
  \right]_{2(6M) \times 1}
  =
  \left[
    \begin{array}{@{\extracolsep{\fill}}c@{\extracolsep{\fill}}}
      \mathbf{b} \\\hdashline[1pt/1pt]
      \mathbf{w}
    \end{array}
  \right]_{6N \times 1}
  \quad\text{or}\quad
  \left[\mathbf{A}\right]%_{6N \times (6M)}
  \left[\,\boldsymbol{\mathfrak{f}}\,\right]
  =
  \left[\mathbf{p}\right]
\end{equation}
where $[\mathbf{A}]$ is the statics matrix.
At $t+dt$,
the new positions, orientations, and forces are
the sums
$\mathbf{u}^{p}+d\mathbf{u}^{p}$,
$\mathbf{b}^{p}+d\mathbf{b}^{p}$,
$\mathbf{x}^{p}+d\mathbf{x}^{p}$,
$\mathbf{r}^{pq}+d\mathbf{r}^{pq}$,
$\mathbf{f}^{pq}+d\mathbf{f}^{pq}$,
etc.
By substituting these sums in Eq.~(\ref{eq:Equilibrium0})
and then subtracting Eq.~(\ref{eq:Equilibrium0}),
we find the incremental conditions for continued equilibrium:
\begin{equation}\label{eq:Equilibrium1}
  -\sum_{\mathclap{c\in\mathcal{C}^{c}_{p}}}
    d\mathbf{f}^{pq} =
  d\mathbf{b}^{p}, \qquad
  -\sum_{\mathclap{c\in\mathcal{C}^{c}_{p}}}
    \left(
      d\mathbf{m}^{pq} +
      \mathbf{r}^{pq,p}\times d\mathbf{f}^{pq} +
      d\mathbf{r}^{pq,p}\times\mathbf{f}^{pq}
    \right)=
  d\mathbf{w}^{p}
\end{equation}
Note that the statics matrix $[\mathbf{A}]$ depends
upon the positions and orientations of the particles and
contacts, $[\mathbf{A}]=[\mathbf{A}(\mathbf{x})]$,
and is altered by the movements $d\mathbf{x}$.
With this understanding, Eq.~(\ref{eq:Equilibrium0})
is equivalent
to the differential of Eq.~(\ref{eq:Equilbrium0a}),
when applied to all particles of the system:
\begin{equation}\label{eq:parAequilibrium}
  \left[\rule{0ex}{2.2ex}(\partial\mathbf{A}/\partial\mathbf{x})\cdot d\mathbf{x}\right]
  \def\arraystretch{1.00}
  \left[
    \begin{array}{@{\extracolsep{\fill}}c@{\extracolsep{\fill}}}
      \mathbf{f} \\\hdashline[1pt/1pt]
      \mathbf{m}
    \end{array}
  \right]%_{(6M) \times 1}
  +
  \left[\mathbf{A}\right]
  \def\arraystretch{1.00}
  \left[
    \begin{array}{@{\extracolsep{\fill}}c@{\extracolsep{\fill}}}
      d\mathbf{f} \\\hdashline[1pt/1pt]
      d\mathbf{m}
    \end{array}
  \right]%_{(6M) \times 1}
  =
  \def\arraystretch{1.00}
  \left[
    \begin{array}{@{\extracolsep{\fill}}c@{\extracolsep{\fill}}}
      d\mathbf{b} \\\hdashline[1pt/1pt]
      d\mathbf{w}
    \end{array}
  \right]%_{(6M) \times 1}
  \quad\text{or}\quad
  \left[\rule{0ex}{2.2ex}(\partial\mathbf{A}/\partial\mathbf{x})\cdot d\mathbf{x}\right]
  \left[\,\boldsymbol{\mathfrak{f}}\,\right]
  +
  \left[\mathbf{A}\right]
  \left[\,d\boldsymbol{\mathfrak{f}}\,\right]
  =\left[ d\mathbf{p}\right]
\end{equation}
%
%where $[d\mathbf{x}]=[d\mathbf{u}/d\boldsymbol{\theta}]$.
The first term on the left can be expressed with index
notation as $A_{ij,k}dx_{k}\mathfrak{f}_{j}$,
where $\mathfrak{f}_{j}$ is an element of vector
$[\boldsymbol{\mathfrak{f}} ] = [\mathbf{f}/\mathbf{m}]$.
As was noted near the start of Section~\ref{sec:notation},
a granular system can include non-contacting rattler particles, 
which can become incorporated in the the load-bearing network
of particles during $dt$, even as some load-bearing particles
become disengaged rattlers.
If we choose to model the entire system of particles~---
both load-bearing and rattler~--- and allow all possible
changes to the contact topology, then matrix $[\mathbf{A}]$
must model the \emph{complete graph} of size $N(N-1)\times N(N-1)$,
so that all potential contacts are considered.
We prefer, however, to model only those contacts that exist
at time $t$, so that $[\mathbf{A}]$ must occasionally be altered
as new contacts are established and existing contacts are broken.
In this approach, the derivative in Eq.~(\ref{eq:parAequilibrium})
does not account for these abrupt changes in the contact topology.
\par
The various differential quantities,
$d\mathbf{r}^{pq}$, $d\mathbf{f}^{pq}$, etc.,
depend upon the movements
$[d\mathbf{u}]$ and $[d\boldsymbol{\theta}]$.
These movement are assumed small when compared with the
particles' sizes,
so that the relations between movement and force
can be linearized in the vicinity of time $t$.
In this section we derive these linear stiffness relationships
between increments of movements and increments of external force:
\begin{equation}\label{eq:Hdudb}
  \left[
    \mathbf{H}
  \right]_{6N\times 6N}
  \def\arraystretch{1.00}
  \left[
    \begin{array}{@{\extracolsep{\fill}}c@{\extracolsep{\fill}}}
      d\mathbf{u} \\\hdashline[1pt/1pt]
      d\boldsymbol{\theta}
    \end{array}
  \right]_{6N \times 1}
  =
  \left[
    \begin{array}{@{\extracolsep{\fill}}c@{\extracolsep{\fill}}}
      d\mathbf{b} \\\hdashline[1pt/1pt]
      d\mathbf{w}
    \end{array}
  \right]_{6N \times 1}
  \quad\text{or}\quad
  \left[
    \mathbf{H}
  \right]
  \left[
    d\mathbf{x}
  \right]_{6N\times 1}
  =
  \left[
    d\mathbf{p}
  \right]_{6N\times 1}
\end{equation}
The stiffness matrix $[\mathbf{H}]$ is shown to
be the sum of several parts,
with each part having either a
mechanical or geometric origin.
Equation~(\ref{eq:Hdudb}) is central in addressing
questions of uniqueness and controllability,
although a rearrangement of the parts of $[\mathbf{H}]$
is required to address stability (Section \ref{sec:stability}).
%whereas issues of stability and sustainability are addressed with
%the alternative approach of Section~\ref{sec:stability},
%which involves a rearrangement of the
%various parts of $[\mathbf{H}]$.
Near the end of this section (Eq.~\ref{eq:Hdxfrakp}),
we will replace
Eq.~(\ref{eq:Hdudb}) with a more general form,
by replacing the loading
vector $[d\mathbf{p}]=[d\mathbf{b}/d\mathbf{w}]$
with a vector $[d\mathfrak{p}]$ that depends on the $L$
loading parameters $[\mathbf{q}]$, as in Eq.~(\ref{eq:pGx}).
\par
As was noted, the incremental quantities in Eq.~(\ref{eq:Equilibrium1})
depend upon the movements
$d\mathbf{u}^{p}$ and $d\boldsymbol{\theta}^{p}$ of particle $p$
and of its neighboring particles.
This dependence presents a difficulty when deriving
the stiffness in Eq.~(\ref{eq:Hdudb}) for an entire assembly:
the particles within the assembly will likely rotate at different rates
and in different directions,
yet one must find the assembly's
stiffness $[\mathbf{H}]$
relative to a stationary coordinate frame.
Kuhn and Chang \cite{Kuhn:2005b}
rewrote the incremental equilibrium equations of particle $p$
with an intermediate set of equations that
use objective, co-rotated
``$\delta$'' increments that are referenced to the
particle's rotation.
In general, an increment $d\mathbf{y}$ is related
to its \emph{co-rotated} increment $\delta\mathbf{y}$ as
\begin{equation}\label{eq:corotate}
d\mathbf{y} = \delta\mathbf{y} + d\boldsymbol{\theta}^{p}\times\mathbf{y}
\end{equation}
The increment $\delta\mathbf{y}$ is objective, since
two observers who are rotating relative to each other (for example,
one attached to a neighboring particle $q$ and the other attached to a boundary
platen)
would observe different increments
$d\mathbf{y}$ and $d\boldsymbol{\theta}^{p}$,
but they would both compute the same $\delta\mathbf{y}$.
The increment $\delta\mathbf{y}$ is simply
the change in $\mathbf{y}$
seen by an observer attached to (and rotating with) the particle $p$,
as this observer would observe no rotation $d\boldsymbol{\theta}^{p}$.
In the following, we will derive incremental stiffnesses
in the ``$\delta$'' systems of individual contacts,
and then later convert this disparate system
into the common ``$d$'' system shared
by all particles in the assembly.
\par
As an intermediate step, we rewrite the equilibrium equations for
particle $p$ in terms of objective $\delta$-differentials:
\begin{equation}\label{eq:Equilibrium1b}
  -\sum_{\mathclap{c\in\mathcal{C}^{c}_{p}}}
    \delta\mathbf{f}^{pq} =
  \delta\mathbf{b}^{p}, \qquad
  -\sum_{\mathclap{c\in\mathcal{C}^{c}_{p}}}
    \left(
      \delta\mathbf{m}^{pq} +
      \mathbf{r}^{pq,p}\times \delta\mathbf{f}^{pq} +
      \delta\mathbf{r}^{pq,p}\times\mathbf{f}^{pq}
    \right)=
  \delta\mathbf{w}^{p}
\end{equation}
which are shown in the appendix of \cite{Kuhn:2005b} to be equivalent to
Eq.~(\ref{eq:Equilibrium1}).
This form permits the simpler analysis of increments
$\delta\mathbf{f}^{pq}$, $\delta\mathbf{m}^{pq}$,
and $\delta\mathbf{r}^{pq,p}$.
Applying Eq.~(\ref{eq:corotate}), the increments $d\mathbf{f}^{pq}$
and $d\mathbf{m}^{pq}$ are the sums
\begin{alignat}{2}\label{eq:dfpqall}
  d\mathbf{f}^{pq} &= \delta\mathbf{f}^{pq} + d\boldsymbol{\theta}^{p}\times\mathbf{f}^{pq}
                   &&= \mathfrak{d}\mathbf{f}^{pq}
                     + \delta\hat{\mathbf{f}}^{pq}
                     + d\boldsymbol{\theta}^{p}\times\mathbf{f}^{pq}\\
  \label{eq:dmpqall}
  d\mathbf{m}^{pq} &= \delta\mathbf{m}^{pq} + d\boldsymbol{\theta}^{p}\times\mathbf{m}^{pq}
                   &&= \mathfrak{d}\mathbf{m}^{pq}
                     + \delta\hat{\mathbf{m}}^{pq}
                     + d\boldsymbol{\theta}^{p}\times\mathbf{m}^{pq}
\end{alignat}
where the ``$d$'' increments are the
sum of ``$\,\mathfrak{d}$'' and ``$\,\delta\:\widehat{\ }\:$'' parts.
These parts are described below, but briefly,
the ``$\,\mathfrak{d}$'' increments are produced
by contact deformation;
whereas, the ``$\,\delta\:\widehat{\ }\:$'' increments
result from the rolling and twirling of the
full contact forces and moments.
Because the force increments
$\delta\mathbf{f}^{pq}$ and $\delta\mathbf{f}^{qp}$
are viewed by observers attached to the two different particles
(as are $\delta\mathbf{m}^{pq}$ and $\delta\mathbf{m}^{qp}$),
$\delta\mathbf{f}^{pq}$ and $\delta\mathbf{m}^{pq}$ are
not necessarily equal to the negatives of their
counterparts,
$-\delta\mathbf{f}^{qp}$ and $-\delta\mathbf{m}^{qp}$.
It is for this reason that we distinguish the $pq$ and $qp$
variants of a contact, which leads to assembly vectors and
matrices of size $2(6M)$,
as in Eqs.~(\ref{eq:gather}\textsubscript{3})
and~(\ref{eq:gather2}\textsubscript{3}).
\par
By substituting Eqs.~(\ref{eq:dfpqall})--(\ref{eq:dmpqall}),
Eqs.~(\ref{eq:Equilibrium1})
and~(\ref{eq:Equilibrium1b})
can be arranged with the multiple contributions that
produce the external force and moment increments,
$d\mathbf{b}^{p}$ and $d\mathbf{w}^{p}$,
of Eq.~(\ref{eq:Hdudb}),
\begin{align} \label{eq:db1}
  &-\sum_{\mathclap{c\in\mathcal{C}^{c}_{p}}}
  \left(
  \underbrace{\mathfrak{d}\mathbf{f}^{pq}}_{\text{m}}
  +
  \underbrace{\delta\hat{\mathbf{f}}^{pq}}_{\text{g-2}}
  +
  \underbrace{d\boldsymbol{\theta}^{p} \times \mathbf{f}^{pq}}_{\text{g-3}}
  \right)
  =
  d\mathbf{b}^{p}
  \\ \label{eq:dw1}
  &
  \begin{aligned}
  -\sum_{\mathclap{c\in\mathcal{C}^{c}_{p}}}
  &\left(\rule{0ex}{2.5ex}
  \underbrace{\mathfrak{d}\mathbf{m}^{pq}
              + \mathbf{r}^{pq,p}\times \mathfrak{d}\mathbf{f}^{pq}}_{\text{m}}
  +
  \underbrace{\delta\mathbf{r}^{pq,p} \times \mathbf{f}^{pq}}_{\text{g-1}}
  \right.
  \\
  &\left.
  \quad +
  \underbrace{\delta\hat{\mathbf{m}}^{pq}
              +\; \mathbf{r}^{pq,p}\times \delta\hat{\mathbf{f}}^{pq}}_{\text{g-2}}
  +
  \underbrace{d\boldsymbol{\theta}^{p} \times \mathbf{m}^{pq}}_{\text{g-3}}
  \right)
  =
  d\mathbf{w}^{p}
  \end{aligned}
\end{align}
The terms on the left of these equilibrium equations contribute to
four parts of the stiffness matrix $[\mathbf{H}]$:
a mechanical part $[\mathbf{H}^{\text{m}}]$ and
three geometric stiffness parts, $[\mathbf{H}^{\text{g-1}}]$,
$[\mathbf{H}^{\text{g-2}}]$, and $[\mathbf{H}^{\text{g-3}}]$.
Furthermore,
a dependence of the external forces on the particles'
positions, as in Eq.~(\ref{eq:pGx}), can produce a fourth
geometric stiffness $[\mathbf{H}^{\text{g-4}}]$.
We now consider these separate parts of
the assembly stiffness.
\par
\begin{figure}
  \centering
  \includegraphics{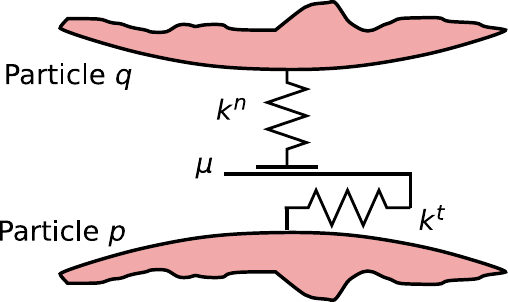}
  \caption{Spring--slider system at the contact between particles $p$ and $q$.
           \label{fig:Springs}}
\end{figure}

The mechanical stiffness $[\mathbf{H}^{\text{m}}]$ arises from
changes in the contact forces and moments,
$\mathfrak{d}\mathbf{f}^{pq}$ and $\mathfrak{d}\mathbf{m}^{pq}$,
that are produced
by deformations of the grains at their contacts.
Although small, the contact deformations produce a non-rigid,
compliant bulk behavior, and such granular systems are termed
``quasirigid'' in the physics community \cite{McNamara:2006a}.
In DEM simulations, the contact deformations are idealized as the
movements of contact ``springs and sliders,''
which alter the contact
forces, as in the system of Fig.~\ref{fig:Springs}.
These local mechanical force increments, as well as the entire
mechanical stiffness $[\mathbf{H}^{\text{m}}]$, depend,
of course, on the contacts' stiffnesses.
The stiffness of a contact $pq$ is expressed with
a force--displacement mapping that
gives the objective increment
of contact force $\mathfrak{d}\mathbf{f}^{pq}$
(and moment $\mathfrak{d}\mathbf{m}^{pq}$) as a function
of the objective increments of
the relative contact displacement $\delta\mathbf{u}^{\text{def},pq}$
(and the relative contact rotation
$\delta\boldsymbol{\theta}^{\text{def},pq}$).
The relative displacement and relative rotation
will deform the two particles
at their contact and are defined by the
kinematic relations
\begin{align}\label{eq:dudef}
  \delta\mathbf{u}^{\text{def},pq} &=
  d\mathbf{u}^{q} - d\mathbf{u}^{p} +
  \left(
    d\boldsymbol{\theta}^{q}\times\mathbf{r}^{pq,q} -
    d\boldsymbol{\theta}^{p}\times\mathbf{r}^{pq,p}
  \right) \\ \label{eq:dthetadef}
  \delta\boldsymbol{\theta}^{\text{def},pq} &=
  d\boldsymbol{\theta}^{q} - d\boldsymbol{\theta}^{p}
\end{align}
where $\mathbf{r}^{pq,p}$ and $\mathbf{r}^{pq,q}$ are the contact vectors
from the material points $\chi^{p}$ and $\chi^{q}$ to the contact point $c$
(note that $\delta\mathbf{u}^{\text{def},pq}=-\delta\mathbf{u}^{\text{def},qp}$
and
$\delta\boldsymbol{\theta}^{\text{def},pq}=-\delta\boldsymbol{\theta}^{\text{def},qp}$).
%These relative displacements are kinematically related
%to the particle movements,
%as in Eqs.~(\ref{eq:dudef}) and (\ref{eq:dthetadef}),
%which can be gathered into the matrix form
These relations between contact movements and particle movements can
be gathered into the matrix form
\begin{equation}\label{eq:Bmatrix}
  \def\arraystretch{1.20}
  \left[
    \begin{array}{@{\extracolsep{\fill}}c@{\extracolsep{\fill}}}
      \delta\mathbf{u}^{\text{def}} \\\hdashline[1pt/1pt]
      \delta\boldsymbol{\theta}^{\text{def}}
    \end{array}
  \right]_{2(6M) \times 1}
  =
  \left[ \mathbf{B} \right]_{2(6M)\times 6N}
  \def\arraystretch{1.00}
  \left[
    \begin{array}{@{\extracolsep{\fill}}c@{\extracolsep{\fill}}}
      d\mathbf{u} \\\hdashline[1pt/1pt]
      d\boldsymbol{\theta}
    \end{array}
  \right]_{6N \times 1}
\end{equation}
where $[\mathbf{B}]$ is the kinematic matrix
(or the rigidity matrix, as in \cite{Agnolin:2007a,Agnolin:2007c}),
and the ``$\delta$'' vector on the left is of size $2(6)$
and contains both
$pq$ and $qp$ variants.
As a condition of equilibrium, $[\mathbf{A}]$ and $[\mathbf{B}]$
are dual, with $[\mathbf{A}]=[\mathbf{B}]^{\text{T}}$.
\par
Both $\delta\mathbf{u}^{\text{def},pq}$ and
$\delta\boldsymbol{\theta}^{\text{def},pq}$ are objective quantities
and, as such, can be
used in computing the objective force
increments $\mathfrak{d}\mathbf{f}^{pq}$ and $\mathfrak{d}\mathbf{m}^{pq}$.
We assume a surjective mapping from
the full $\mathbb{R}^{6}$ space of a contact's incremental
deformation
(in 3D, the 3-component %relative displacement
$\delta\mathbf{u}^{\text{def},pq}$ and the 3-component
%relative rotation
$\delta\boldsymbol{\theta}^{\text{def},pq}$)
into the possibly smaller space of incremental contact force and moment
(the 3-component
%contact force increments
$\mathfrak{d}\mathbf{f}^{pq}$
%in Eq.~\ref{eq:W2a}
and 3-component
%contact moment complement,
$\mathfrak{d}\mathbf{m}^{pq}$).
This condition excludes Signorini and rigid-frictional models
of contact behavior.
We will assume, however, that the force--displacement relation
is rate-independent, so that the mapping of displacement to
force is homogeneous of degree one with respect to
the relative contact displacement
$\delta\mathbf{u}^{\text{def},pq}$
and rotation $\delta\boldsymbol{\theta}^{\text{def},pq}$,
perhaps in the restricted functional form
\begin{align}\label{eq:constitutive}
  \mathfrak{d}\mathbf{f}^{pq}
  &=
  \mathbf{F}^{pq}
  \left(
  \frac{\delta\mathbf{u}^{\text{def},pq}}
       {|\delta\mathbf{u}^{\text{def},pq} |}\,,\, \mathbf{f}^{pq}
  \right)
  \cdot\delta\mathbf{u}^{\text{def},pq}
  \\[0.5ex]
  \label{eq:constitutiveM}
  \mathfrak{d}\mathbf{m}^{pq}
  &=
  \mathbf{M}^{pq}
  \left(
  \frac{\delta\boldsymbol{\theta}^{\text{def},pq}}
       {|\delta\boldsymbol{\theta}^{\text{def},pq} |}\,,\, \mathbf{m}^{pq}
  \right)
  \cdot\delta\boldsymbol{\theta}^{\text{def},pq}
\end{align}
noting that the stiffness
matrices $\mathbf{F}^{pq}$ and $\mathbf{M}^{pq}$ might
depend upon the direction of the relative displacement
(and rotation) and on the current contact force
(and moment).
That is, in contrast with elasticity or with the smooth
hypoelasticity of
Truesdell \cite{Truesdell:1963a,Truesdell:2004a},
the incremental force relation in
Eq.~(\ref{eq:constitutive}) is non-smooth, as it
depends % on both the current force $\mathbf{f}^{pq}$ and
on the movement $\delta\mathbf{u}^{\text{def},pq}$
\emph{and} its direction
$\delta\mathbf{u}^{\text{def},pq}/|\delta\mathbf{u}^{\text{def},pq}|$.
The force--displacement can be irreversible, and
an example two-branch frictional contact model is reviewed
in Section~\ref{sec:contacts} and is applied in the examples
of Section~\ref{sec:threedisks}.
Other contact models, however, are even more general 
than Eqs.~(\ref{eq:constitutive})--(\ref{eq:constitutiveM}):
for example, the force increment
in a Cattaneo--Mindlin contact depends
on the entire history of the force $\mathbf{f}^{pq}$ and not
just on its current value \cite{Mindlin:1953a}.
\par
Also embedded in
Eqs.~(\ref{eq:constitutive})--(\ref{eq:constitutiveM})
is an assumption of locality in the contacts' behaviors
(e.g. \cite{Truesdell:2004a}, \S26):
the force of a contact $pq$ depends only on the movement
$\delta\mathbf{u}^{\text{def},pq}$
of this contact and not on movements at other contacts
within an assembly.
This assumption allows the assembly of these relations
into a block-diagonal matrix of contact stiffnesses,
\begin{equation}
  \left[
    \begin{array}{@{\extracolsep{\fill}}c@{\extracolsep{\fill}}}
      \mathbf{F}^{pq}_{\;3\times 6} \\[0.1ex]\hdashline[1pt/1pt]
      \mathbf{M}^{pq}_{\;3\times 6}
    \end{array}
  \right]_{6\times 6}
  \rightsquigarrow
  \def\arraystretch{1.00}
  \left[
    \begin{array}{@{\extracolsep{\fill}}c@{\extracolsep{\fill}}}
      \mathbf{F} \\\hdashline[1pt/1pt]
      \mathbf{M}
    \end{array}
  \right]_{2(6M)\times 2(6M)}
\end{equation}
%
%The stiffness relations of all $M$ contacts are collected
%as the matrix relation
such that
\begin{equation}\label{eq:dfrakB}
  \def\arraystretch{1.00}
  \left[
    \begin{array}{@{\extracolsep{\fill}}c@{\extracolsep{\fill}}}
      \mathfrak{d}\mathbf{f} \\\hdashline[1pt/1pt]
      \mathfrak{d}\mathbf{m}
    \end{array}
  \right]_{2(6M)\times 1}
  =
%  \left[\begin{MAT}(r){l}
%        \mathfrak{d}\mathbf{f}\\:\mathfrak{d}\mathbf{m}\\
%        \end{MAT}\right]_{(6M)\times 1} =
  \def\arraystretch{1.00}
  \left[
    \begin{array}{@{\extracolsep{\fill}}c@{\extracolsep{\fill}}}
      \mathbf{F} \\\hdashline[1pt/1pt]
      \mathbf{M}
    \end{array}
  \right]%_{(6M)\times (6M)}
%  \left[\begin{MAT}(r){l}
%        \delta\mathbf{u}^{\text{def}}\\:
%        \delta\boldsymbol{\theta}^{\text{def}}\\
%        \end{MAT}\right]%_{(6M)\times 1}
  \def\arraystretch{1.00}
  \left[
    \begin{array}{@{\extracolsep{\fill}}c@{\extracolsep{\fill}}}
      \delta\mathbf{u}^{\text{def}} \\\hdashline[1pt/1pt]
      \delta\boldsymbol{\theta}^{\text{def}}
    \end{array}
  \right]%_{(6M)\times 1}
  =
  \def\arraystretch{1.00}
  \left[
    \begin{array}{@{\extracolsep{\fill}}c@{\extracolsep{\fill}}}
      \mathbf{F} \\\hdashline[1pt/1pt]
      \mathbf{M}
    \end{array}
  \right]%_{(6M)\times (6M)}
  \left[\mathbf{B}\right]
  \def\arraystretch{1.00}
  \left[
    \begin{array}{@{\extracolsep{\fill}}c@{\extracolsep{\fill}}}
      d\mathbf{u} \\\hdashline[1pt/1pt]
      d\boldsymbol{\theta}
    \end{array}
  \right]%_{6N \times 1}
\end{equation}
noting again that the contents of
$[\mathbf{F}/\mathbf{M}]$
may depend upon the current contact forces, $\mathbf{f}$ and
$\mathbf{m}$,
and on the \emph{directions}
of the incremental contact deformations,
$\delta\mathbf{u}^{pq,\text{def}}$
and $\delta\boldsymbol{\theta}^{pq,\text{def}}$.
%statics matrix $[\mathbf{A}]$,
\par
Combining Eqs.~(\ref{eq:Equilbrium0a}), (\ref{eq:dfpqall}),
(\ref{eq:dmpqall}), and~(\ref{eq:dfrakB}) yields the
mechanical stiffness
$[\mathbf{H}^{\text{m}}]$, which gives the contributions of
the contact forces,
$\mathfrak{d}\mathbf{f}$ and $\mathfrak{d}\mathbf{m}$,
to the external forces,
$d\mathbf{b}$ and $d\mathbf{w}$:
\begin{equation}\label{eq:Hmprod}
  \left[ \mathbf{H}^{\text{m}} \right]_{6N\times 6N}
  =
  \left[ \mathbf{A} \right]_{6N\times 2(6M)}
  \def\arraystretch{1.00}
  \left[
    \begin{array}{@{\extracolsep{\fill}}c@{\extracolsep{\fill}}}
      \mathbf{F} \\\hdashline[1pt/1pt]
      \mathbf{M}
    \end{array}
  \right]_{2(6M)\times 2(6M)}
  \left[ \mathbf{B} \right]_{2(6M)\times 6N}
\end{equation}
%where $[\mathbf{A}_{1}] = [\mathbf{B}]^{\text{T}}$ is the
%statics matrix.
in which the contact forces
are collected with the statics matrix $[\mathbf{A}]$,
as in Eq.~(\ref{eq:Equilbrium0a}).
\par
\begin{figure}
  \centering
  \includegraphics{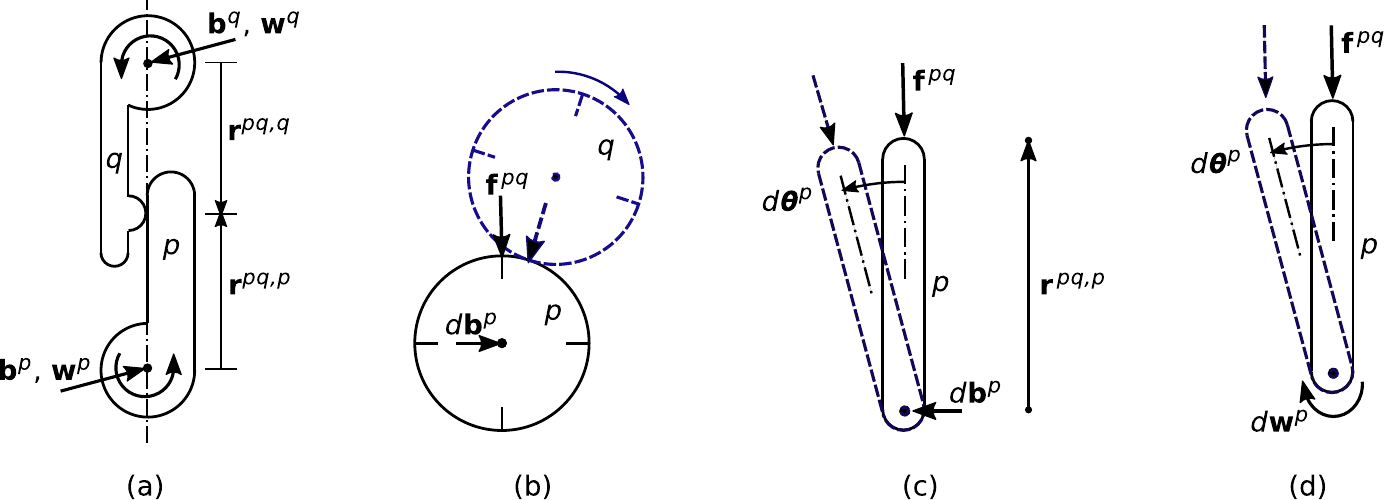}
  \caption{Geometric effects on internal forces:
           (a)~alterations of $\mathbf{r}^{pq}$
           and $\mathbf{r}^{qp}$ depend upon the curvatures
           of the particles at their contacts
           (see the discussion that precedes
           Eq.~\ref{eq:delrpq});
           (b)~rolling of particle $q$ across $p$ alters
           the direction of the contact force
           (see the discussion that precedes
           Eq.~\ref{eq:dfhat});
           (c)~rotations of $p$ and its contact force
           $\mathbf{f}^{pq}$ require an alteration of the
           external force, $d\mathbf{b}^{p}$
           (see the discussion that precedes
           Eq.~\ref{eq:Hg3}); and
           (d)~a rotation of $p$ but an unrotated contact
           force $\mathbf{f}^{pq}$ requires an alteration
           of the external moment, $d\mathbf{w}^{p}$
           (see the discussion that precedes
           Eq.~\ref{eq:Hg3}).
           \label{fig:g2}}
\end{figure}

The geometric ``g-1'' term in the equilibrium Eq.~(\ref{eq:dw1})
%and an entire assembly's $[\mathbf{H}^{g-1}]$ stiffness
arises from changes in the
contact vector $\delta\mathbf{r}^{pq,p}$ when its contact point
shifts across the particle $p$.
This effect is illustrated in Fig.~\ref{fig:g2}a,
in which the flat surface of $p$ touches the round surface
of $q$.
The contact vector $\mathbf{r}^{pq,p}$ is lengthened when $q$ moves
upward; whereas, vector $\mathbf{r}^{pq,q}$ is unchanged.
These changes depend upon the curvatures of the two particles at
their contact (e.g., with an opposite counterpart of
Fig.~\ref{fig:g2}a, in which $q$ is flat and $p$ is rounded,
the $\mathbf{r}^{pq,q}$ would change with an upward
movement of $q$, but $\mathbf{r}^{pq,p}$ would not, thus
indicating the non-symmetry of this geometric effect).
The increment $\delta\mathbf{r}^{pq,p}$ is a
sum of normal and tangential parts:
\begin{align}\label{eq:delrpq}
  \delta\mathbf{r}^{pq,p} &=
  \delta\mathbf{r}^{\text{n},pq,p} +
  \delta\mathbf{r}^{\text{t},pq,p} \\
  \delta\mathbf{r}^{\text{n},pq,p} &=
  \frac{1}{2}\left(
    \delta\mathbf{u}^{\text{def},pq}
    \cdot
    \mathbf{n}^{pq}
  \right)
  \mathbf{n}^{pq} \\
  \label{eq:drKPKQ}
  \delta\mathbf{r}^{\text{t},pq,p} &=
  -\left(
    \mathbf{K}^{pq,p} + \mathbf{K}^{pq,q}
  \right)^{\dagger} \cdot
  \left[
    \delta\boldsymbol{\theta}^{\text{def},pq} \times \mathbf{n}^{pq}
    - \mathbf{K}^{pq,q}\cdot
      \left(
        \delta\mathbf{u}^{\text{def},pq}
        - \left(
          \delta\mathbf{u}^{\text{def},pq} \cdot
          \mathbf{n}^{pq}
        \right) \mathbf{n}^{pq}
      \right)
  \right]
\end{align}
where $\mathbf{n}^{pq}$ is the contact normal vector
directed outward from $p$ (Fig.~\ref{fig:ContactVectors}),
and
$\mathbf{K}^{pq,p}$ and $\mathbf{K}^{pq,q}$ are the curvatures
of the two particles' surfaces at their shared contact point $pq$
(see \cite{Kuhn:2004b,Kuhn:2005b}).
For the $pq$ and $qp$ variants of a contact,
$\mathbf{n}^{pq}=-\mathbf{n}^{pq}$,
$\mathbf{K}^{pq,p}=\mathbf{K}^{qp,p}$, and
$\mathbf{K}^{qp,q}=\mathbf{K}^{qp,q}$.
The curvature matrices are singular, as they are surjective
mappings from the three-dimensional space of contact movements onto
the two-dimensional contact tangent plane,
and a generalized inverse, such as
the Moore-Penrose ``$\,\dagger\,$'' inverse must be used
in Eq.~(\ref{eq:drKPKQ}) \cite{Kuhn:2004b}.
The ``g-1'' terms in Eq.~(\ref{eq:dw1}) only affect
moment equilibrium and depend linearly on
the movements $d\mathbf{u}^{p}$, $d\mathbf{u}^{q}$,
$d\boldsymbol{\theta}^{p}$, and $d\boldsymbol{\theta}^{q}$.
These terms can be gathered into a contact stiffness
$[\mathbf{H}^{\text{g-1}}]$:
\begin{equation}\label{eq:Hg1}
  \left.\begin{array}{l}
  \mathbf{0}\\
  -\delta\mathbf{r}^{pq,p} \times \mathbf{f}^{pq}
  \end{array}\right\}_{6\times 1}
%  \delta\mathbf{r}^{pq} \times \mathbf{f}^{pq}
  \rightsquigarrow
  \left[\mathbf{H}^{\text{g-1}}\right]_{6N\times 6N}
  \left[
    \begin{array}{@{\extracolsep{\fill}}c@{\extracolsep{\fill}}}
      d\mathbf{u} \\\hdashline[1pt/1pt]
      d\boldsymbol{\theta}
    \end{array}
  \right]_{6N \times 1}
\end{equation}
where the stiffness gives changes
%$[\mathbf{H}^{\text{g-1}}]$ gives changes
in the forces at the $M$ contacts
produced by movements of the $N$ particles.
The $3\times 1$ zero vector on the left represents the
nil contribution of this geometric effect on the force
equilibrium of $p$.
The unsymmetric arrangement on the left of
Eq.~(\ref{eq:Hg1}) leads to a non-symmetric
stiffness $[\mathbf{H}^{\text{g-1}}]$.
%The contact changes are multiplied by the statics matrix
%$[\mathbf{A}]$ to collect the contact forces
%into particles forces
%(i.e., the summations in Eqs.~\ref{eq:Equilibrium0} and~\ref{eq:Equilbrium0a}),
%resulting in the first geometric stiffness $[\mathbf{H}^{\text{g-1}}]$:
%
%\begin{equation}\label{eq:Hg2prod}
%  \left[ \mathbf{H}^{\text{g-1}}\right]_{6N\times 6N}
%  =
%   \left[ \mathbf{A}\right]_{6N\times 2(6M)}
%   \left[ \mathbf{E}^{\text{g-1}}\right]_{2(6M)\times 6N}
%\end{equation}
%
\par
Comparing Eqs.~(\ref{eq:parAequilibrium}) and~(\ref{eq:Hg1}),
we see that $[\mathbf{H}^{\text{g-1}}]$ expresses the effect
of changes in the static matrix upon the incremental
equilibrium:
\begin{equation}\label{eq:Hg1Af}
  \left[\mathbf{H}^{\text{g-1}}\right]
  \left[d\mathbf{x}\right]
  =
  \left[\rule{0ex}{2.2ex}(\partial\mathbf{A}/\partial\mathbf{x})\cdot d\mathbf{x}\right]
  \left[\,\boldsymbol{\mathfrak{f}}\,\right]
  \quad\text{or}\quad
  H_{ij}^{\text{g-1}} dx_{j}
  =
  A_{ik,j}\mathfrak{f}_{k} dx_{j}
\end{equation}
with $H_{ij}^{\text{g-1}}=A_{ik,j}\mathfrak{f}_{k}$.
\par
The incremental change in the contact vector
$\delta\mathbf{r}^{pq,p}$ in Eq.~(\ref{eq:delrpq})
is composed of two parts:
a normal increment $\delta\mathbf{r}^{\text{n},pq,p}$
and a tangential increment $\delta\mathbf{r}^{\text{t},pq,p}$.
Considering the two parts, the geometric stiffness
$[\mathbf{H}^{\text{g-1}}]$ that originates from the normal
increment will usually be insignificant,
since the stiffness that is associated with this part
is of order $|\boldsymbol{\mathfrak{f}}^{pq}|/(k\,|\mathbf{r}^{pq,p}|)$,
where $k$ is the contact stiffness
(Section~{\ref{sec:contacts}) and
$\boldsymbol{\mathfrak{f}}^{pq}$ is the contact force.
This ratio is usually quite small for most materials
(an exception might be gel-like particles with very soft
contacts).
However,
the stiffness that originates from the tangential
increment $\delta\mathbf{r}^{\text{t},pq,p}$ is significant
whenever rolling movements are large.
\par
The ``g-2'' terms in Eqs.~(\ref{eq:db1}) and~(\ref{eq:dw1})
account for the changes,
$\delta\hat{\mathbf{f}}^{pq}$ and $\delta\hat{\mathbf{m}}^{pq}$,
in the contact forces %$\mathbf{f}^{pq}$ and $\mathbf{m}^{pq}$
that are produced by rotations of the contact normal $d\mathbf{n}^{pq}$.
In Fig.~\ref{fig:g2}b, particle $q$ rolls across $p$ while maintaining the magnitude
of its contact force, although the rolling of $q$
rotates the force.
Even though the relative contact
movements $\delta\mathbf{u}^{\text{def},pq}$
and $\delta\boldsymbol{\theta}^{\text{def},pq}$ are zero
during a rolling movement
(thus producing no deformation forces
$\mathfrak{d}\mathbf{f}^{pq}$ and $\mathfrak{d}\mathbf{m}^{pq}$),
the contact force is altered by its rotation,
and this change in the contact force must be counteracted
by a change in the external force, $d\mathbf{b}^{p}$.
The internal force $\mathbf{f}^{pq}$ is a circulatory, follower
force, and one would expect the stiffness matrix $[\mathbf{H}]$ of
this two-particle system to be non-symmetric.
A similar condition is produced between two particles that twirl
and waltz as a joined pair, since their contact force
will rotate in unison with the pair's motions.
These force alterations are expressed as
\begin{align} \label{eq:dfhat}
  \delta\hat{\mathbf{f}}^{pq} &=
    \mathbf{f}^{pq} \times
      \left( \delta\mathbf{n}^{pq} \times \mathbf{n}^{pq} \right)
    - \frac{1}{2}\left(
                 \delta\boldsymbol{\theta}^{\text{def},pq}
                 \cdot \mathbf{n}^{pq}\right) \mathbf{f}^{pq}
                 \times \mathbf{n}^{pq}
  \\ \label{eq:dmhat}
  \delta\hat{\mathbf{m}}^{pq} &=
    \mathbf{m}^{pq} \times
      \left( \delta\mathbf{n}^{pq} \times \mathbf{n}^{pq} \right)
    - \frac{1}{2}\left(
                 \delta\boldsymbol{\theta}^{\text{def},pq}
                 \cdot \mathbf{n}^{pq}\right) \mathbf{m}^{pq}
                 \times \mathbf{n}^{pq}
\end{align}
where rolling produces a rotation of the contact normal
$\delta\mathbf{n}^{pq}$ (as seen from the perspective
of an observer attached to $p$) that depends
upon the particles' curvatures \cite{Kuhn:2004b},
\begin{equation} \label{eq:deltan}
  \delta\mathbf{n}^{pq} =
  - \mathbf{K}^{pq,p} \cdot
    \delta\mathbf{r}^{\text{t},pq,p}
\end{equation}
where $\delta\mathbf{r}^{\text{t},pq,p}$ is defined
in Eq.~(\ref{eq:drKPKQ}).
After substituting
Eqs.~(\ref{eq:dfhat}), (\ref{eq:dmhat}), and (\ref{eq:deltan}),
the ``g-2'' terms in Eqs.~(\ref{eq:db1}) and (\ref{eq:dw1})
depend linearly on
the movements $d\mathbf{u}^{p}$, $d\mathbf{u}^{q}$,
$d\boldsymbol{\theta}^{p}$, and $d\boldsymbol{\theta}^{q}$,
and these terms can be gathered into a contact stiffness matrix
$[\mathbf{E}^{\text{g-2}}]$ and an assembly
stiffness matrix $[\mathbf{H}^{\text{g-2}}]$:
\begin{equation}\label{eq:Hg2}
  \left.
    \begin{aligned}
      &\,\delta\hat{\mathbf{f}}^{pq}\\
      &%\left(
      \delta\hat{\mathbf{m}}^{pq}
        %+\; \mathbf{r}^{pq}\times \delta\hat{\mathbf{f}}^{pq}\right)
    \end{aligned}
  \right\}_{6\times 1}
  \rightsquigarrow
   \left[ \mathbf{E}^{\text{g-2}}\right]_{2(6M)\times 6N}
  \left[
    \begin{array}{@{\extracolsep{\fill}}c@{\extracolsep{\fill}}}
      d\mathbf{u} \\\hdashline[1pt/1pt]
      d\boldsymbol{\theta}
    \end{array}
  \right]%_{6N \times 1}
  \quad\text{and}\quad
  \left[\mathbf{H}^{\text{g-2}}\right]_{6N\times 6N}
  =
  \left[\mathbf{A}\right]
  \left[\mathbf{E}^{\text{g-2}}\right]
\end{equation}
where $[\mathbf{A}]$ is the statics matrix of Eq.~(\ref{eq:Equilbrium0a}).
Stiffness $[\mathbf{H}^{\text{g-2}}]$ is clearly non-symmetric.
%where $[\mathbf{E}^{\text{g-2}}]$ gives the changes in force at the $M$ contacts
%produced by movements of the $N$ particles.
%The contact changes are multiplied by the statics matrix
%$[\mathbf{A}]$ to collect the contact forces
%into particles forces
%(i.e., the summations in Eqs.~\ref{eq:Equilibrium0} and~\ref{eq:Equilbrium0a}),
%resulting in the second geometric stiffness $[\mathbf{H}^{\text{g-2}}]$:
%
%\begin{equation}\label{eq:Hg2prod}
%  \left[ \mathbf{H}^{\text{g-2}}\right]_{6N\times 6N}
%  =
%   -\left[ \mathbf{A}\right]_{6N\times (6M)}
%   \left[ \mathbf{E}^{\text{g-2}}\right]_{(6M)\times 6N}
%\end{equation}
%
\par
Equations~(\ref{eq:Equilibrium1b}\textsubscript{1})
and~(\ref{eq:Equilibrium1b}\textsubscript{2})
yield the co-rotated external
force increments $\delta\mathbf{b}^{p}$ and $\delta\mathbf{w}^{p}$,
as would be seen by an observer attached to $p$.
The geometric ``g-3'' terms in Eqs.~(\ref{eq:db1}) and (\ref{eq:dw1})
recover the global increments
$d\mathbf{b}^{p}$ and $d\mathbf{w}^{p}$ by
applying Eq.~(\ref{eq:corotate}) to the forces
$\mathbf{b}^{p}$ and $\mathbf{w}^{p}$.
Figure~\ref{fig:g2}c shows a single particle $p$ with
external force $\mathbf{b}^{p}$
and a counteracting contact follower
force $\mathbf{f}^{pq}$
that rotates with the particle.
If the contact is undeformed,
the material increment $\mathfrak{d}\mathbf{f}^{pq}$ is zero,
and if no rolling occurs at the contact,
the directions of the contact normal and contact force
%the contact normal and the contact force
will appear unchanged when viewed by an
observer who rotates with the particle
(i.e., $\delta\mathbf{n}^{pq}=\delta\mathbf{f}^{pq}=\mathbf{0}$).
Yet, the force $\mathbf{f}^{pq}$ rotates within the global frame.
This rotation is produced by the ``g-3'' term
$d\boldsymbol{\theta}^{p} \times \mathbf{f}^{pq}$:
that is,
to maintain equilibrium after the small
rotation $d\boldsymbol{\theta}^{p}$,
the external force $\mathbf{b}^{p}$ must rotate
to $\mathbf{b}^{p}+d\mathbf{b}^{p}$.
\par
As another example,
Fig.~\ref{fig:g2}d shows a contact force $\mathbf{f}^{pq}$
that remains vertical:
perhaps a perfect rolling has shifted the contact point
while inducing no contact deformation, so that
$\mathfrak{d}\mathbf{f}^{pq}=d\mathbf{f}^{pq}=\mathbf{0}$.
However,
an observer attached to the particle would see a
rotation of the contact force, with
$\delta\hat{\mathbf{f}}^{pq}\neq 0$.
The $d\boldsymbol{\theta}^{p} \times \mathbf{f}^{pq}$ term
in Eq.~(\ref{eq:db1}) nullifies this increment, so that
$d\mathbf{f}^{pq}=d\mathbf{b}^{p}=0$.
Yet, the
$\mathbf{r}^{pq}\times \delta\hat{\mathbf{f}}^{pq}$
term in Eq.~(\ref{eq:dw1}) yields the counteracting
external moment $d\mathbf{w}^{p}$ that is required
to maintain equilibrium.
\par
The ``g-3'' terms in Eqs.~(\ref{eq:db1})
and~(\ref{eq:dw1}) depend linearly on
the movements $d\mathbf{u}^{p}$, $d\mathbf{u}^{q}$,
$d\boldsymbol{\theta}^{p}$, and $d\boldsymbol{\theta}^{q}$,
and these terms can be gathered into the %contact stiffness matrix
%$[\mathbf{E}^{\text{g-3}}]$ and the
assembly stiffness matrix $[\mathbf{H}^{\text{g-3}}]$:
\begin{equation}\label{eq:Hg3}
  \left.
    \begin{aligned}
      &-d\boldsymbol{\theta}^{p} \times \mathbf{f}^{pq}\\
      &-d\boldsymbol{\theta}^{p} \times \mathbf{m}^{pq}
    \end{aligned}
  \right\}_{6\times 1}
  \rightsquigarrow
  \left[\mathbf{H}^{\text{g-3}}\right]_{6N\times 6N}
  \left[
    \begin{array}{@{\extracolsep{\fill}}c@{\extracolsep{\fill}}}
      d\mathbf{u} \\\hdashline[1pt/1pt]
      d\boldsymbol{\theta}
    \end{array}
  \right]%_{6N \times 1}
%  \quad\text{and}\quad
%  \left[\mathbf{H}^{\text{g-3}}\right]
%  =
%  \left[\mathbf{A}\right]
%  \left[\mathbf{E}^{\text{g-3}}\right]
\end{equation}
The matrix $[\mathbf{H}^{\text{g-3}}]$ is non-symmetric.
\par
We complete the derivation of stiffness $[\mathbf{H}]$
by considering a possible dependence of the
external forces $[\mathbf{b}]$ and $[\mathbf{w}]$
on the particles' positions $[\mathbf{u}]$ and $[\boldsymbol{\theta}]$,
as in Eq.~(\ref{eq:pGx})--(\ref{eq:frakp}) and Fig.~\ref{fig:Follower}.
The matrix $[\partial \mathbf{p}/\partial\mathbf{x}]$
in Eq.~(\ref{eq:dbG})
applies to position-dependent forces and
is the fourth geometric stiffness,
%since the partial derivatives $\partial \mathbf{q}/\partial\mathbf{x}$
%are functions of the current positions $\mathbf{x}$
%and of the current
%external forces $\mathbf{q}$, the latter expressed as loading
%parameters.
%The fourth geometric stiffness is defined as
%
\begin{equation}\label{eq:Hg4}
  \left[\mathbf{H}^{\text{g-4}}\right]_{6N\times 6N}
  =
  - \left[
      \partial \mathbf{p}/\partial \mathbf{x}
    \right]_{6N\times 6N}
\end{equation}
On the other hand,
when all external loads are independent of the particles'
positions (as with non-follower dead loads),
the problem is simplified, with
$[\mathbf{H}^{\text{g-4}}]=[\mathbf{0}]$,
$[\mathbf{Q}]=[\mathbf{I}]$, and
$[d\boldsymbol{\mathfrak{p}}]=[d\mathbf{q}]=[d\mathbf{p}]$.
\par
Combining Eqs.~(\ref{eq:parAequilibrium}), (\ref{eq:Hdudb}),
(\ref{eq:dfpqall}), (\ref{eq:dmpqall}), (\ref{eq:dfrakB}),
(\ref{eq:Hmprod}),
and (\ref{eq:Hg2})--(\ref{eq:frakp}),
the stiffness matrix $[\mathbf{H}]$
in Eq.~(\ref{eq:Hdudb}) is the sum of mechanical
and geometric parts, the latter being the sum of four influences:
\begin{align}\label{eq:Hdxfrakp}
  &\left[\mathbf{H}(\boldsymbol{\nu})\right]
  \left[d\mathbf{x}\right] =
  \left[d\boldsymbol{\mathfrak{p}}\right]\\
  \label{eq:GeomMech}
  &\left[ \mathbf{H} (\boldsymbol{\nu})\right]
    =   \left[ \mathbf{H}^{\text{m}}(\boldsymbol{\nu}) \right]
       + \left[ \mathbf{H}^{\text{g}} \right]\\
  \label{eq:GeomMech2}
  &\left[ \mathbf{H}^{\text{g}} \right]
    =  \left[ \mathbf{H}^{\text{g-1}} \right]
      + \left[ \mathbf{H}^{\text{g-2}} \right]
      + \left[ \mathbf{H}^{\text{g-3}} \right]
      + \left[ \mathbf{H}^{\text{g-4}} \right]
\end{align}
%
%If the forces $[\mathbf{p}]$ depend upon positions $[\mathbf{x}]$
%and loading parameters $[\mathbf{q}]$,
%as in Eq.~(\ref{eq:pGx}),
%then Eq.~(\ref{eq:Hdudb}) has the alternative form
%
%\begin{equation}\label{eq:Hdudb2}
%  \left[
%    \mathbf{H}^{q}
%  \right]_{6N\times 6N}
%  \def\arraystretch{1.00}
%  \left[
%    \begin{array}{@{\extracolsep{\fill}}c@{\extracolsep{\fill}}}
%      d\mathbf{u} \\\hdashline[1pt/1pt]
%      d\boldsymbol{\theta}
%    \end{array}
%  \right]_{6N \times 1}
%  =
%  \left[
%    \mathbf{Q}
%  % \frac{\partial \mathbf{p}}{\partial \mathbf{q}}
%  \right]_{6N\times L}
%  \left[ d\mathbf{q}\right]_{L\times 1}
%  \quad\text{or}\quad
%  \left[
%    \mathbf{H}^{q}
%  \right]
%  \left[
%    d\mathbf{x}
%  \right]_{6N\times 1}
%  =
%  \left[
%    d\mathbf{p}
%  \right]_{6N\times 1}
%\end{equation}
%
%where
%
%\begin{align}\label{eq:GeomMech3}
%  \left[ \mathbf{H}^{q} \right]
%    &=   \left[ \mathbf{H}^{\text{m}} \right]
%       + \left[ \mathbf{H}^{\text{g},q} \right]\\
%  \label{eq:GeomMech4}
%  \left[ \mathbf{H}^{\text{g},q} \right]
%    &=  \left[ \mathbf{H}^{\text{g-1}} \right]
%      + \left[ \mathbf{H}^{\text{g-2}} \right]
%      + \left[ \mathbf{H}^{\text{g-3}} \right]
%      + \left[ \mathbf{H}^{\text{g-4}} \right]
%\end{align}
%
%and the loading matrix $[\mathbf{Q}]$ is
%\begin{equation}\label{eq:Qdef2}
%  \left[\mathbf{Q}\right]
%  =
%  \left[
%%   \mathbf{Q}
%    \partial \mathbf{p}/\partial \mathbf{q}
%  \right]_{6N\times L}
%\end{equation}
%
In the first two equations, we allow for inelastic and
incrementally nonlinear contact
behavior, in which %a dependence of
the mechanical stiffness $[\mathbf{H}^{\text{m}}(\boldsymbol{\nu})]$
depends
on the direction of loading, as expressed with
the direction cosines
$[\boldsymbol{\nu}]=[d\mathbf{x}]/\sqrt{[d\mathbf{x}]^{\text{T}}[d\mathbf{x}]}$.
This possibility is expressed in
Eqs.~(\ref{eq:constitutive})--(\ref{eq:constitutiveM}),
where the increment in contact force depends on the
direction of the contact movement
%as explained in
(this dependence is developed in
Sections~\ref{sec:contacts} and~\ref{sec:branches}).
%
%\par
%The mechanical stiffness $[\mathbf{H}^{\text{m}}]$
%in Eq.~(\ref{eq:GeomMech})
%and~(\ref{eq:GeomMech3})
%derives from the contacts' stiffnesses.
%For inelastic contacts, such as those with friction,
%this stiffness will depend upon the loading
%direction $[\boldsymbol{\nu}]$
%and can be non-symmetric
%%for when such contacts are plastic or are actively slipping
%(Sections~\ref{sec:contacts} and~\ref{sec:branches}).
\par
The total stiffness is altered by geometric changes of
the assembly that accompany the particle movements.
The use of separate mechanical and geometric stiffnesses is an
established concept in finite element analysis
\cite{Argyris:1965a,Maier:1973a}, and Maier \cite{Maier:1971a}
designated their separate tendencies for instability
as ``physical instabilizing effects''
and ``geometrical instabilizing effects.''
The three stiffnesses ``{g-1}'', ``g-2'', and ``g-3''
have an internal origin, as they arise from the
%originate from internal
shifting and rotation of the current contact forces
among particles.
These three stiffnesses
depend upon (and are proportional to)
the current forces and the particles' sizes and
surface curvatures.
Stiffness ``g-4'' accounts for
any position-dependent rotations of external
follower forces.
All four geometric stiffness matrices
can be non-symmetric, which, by itself,
would lead to a non-symmetric total stiffness $[\mathbf{H}]$.
When one or more contacts reach the friction limit
(see Section~\ref{sec:contacts}),
non-symmetry of the mechanical stiffness $[\mathbf{H}^{\text{m}}]$
also contributes to non-symmetry of the total stiffness.
%In principle,
%the geometric stiffness $[\mathbf{H}^{\text{g}}]$
%can depend upon the
%loading direction,
%although we will not consider this possibility.
%\todo[inline]{
%These geometric stiffnesses
%are independent of the
%loading direction,
%provided that %all particle pairs share, at most, a single contact
%%and that
%the particles' surfaces have continuous
%curvatures at the contact point,
%thus precluding the rocking modes that can
%occur with angular contact surfaces.
%%or between
%%particles that share multiple contact points).
%We will not consider such direction-dependence
%in this work.}
%
\subsection{\normalsize Stability framework}\label{sec:stability}
We use the thermodynamic
approach of Ba\v{z}ant \cite{Bazant:1988b,Bazant:1991a},
Petryk \cite{Petryk:1991a}, and
Nicot et al. \cite{Nicot:2012a} to investigate
the stability of granular systems,
an approach that will place stability in the context of the
stiffness matrix $[\mathbf{H}]$ defined in
Eqs.~(\ref{eq:Hdudb}) and (\ref{eq:Hdxfrakp}).
Unlike previous works, we apply these principles
in the context of a granular system,
by explicitly accounting
for the internal geometric effects that are due to the
shapes of particles at their contact points,
as embodied in the $[\mathbf{H}^{\text{g}}]$
geometric stiffness.
Our restrictive
approach is a systematic application of the principle of
stationary total potential energy, %(and its second variation)
commonly used in structural analysis
(e.g., \cite{Ziegler:1968a}).
The approach considers the kinetic energy
$E$ of a system that is initially in equilibrium
in the reference configuration
$[\mathbf{x}^{\ast}]=[\mathbf{u}^{\ast}/\boldsymbol{\theta}^{\ast}]$
at time $t^{\ast}$ and with no initial rate,
$d\mathbf{x}/dt=0$ at $t^{\ast}$.
As such, the analysis leads to a criterion for initial,
incipient instability but does not fully characterize the
subsequent dynamics, such as the flutter instability of
non-conservative systems.
\par
Following the notation of Nicot et al. \cite{Nicot:2012a},
at time $t\ge t^{\ast}$,
the rate of change of kinetic energy
(i.e., the first variation of the total potential energy)
is the difference
in the work rates of the external forces $\mathbf{p}$
and internal forces $\boldsymbol{\mathfrak{f}}$,
\begin{align}\label{eq:dE1}
  \dot{E}(t) = \dot{W}_{\text{ext}}(t) - \dot{W}_{\text{int}}(t)
  &=
  \left[\mathbf{p}(t)\right]^{\text{T}}\left[\dot{\mathbf{x}}\right]
  -
  \left[\boldsymbol{\mathfrak{f}}(t)\right]^{\text{T}}_{2(6M)\times 1}
  \left[\dot{\mathbf{u}}^{\text{def}}\right]\\\label{eq:dE2}
  &=
  \left[\mathbf{p}(t)\right]^{\text{T}}\left[\dot{\mathbf{x}}\right]
  -
  \left[\boldsymbol{\mathfrak{f}}(t)\right]^{\text{T}}
  \left[\mathbf{B}\right]
  \left[\dot{\mathbf{x}}\right]
\end{align}
%
%where $\dot{W}_{\text{ext}}$ and $\dot{W}_{\text{int}}$ are the
%work rates of the external forces $\mathbf{p}$
%and internal forces $\mathbf{f}$.
%In these equations, we write the contact
%forces with the abbreviated notation $[\mathbf{f}]=[\mathbf{f}/\mathbf{m}]$,
%which collects both contact forces $\mathbf{f}^{pq}$
%and contact moments $\mathbf{m}^{pq}$
%(Eq.~\ref{eq:gather}).
%Likewise,
where the vector of contact movements
$[\dot{\mathbf{u}}^{\text{def}}]$
is abbreviated to include the displacement and rotation rates,
$\dot{\mathbf{u}}^{\text{def},pq}$ and
$\dot{\boldsymbol{\theta}}^{\text{def},pq}$,
of Eqs.~(\ref{eq:dudef}) and~(\ref{eq:dthetadef}).
In Eq.~(\ref{eq:dE2}),
contact velocities $\dot{\mathbf{u}}^{\text{def}}$
have been replaced with particle velocities
$\dot{\mathbf{x}}$, by applying
Eq.~(\ref{eq:Bmatrix}).
%In the following,
%we remove the restriction to elastic systems, so that
%the internal work can be either elastic or frictional.
The work done by internal contact forces can include both
elastic and frictional (reversible and irreversible)
parts, and these forces need not be derived from an elastic
energy (or state) function,
provided that the forces
$\mathbf{p}(t)$ and $\boldsymbol{\mathfrak{f}}(t)$ are consistent
with the directions of the movements $\dot{\mathbf{x}}$
(see Section~\ref{sec:branches}).
\par
Although Eq.~(\ref{eq:dE1}) takes
$\dot{E}$ as a rate of change of kinetic energy
(e.g. \cite{Nicot:2012a}),
similar results are found by taking $\dot{E}$ as the rate
of the internal or free energy of a isotropic or adiabatic system
(as in \cite{Bazant:1988b}) or as the rate
of total potential energy associated with a
deformed body and its loading system
(e.g. \cite{Petryk:1991a}).
\par
We will follow a general approach and allow
external follower forces, using Eq.~(\ref{eq:Hdxfrakp})
instead of Eq.~(\ref{eq:Hdudb}).
If the external forces are rate-independent,
then their values
at $t=t^{\ast} + \Delta t$ depend only on
the particles' positions and on the loading increments,
and they can be approximated with the series
\begin{align}\label{eq:pstar}
  [\mathbf{p}(t^{\ast}+\Delta t)]_{6N\times 1} &=
  [\mathbf{p}(t^{\ast})] +
  \left[
    \frac{\partial\mathbf{p}}{\partial\mathbf{x}}(t^{\ast}) \right]
  \left[\Delta \mathbf{x}\right]_{6N\times 1}
  + \left[
    \frac{\partial\mathbf{p}}{\partial\mathbf{q}}(t^{\ast}) \right]
    \left[\Delta\mathbf{q}\right]_{L\times 1}
  + \ldots
  \\ \label{eq:fstar}
  \left[ \,\boldsymbol{\mathfrak{f}}(t^{\ast}+\Delta t)\right]_{2(6M)\times 1}&=
  \left[ \,\boldsymbol{\mathfrak{f}} (t^{\ast})\right] +
  \left[
    \frac{\partial\boldsymbol{\mathfrak{f}}}{\partial\mathbf{x}}(t^{\ast}) \right]
    \left[\Delta \mathbf{x}\right]_{6N\times 1} + \ldots
%    \left[\Delta \mathbf{u}^{\text{def}}\right]_{(6M)\times 1} + \ldots
  \\ \label{eq:Fstar}
  [\mathbf{B}(t^{\ast}+\Delta t)]_{2(6M)\times 6N} &=
  [\mathbf{B}(t^{\ast})] +
  \left[
    \frac{\partial\mathbf{B}}{\partial\mathbf{x}}(t^{\ast}) \right]
  \left[\Delta \mathbf{x}\right]_{6N\times 1}
  + \ldots
\end{align}
which we truncate at the second-order terms.% (see Eq.~\ref{eq:pGx}).
%\par
The particle displacements
$[\Delta\mathbf{x}]=[\Delta\mathbf{u}/\Delta\boldsymbol{\theta}]$
%in Eq.~(\ref{eq:pstar}) and
%the contact movements
%$[\Delta\mathbf{u}^{\text{def}}]=[\Delta\mathbf{u}^{\text{def}}/\Delta\boldsymbol{\theta}^{\text{def}}]$
%in Eq.~(\ref{eq:fstar})
occur during the interval $\Delta t$,
and the forces at $t^{\ast}+\Delta t$
\emph{are assumed to satisfy equilibrium in the displaced configuration}
(note the absence of both mass and damping).
%With the internal forces of Eq.~(\ref{eq:fstar}),
The last term in Eq.~(\ref{eq:fstar})
corresponds to the internal force increments $d\mathbf{f}^{pq}$
and $d\mathbf{m}^{pq}$ that appear in
Eqs.~(\ref{eq:dfpqall}) and~(\ref{eq:dmpqall}),
and these increments are produced
by a combination of contact deformations
(i.e., the mechanical changes $\mathfrak{d}\mathbf{f}^{pq}$
and $\mathfrak{d}\mathbf{m}^{pq}$),
of force rotations that accompany
the contact movements (the geometric changes
$\delta\hat{\mathbf{f}}^{pq}$ and $\delta\hat{\mathbf{m}}^{pq}$),
and of frame rotations,
$d\boldsymbol{\theta}^{p}\times\mathbf{f}^{pq}$ and
$d\boldsymbol{\theta}^{p}\times\mathbf{m}^{pq}$.
Combining these parts from Eqs.~(\ref{eq:dfrakB}),
(\ref{eq:Hg2}), and~(\ref{eq:Hg3}),
\begin{equation}
  \left[
    \frac{\partial\boldsymbol{\mathfrak{f}}}{\partial\mathbf{x}}(t^{\ast})
  \right]
  \left[\Delta \mathbf{x}\right]
  =
  \def\arraystretch{1.00}
  \left[
    \begin{array}{@{\extracolsep{\fill}}c@{\extracolsep{\fill}}}
      \mathbf{F}(t^{\ast}) \\\hdashline[1pt/1pt]
      \mathbf{M}(t^{\ast})
    \end{array}
  \right]
  \left[
    \mathbf{B}(t^{\ast})
  \right]
  \left[
    \Delta\mathbf{x}
  \right]
  +
  \left[
    \mathbf{E}^{\text{g-2}}(t^{\ast})
  \right]
  \left[
    \Delta\mathbf{x}
  \right]
  +
  \left[
    \mathbf{E}^{\text{g-3}}(t^{\ast})
  \right]
  \left[
    \Delta\mathbf{x}
  \right]
\end{equation}
where $[\mathbf{E}^{\text{g-3}}]_{2(6M)\times 6N}$ is an intermediate matrix,
such that $[\mathbf{H}^{\text{g-3}}] = [\mathbf{A}][\mathbf{E}^{\text{g-3}}]$.
On the right side of Eq.~(\ref{eq:pstar}),
the second and third terms include the explicit
loading changes and geometric alterations
in Eq.~(\ref{eq:dbG}),
%force derivatives in Eq.~(\ref{eq:dbG})
%and the changes in the external forces that are necessary to
such that
%but these terms must also account for
%changes in the external forces that are necessary to
%balance alterations in the directions and positions of the internal
%contact forces, as in Eqs.(\ref{eq:Hg1}) and~(\ref{eq:Hg3}):
%
\begin{equation}\label{eq:dpdxdelx}
  \left[
    \frac{\partial\mathbf{p}}{\partial\mathbf{x}}(t^{\ast}) \right]
  \left[\Delta \mathbf{x}\right]
  =
  -\left[ \mathbf{H}^{\text{g-4}}(t^{\ast})\right]
   \left[\Delta \mathbf{x}\right]
   \quad\text{and}\quad
   \left[
    \frac{\partial\mathbf{p}}{\partial\mathbf{q}}(t^{\ast}) \right]
    \left[\Delta\mathbf{q}\right]
   =
   \left[\mathbf{Q}(t^{\ast})\right]
   \left[\Delta\mathbf{q}\right]
%  -\left[ \mathbf{H}^{\text{g-1}}(t^{\ast})\right]
%   \left[\Delta \mathbf{x}\right]
%  -\left[ \mathbf{H}^{\text{g-3}}(t^{\ast})\right]
%   \left[\Delta \mathbf{x}\right]
\end{equation}
%
%The additional loading increment in Eq.~(\ref{eq:pstar}) is produced
%by changes in the imposed loading parameters $[\Delta\mathbf{q}]$
%(see Eq.~\ref{eq:dbG}):
%
%\begin{equation}\label{eq:dptG}
%    \left[
%    \frac{\partial\mathbf{p}}{\partial t}(t^{\ast}) \right]
%    \Delta t
%    =
%    \left[
%      \frac{\partial\mathbf{G}}{\partial\mathbf{q}}
%      (t^{\ast})
%    \right]
%    \left[
%      \Delta\mathbf{q}
%    \right]
%\end{equation}
%
\par
Continuing with the notation of Nicot et al. \cite{Nicot:2012a},
the second time-derivative of the kinetic energy in
Eq.~(\ref{eq:dE1}) is
%(i.e., the second variation of the total potential energy for
%isentropic conditions or of the total Helmholtz energy for
%isothermal conditions) is
%
\begin{equation}
  \ddot{E} =
  \left[\mathbf{p}(t)\right]^{\text{T}}\left[\ddot{\mathbf{x}}\right]
  -
  \left[\boldsymbol{\mathfrak{f}}(t)\right]^{\text{T}}\left[\mathbf{B}\right]\left[\ddot{\mathbf{x}}\right]
  +
  \left[\frac{d}{dt}
  \mathbf{p}(t)\right]^{\text{T}}\left[\dot{\mathbf{x}}\right]
  -
  \left[\frac{d}{dt}
  \boldsymbol{\mathfrak{f}}(t)\right]^{\text{T}}\left[\mathbf{B}\right]\left[\dot{\mathbf{x}}\right]
  -
  \left[\boldsymbol{\mathfrak{f}}(t)\right]^{\text{T}}
  \left[\frac{d}{dt}\mathbf{B}\right]
  \left[\dot{\mathbf{x}}\right]
\end{equation}
This second derivative can also be viewed
as the second variation of the total potential energy for
isentropic conditions or of the total Helmholtz energy for
isothermal conditions \cite{Bazant:1988b}.
Substituting Eqs.~(\ref{eq:pstar})--(\ref{eq:dpdxdelx}),
\begin{equation}\label{eq:Eddot2}
  \begin{aligned}
  \ddot{E}
  =
  &\left(
  \left[\mathbf{p}(t^{\ast})\right]^{\text{T}}
  -
  \left[\boldsymbol{\mathfrak{f}}(t^{\ast})\right]^{\text{T}}\left[\mathbf{B}\right]\right)
  \left[\ddot{\mathbf{x}}\right]\\
  &-
  \left[
    \Delta \mathbf{x}
  \right]^{\text{T}}
% \left(
   \left[ \mathbf{H}^{\text{g-4}}(t^{\ast})\right]^{\text{T}}
%  +\left[ \mathbf{H}^{\text{g-1}}(t^{\ast})\right]^{\text{T}}
%  +\left[ \mathbf{H}^{\text{g-3}}(t^{\ast})\right]^{\text{T}}
% \right)
  \left[\ddot{\mathbf{x}}\right]
  -
  \frac{d}{dt}\left[\Delta\mathbf{x}\right]^{\text{T}}
  %\left(
   \left[ \mathbf{H}^{\text{g-4}}(t^{\ast})\right]^{\text{T}}
  %+\left[ \mathbf{H}^{\text{g-1}}(t^{\ast})\right]^{\text{T}}
  %+\left[ \mathbf{H}^{\text{g-3}}(t^{\ast})\right]^{\text{T}}
  %\right)
  \left[\dot{\mathbf{x}}\right]\\
  &-
  \left[
    \Delta \mathbf{x}
  \right]^{\text{T}}
  \left(
    \left[\mathbf{B}(t^{\ast})\right]^{\text{T}}
  \left[
    \begin{array}{@{\extracolsep{\fill}}c@{\extracolsep{\fill}}}
      \mathbf{F}(t^{\ast}) \\\hdashline[1pt/1pt]
      \mathbf{M}(t^{\ast})
    \end{array}
  \right]^{\text{T}}
  +
  \left[ \mathbf{E}^{\text{g-2}}(t^{\ast}) \right]^{\text{T}}
  +
  \left[ \mathbf{E}^{\text{g-3}}(t^{\ast}) \right]^{\text{T}}
  \right)
  \left[\mathbf{B}\right]
  \left[\ddot{\mathbf{x}}\right]\\
  %&-
  %\frac{d}{dt}\left[\Delta\mathbf{x}\right]^{\text{T}}
  %\left(
  % \left[ \mathbf{H}^{\text{g-4}}(t^{\ast})\right]^{\text{T}}
  %+\left[ \mathbf{H}^{\text{g-1}}(t^{\ast})\right]^{\text{T}}
  %+\left[ \mathbf{H}^{\text{g-3}}(t^{\ast})\right]^{\text{T}}
  %\right)
  %\left[\dot{\mathbf{x}}\right]\\
  &-
  \frac{d}{dt}
  \left[
    \Delta \mathbf{x}
  \right]^{\text{T}}
  \left(
    \left[\mathbf{B}(t^{\ast})\right]^{\text{T}}
  \left[
    \begin{array}{@{\extracolsep{\fill}}c@{\extracolsep{\fill}}}
      \mathbf{F}(t^{\ast}) \\\hdashline[1pt/1pt]
      \mathbf{M}(t^{\ast})
    \end{array}
  \right]^{\text{T}}
  \!\!\!+
  \left[ \mathbf{E}^{\text{g-2}}(t^{\ast}) \right]^{\text{T}}
  \!\!\!+
  \left[ \mathbf{E}^{\text{g-3}}(t^{\ast}) \right]^{\text{T}}
  \right)
  \left[\mathbf{B}\right]
  \left[\dot{\mathbf{x}}\right]\\
  &+
  \left[\Delta\mathbf{q}\right]^{\text{T}}
% \left[\frac{\partial\mathbf{p}}{\partial\mathbf{q}}(t^{\ast}) \right]^{\text{T}}
  \left[\mathbf{Q}(t^{\ast}) \right]^{\text{T}}
  \left[\ddot{\mathbf{x}}\right]
  +
  \frac{d}{dt}\left[\Delta\mathbf{q}\right]^{\text{T}}
% \left[\frac{\partial\mathbf{p}}{\partial\mathbf{q}}(t^{\ast}) \right]^{\text{T}}
  \left[\mathbf{Q}(t^{\ast}) \right]^{\text{T}}
  \left[\dot{\mathbf{x}}\right] \\
  &- \left[\,\boldsymbol{\mathfrak{f}}(t^{\ast})\right]^{\text{T}}
    \left[\frac{\partial\mathbf{B}}{\partial\mathbf{x}}
    (t^{\ast})
     \cdot \frac{d}{dt}\Delta\mathbf{x}
    \right]
    \left[\dot{\mathbf{x}}\right]
  \end{aligned}
\end{equation}
The system is in equilibrium at time $t^{\ast}$,
as in Eq.~(\ref{eq:Equilbrium0a}),
which eliminates the right side of the first line in this equation.
Because $[\mathbf{A}]=[\mathbf{B}]^{\text{T}}$, the various products
$[\mathbf{E}]^{\text{T}}[\mathbf{B}]$ can be replaced with
their $[\mathbf{H}]^{\text{T}}$ counterparts,
as in %Eq.~(\ref{eq:Hg2prod}),
Eq.~(\ref{eq:Hg2}\textsubscript{2}).
%, and~(\ref{eq:Hg3}\textsubscript{2}).
%the relations in Eqs.~(\ref{eq:Hmprod}), (\ref{eq:Hg2prod}),
%(\ref{eq:GeomMech3}), and~(\ref{eq:GeomMech4}).
The last term in Eq.~(\ref{eq:Eddot2}) is
%written in index form as
$B_{ij,k}f_{i}\frac{d}{dt}\Delta x_{k}\dot{x}_{j}$,
but because $A_{ij}=B_{ji}$,
the last term is equivalent to
$A_{ij,k}f_{j}\frac{d}{dt}\Delta x_{k}\dot{x}_{i}=H^{\text{g-1}}_{ik}\frac{d}{dt}\Delta x_{k}\dot{x}_{i}$
(see Eq.~\ref{eq:Hg1Af}), or
\begin{equation}
  \left[\,\boldsymbol{\mathfrak{f}}(t^{\ast})\right]^{\text{T}}
    \left[\frac{\partial\mathbf{B}}{\partial\mathbf{x}}
    (t^{\ast})
     \cdot \frac{d}{dt}\Delta\mathbf{x}
    \right]
    \left[\dot{\mathbf{x}}\right]
    =
    \frac{d}{dt}\left[\Delta\mathbf{x}\right]^{\text{T}}
    \left[\mathbf{H}^{\text{g-1}}\right]^{\text{T}}
    \left[\dot{\mathbf{x}}\right]
\end{equation}
Equation~(\ref{eq:Eddot2}) applies at $t=t^{\ast}$, at which
$[\Delta\mathbf{x}]=[\Delta\mathbf{q}]=[\mathbf{0}]$,
which eliminates the $[\ddot{\mathbf{x}}]$ terms in
Eq.~(\ref{eq:Eddot2}),
but we allow perturbations of the velocities and loading rates,
with
$(d/dt)[\Delta\mathbf{x}]=(1/\Delta t)[\Delta\mathbf{x}]=[\dot{\mathbf{x}}]\neq[\mathbf{0}]$
and
$(d/dt)[\Delta\mathbf{q}]=(1/\Delta t)[\Delta\mathbf{q}]=[\dot{\mathbf{q}}]\neq[\mathbf{0}]$.
As such, Eq.~(\ref{eq:Eddot2}) is simplified as
\begin{equation}\label{eq:Eddot3}
    \ddot{E}(t^{\ast})=
    \left[\dot{\mathbf{q}}\right]^{\text{T}}
    \left[\mathbf{Q}(t^{\ast}) \right]^{\text{T}}
    \left[\dot{\mathbf{x}}\right]
    -
    \left[\dot{\mathbf{x}}\right]^{\text{T}}
    \left[\mathbf{H}(\boldsymbol{\nu},t^{\ast})\right]^{\text{T}}
    \left[\dot{\mathbf{x}}\right]
\end{equation}
where the stiffness matrix $[\mathbf{H}(\boldsymbol{\nu})]$ is the sum of
the mechanical stiffness $[\mathbf{H}^{\text{m}}(\boldsymbol{\nu})]$
and four non-conservative geometric contributions,
as in Eqs.~(\ref{eq:GeomMech})--(\ref{eq:GeomMech2}).
Note that the possible dependence of stiffness $[\mathbf{H}]$
on loading direction $[\boldsymbol{\nu}]$ is included
in this equation.
\par
The kinetic energy at $t^{\ast}+\Delta t$ is estimated with the
series
\begin{equation}\label{eq:Etplusdt}
  E(t^{\ast}+\Delta t) = E(t^{\ast})
                         + \dot{E}(t^{\ast})
                         + \frac{1}{2}(\Delta t)^{2}\ddot{E}(t^{\ast})
                         + \ldots
\end{equation}
and because the system is assumed in equilibrium at $t^{\ast}$,
with $\dot{E}(t^{\ast})=0$,
the change in
kinetic energy during increment $\Delta t$,
excluding terms of order $(\Delta t)^{3}$ and higher,
is due to the velocity
perturbation $[\dot{\mathbf{x}}]$ in Eq.~(\ref{eq:Eddot3}):
\begin{equation}\label{eq:quadratic}
  E(t^{\ast}+\Delta t) - E(t^{\ast})
  =
  \frac{(\Delta t)^{2}}{2}
  \left(\rule{0ex}{2ex}\mathcal{B}_{2}(\dot{\mathbf{x}}) -
        \mathcal{W}_{2}(\dot{\mathbf{x}})
  \right)
\end{equation}
where
%\begin{alignat}{2}
\begin{gather}
  \label{eq:B2}
  \mathcal{B}_{2}(\dot{\mathbf{x}}) =
  \left[\dot{\mathbf{q}}\right]^{\text{T}}
    \left[\mathbf{Q}(t^{\ast}) \right]^{\text{T}}
    \left[\dot{\mathbf{x}}\right]
    =
    \left[\dot{\boldsymbol{\mathfrak{p}}}\right]^{\text{T}}
    \left[\dot{\mathbf{x}}\right]
    \quad\text{or}\quad
    \mathcal{B}_{2}(\dot{\mathbf{x}}) =
    \left[\dot{\mathbf{p}}\right]^{\text{T}}
    \left[\dot{\mathbf{x}}\right] \\
  \label{eq:W2}
  \mathcal{W}_{2}(\dot{\mathbf{x}}) =
  \left[\dot{\mathbf{x}}\right]^{\text{T}}
    \left[\mathbf{H}(\boldsymbol{\nu},t^{\ast})\right]
    \left[\dot{\mathbf{x}}\right]
%    \quad\text{or}\quad
%    \mathcal{W}_{2}(\dot{\mathbf{x}}) &=
%    \left[\dot{\mathbf{x}}\right]^{\text{T}}
%    \left[\mathbf{H}(t^{\ast})\right]
%    \left[\dot{\mathbf{x}}\right]
\end{gather}
Of the two alternative forms in Eq.~(\ref{eq:B2}),
the more general
Eq.~(\ref{eq:B2}\textsubscript{1}) %and~(\ref{eq:W2}\textsubscript{1})
applies with or without external follower forces;
whereas, the more restrictive
Eq.~(\ref{eq:B2}\textsubscript{2}) %and~(\ref{eq:W2}\textsubscript{2})
only applies with non-follower (dead) force increments.
Ba\v{z}ant \cite{Bazant:1988b,Bazant:1991a} developed a similar
expression %for irreversible systems
by viewing the difference
$\mathcal{B}_{2}-\mathcal{W}_{2}$
as the rate of internal entropy production,
$\dot{S}_{\text{in}}$.
He noted that
the product of this rate and the
(always positive) temperature,
$T\dot{S}_{\text{in}}$,
represents an influx of kinetic energy.
\par
When the difference
$\mathcal{B}_{2}-\mathcal{W}_{2}$
is positive for some $\dot{\mathbf{x}}$,
the system
can suffer a spontaneous infusion~--- a
quadratic increase~--- of kinetic energy when the
particle velocities are perturbed in the particular
direction of $\dot{\mathbf{x}}$.
A negative difference means that small perturbances are
met with a suppression of kinetic energy,
restoring the system to equilibrium.
The second-order work $\mathcal{W}_{2}$ is an indicator of stability,
as negative values allow a quadratic increase in kinetic energy
in the absence of any change in external loading~--- when
$[\dot{\mathbf{q}}]$ or $[\dot{\mathbf{p}}]$ are zero~---
and the system, which was assumed to be in equilibrium,
enters a dynamical domain of behavior \cite{Nguyen:2016a}.
Note that in the absence of dead or follower forces,
this increase in kinetic energy is bounded by the system's stored
internal energy and is suppressed by dissipative friction.
Also note that $\mathcal{W}_{2}$ is not an
objective quantity,
as different values of $\mathcal{W}_{2}$ will
be measured by observers that are moving (rotating)
relative to each other (see the discussion of Type~IV constraint).
Finally, it is possible to show, with a full dynamic treatment,
that flutter can appear before a negative second-order work
appears \cite{Challamel:2010a}.
The question of stability is explored further in
Section~\ref{sec:IofE}.
\par
Prior to the 1960's,
the principle of stationary total potential
energy was usually applied to elastic systems,
in which
the internal energy is a path-independent state
function of the current displacement
(notable exceptions are the Shanley column \cite{Shanley:1947a}
and the linear comparison methods of Hill \cite{Hill:1959a}).
The rate $\ddot{E}$ in Eq.~(\ref{eq:Eddot3}), however,
applies to rate-independent \emph{inelastic} systems and
depends upon the direction
$[\boldsymbol{\nu}]$ of the particles' movements,
and for such systems, the susceptibility
to instability depends upon
the direction of the perturbance $[\dot{\mathbf{x}}]$.
For granular systems,
this
path-dependence originates with inelastic contacts,
%With the contact model of Sections~\ref{} and~\ref{},
which impart an influence of the direction of a contact's
movement $d\mathbf{u}^{pq,\text{def}}$
on the contact stiffness $\mathbf{F}^{pq}$
(Eqs.~\ref{eq:constitutive}--\ref{eq:constitutiveM}).
%and, in turn,
%of the influence of $[\boldsymbol{\nu}]$ on the assembly stiffness
%$[\mathbf{H}(\boldsymbol{\nu})]$.
%in Eq.~(\ref{}).
Rate $\ddot{E}$ is based, therefore, on the notion of a
\emph{tangentially equivalent linear system}
in which a single stiffness $[\mathbf{H}]$
applies only to a particular domain of perturbation directions
$[\boldsymbol{\nu}]=[\dot{\mathbf{x}}]/\sqrt{[\dot{\mathbf{x}}]^{\text{T}}[\dot{\mathbf{x}}]}$
(see \cite{Bazant:1991a}).
The problem is somewhat simpler for
the two-branch, incrementally non-linear
contact model of Sections~\ref{sec:contacts},
as we can determine the stability of an assembly
by investigating a finite number of piece-wise
constant assembly stiffnesses (Section~\ref{sec:branches}).
\subsection{\normalsize Displacement constraints}\label{sec:constraints}
Particle movements can be restricted in various ways,
and we describe four types of constraints, presented in the order
of their increasing
complexity.
With these types of constraints, varying degrees of
restriction are placed upon the
\emph{displacement} increments $[d\mathbf{x}]=[d\mathbf{u}/d\boldsymbol{\theta}]$.
This does not mean that we neglect load-control
(indeed, our fourth type is fully load-controlled),
since the applied loading increments $[d\mathfrak{p}]$
on the right of
Eqs.~(\ref{eq:Hdudb}) and (\ref{eq:Hdxfrakp})
are always assigned, whether they are zero or
otherwise.
However,
displacement constraints will typically require
that certain reaction forces be superposed on the applied loads
to achieve the prescribed displacements.
\par
Rather than using the method of Lagrange multipliers,
we approach %Eqs.~(\ref{eq:Hdudb}) and~(\ref{eq:Hdudb2})
constrained systems
by applying various generalized inverses,
and the reader is referred to  \cite{BenIsrael2003a}
%Ben-Israel and Greville
for a review.
In the following, we shorten the notation
by writing the direction-dependent
$[\mathbf{H}(\boldsymbol{\nu})]$ as simply
$[\mathbf{H}]$, deferring until Section~\ref{sec:branches}
the implications of incremental non-linearity.
\begin{description}
  \item[Type~I constraint:]
    With this simplest type of constraint,
    a subset of $r$ displacements
    and rotations
    %$[d\mathbf{x}]$ %=[d\mathbf{u}/d\boldsymbol{\theta}]$
    are assigned pre-determined values~--- a common
    situation in structural mechanics.
    In the usual manner,
    the full set $[d\mathbf{x}]=[d\mathbf{u}/d\boldsymbol{\theta}]$
    is partitioned into the $r$
    constrained ``c'' movements $[d\mathbf{x}^{\text{c}}]$
    and the $6N-r$ unconstrained
    free ``f'' movements $[d\mathbf{x}^{\text{f}}]$.
    To prevent rigid rotations of 3D (2D) assemblies,
    at least six (three) movements should be constrained,
    and these $[d\mathbf{x}^{c}]$ should preclude rigid translations
    and rotations.
    Likewise,
    the external forces $[d\boldsymbol{\mathfrak{p}}]$ are
    partitioned into lists,
    $[d\boldsymbol{\mathfrak{p}}^{\text{c}}]$ and
    $[d\boldsymbol{\mathfrak{p}}^{\text{f}}]$,
    that correspond
    to the complementary constrained and free
    displacements.
%    Recall that $[d\boldsymbol{\mathfrak{p}}]$
%    are applied loads, such as the $[d\boldsymbol{\mathfrak{p}}]$
%    in Eq.~(\ref{eq:gather2}\textsubscript{2}),
%    but the $[d\boldsymbol{\mathfrak{p}}]$
%    can be coordinated with a set of loading parameters
%    $[d\mathbf{q}]$, as in Eqs.~(\ref{eq:dbG}) and~(\ref{eq:frakp}).
    The forces $[d\boldsymbol{\mathfrak{p}}^{\text{c}}]$ represent
    the reaction forces of constraint.
    %$[d\mathbf{x}^{\text{c}}]$ and $[d\mathbf{x}^{\text{f}}]$.
    Equation~(\ref{eq:Hdudb}) is rearranged in the
    standard partitioned form
    \begin{equation}
      \left[
        \def\arraystretch{1.20}
        \begin{array}{@{\extracolsep{0ex}}c;{1pt/1pt}c@{\extracolsep{\fill}}}
          \mathbf{H}^{\text{cc}} & \mathbf{H}^{\text{cf}}\\\hdashline[1pt/1pt]
          \mathbf{H}^{\text{fc}} & \mathbf{H}^{\text{f{}f}}
        \end{array}
      \right]%_{6N \times 1}
      \left[
        \def\arraystretch{1.20}
        \begin{array}{l@{\extracolsep{\fill}}}
          d\mathbf{x}^{\text{c}}_{r\times 1} \\\hdashline[1pt/1pt]
          d\mathbf{x}^{\text{f}}_{(6N-r)\times 1}
        \end{array}
      \right]%_{6N \times 1}
      =
      \left[
        \def\arraystretch{1.20}
        \begin{array}{l@{\extracolsep{\fill}}}
          d\boldsymbol{\mathfrak{p}}^{\text{c}}_{r\times 1} \\\hdashline[1pt/1pt]
          d\boldsymbol{\mathfrak{p}}^{\text{f}}_{(6N-r)\times 1}
        \end{array}
      \right]%_{6N \times 1}
    \end{equation}
    and after shifting all response quantities to the left,
    \begin{equation}\label{eq:IHH}
      \left[\mathbf{U}\right]
      \left[
        \def\arraystretch{1.20}
        \begin{array}{@{\extracolsep{\fill}}c@{\extracolsep{\fill}}}
          d\boldsymbol{\mathfrak{p}}^{\text{c}} \\\hdashline[1pt/1pt]
          d\mathbf{x}^{\text{f}}
        \end{array}
      \right]%_{6N \times 1}
      =
      \left[\mathbf{V}\right]%_{6N \times 1}
      \left[
        \def\arraystretch{1.20}
        \begin{array}{@{\extracolsep{\fill}}c@{\extracolsep{\fill}}}
          d\mathbf{x}^{\text{c}} \\\hdashline[1pt/1pt]
          d\boldsymbol{\mathfrak{p}}^{\text{f}}
        \end{array}
      \right]%_{6N \times 1}
    \end{equation}
    where
    \begin{equation}\label{eq:IHH2}
      \left[\mathbf{U}\right]=
      \left[
        \def\arraystretch{1.20}
        \begin{array}{@{\extracolsep{0ex}}c;{1pt/1pt}c@{\extracolsep{\fill}}}
          -\mathbf{I} & \mathbf{H}^{\text{cf}}\\\hdashline[1pt/1pt]
           \mathbf{0} & \mathbf{H}^{\text{f{}f}}
        \end{array}
      \right]%_{6N \times 1}
      \quad\text{and}\quad
      \left[\mathbf{V}\right]=
      \left[
        \def\arraystretch{1.20}
        \begin{array}{@{\extracolsep{0ex}}c;{1pt/1pt}c@{\extracolsep{\fill}}}
          -\mathbf{H}^{\text{cc}} & \mathbf{0}\\\hdashline[1pt/1pt]
          -\mathbf{H}^{\text{fc}} & \mathbf{I}
        \end{array}
      \right]%_{6N \times 1}
    \end{equation}
    which is solved with the usual method of static condensation.
    Note that separate $[\mathbf{Q}]$ matrices
    (Eq.~\ref{eq:frakp}) may be required
    for resolving the two sets of forces
    $[d\boldsymbol{\mathfrak{p}}^{\text{c}}]$
    and $[d\boldsymbol{\mathfrak{p}}^{\text{f}\,}]$
    when both sets are position-dependent
    (for example, platen displacements
    might be controlled while also controlling a membrane pressure).
%    Questions of uniqueness and stability are resolved with the
%    submatrices on the left of Eq.~(\ref{eq:IHH}).
  \item[Type~II constraint:]
    As an intermediate case that will be used to
    derive the more general non-homogeneous
    Type~III constraint,
    a set of $r$ linear holonomic
    \emph{homogeneous} constraints is applied to
    the displacements and rotations $[d\mathbf{x}]$,
    \begin{equation}\label{eq:homogeneousC}
      \left[\mathbf{C}\right]_{r\times 6N}
      \left[d\mathbf{x}\right]%_{6N\times 1}
      =
      \left[\mathbf{0}\right]_{r\times 1}
    \end{equation}
    restricting $[d\mathbf{x}]$ to the null-space $\mathcal{L}$
    of $[\mathbf{C}]$,
    such that
    $[d\mathbf{x}]\in \mathcal{L} = \mathcal{N}(\mathbf{[C]})$.
    We will assume that matrix $[\mathbf{C}]$ has full rank $r$.
    In 3D (2D), at least six (three) constraints are required,
    and these constraints should preclude rigid motions of the assembly.
    The matrices $[\mathbf{P}_{\mathcal{L}}]$ and
    $[\mathbf{P}_{\mathcal{L}^\perp}]$ project vectors $[d\mathbf{x}]$
    onto the subspace $\mathcal{L}$ and its orthogonal complement:
    \begin{align}\label{eq:PL}
      \left[\mathbf{P}_{\mathcal{L}}\right] &=
      \left[\,\mathbf{I}\,\right]_{6N} -
      [\mathbf{P}_{\mathcal{L}^\perp}] \\ \label{eq:PLperp}
      [\mathbf{P}_{\mathcal{L}^\perp}] &=
      \left[\,\mathbf{C}\,\right]^{\text{T}}
      \left(
        \left[\,\mathbf{C}\,\right]
        \left[\,\mathbf{C}\,\right]^{\text{T}}
      \right)^{-1}
      \left[\,\mathbf{C}\,\right]
    \end{align}
    where $[\mathbf{I}]_{6N}$ is the
    $6N\times 6N$ identity matrix.
    That is, $[\mathbf{P}_{\mathcal{L}}][d\mathbf{x}]$
    projects vector $[d\mathbf{x}]$ onto the null-space of
    $[\mathbf{C}]$; whereas,
    $[\mathbf{P}_{\mathcal{L}^{\perp}}][d\mathbf{x}]$
    projects $[d\mathbf{x}]$ onto the row-space of $[\mathbf{C}]$.
    Because $[\mathbf{C}]$ is assumed to have full rank,
    the inverse in Eq.~(\ref{eq:PLperp}) exists.
    \par
    Considered together,
    Eqs.~(\ref{eq:Hdxfrakp}) and~(\ref{eq:homogeneousC})
    can be cast in the alternative form
%   Now consider the equation
    \begin{equation}\label{eq:Cxp}
      \begin{aligned}
      \left[\mathbf{H}\right]
      \left[d\mathbf{x}\right] -
      \left[d\mathbf{y}\right] =
      \left[d\boldsymbol{\mathfrak{p}}\right],\;
      &\left[\mathbf{H}\right]\in\mathbb{R}^{6N\times 6N},\;
      \left[d\boldsymbol{\mathfrak{p}}\right]\in\mathbb{R}^{6N},\\
      &\left[d\mathbf{x}\right]\in \mathcal{L}=\mathcal{N}(\left[\mathbf{C}\right]),\;
      \left[d\mathbf{y}\right]\in \mathcal{L}^{\perp}
      \end{aligned}
    \end{equation}
    noting that $[d\boldsymbol{\mathfrak{p}}]$ can lie outside
    the range of $[\mathbf{H}]$, which requires amendment with
    the additional vector $[d\mathbf{y}]$, as explained below.
    The consistency (i.e. the existence of a solution)
    of this system is equivalent to the
    consistency of the equation
    \begin{equation}\label{eq:TypeIIc}
      \left(\rule{0ex}{2.2ex}
      \left[\mathbf{H}\right]
      \left[\mathbf{P}_{\mathcal{L}}\right] +
      \left[\mathbf{P}_{\mathcal{L}^{\perp}}\right]
      \right)
      \left[d\mathbf{z}\right] =
      \left[d\boldsymbol{\mathfrak{p}}\right]
    \end{equation}
    as shown in \cite{BenIsrael2003a,Nikuie:2013a}.
    If matrix
    $([\mathbf{H}][\mathbf{P}_{\mathcal{L}}]+[\mathbf{P}_{\mathcal{L}^{\perp}}])$
    is non-singular, then Eq.~(\ref{eq:Cxp})
    is consistent for all external force
    increments $[d\boldsymbol{\mathfrak{p}}]$ and has
    the unique solution
    \begin{equation}\label{eq:dxdyBott}
      \left[d\mathbf{x}\right] =
      \left[\mathbf{P}_{\mathcal{L}}\right]\left[d\mathbf{z}\right]=
      \left[\mathbf{H}\right]_{(\mathcal{L})}^{(-1)}
      \left[d\boldsymbol{\mathfrak{p}}\right],\quad
      \left[d\mathbf{y}\right] =
      -\left[d\boldsymbol{\mathfrak{p}}\right]
      +\left[\mathbf{H}\right]\left[d\mathbf{x}\right]
    \end{equation}
    where $[\mathbf{H}]_{(\mathcal{L})}^{(-1)}$ is the
    Bott-Duffin constrained inverse of $[\mathbf{H}]$ on $\mathcal{L}$:
    \begin{equation}
      \left[\mathbf{H}\right]_{(\mathcal{L})}^{(-1)} =
      \left[\mathbf{P}_{\mathcal{L}}\right]
      \left(\rule{0ex}{2.2ex}
      \left[\mathbf{H}\right]
      \left[\mathbf{P}_{\mathcal{L}}\right] +
      \left[\mathbf{P}_{\mathcal{L}^{\perp}}\right]
      \right)^{-1}
    \end{equation}
    in which the prefactor $[\mathbf{P}_{\mathcal{L}}]$
    projects the solution
    $[d\mathbf{z}]$ onto the admissible domain $\mathcal{L}$,
    so that $[d\mathbf{x}]$ satisfies Eq.~(\ref{eq:homogeneousC}).
    \par
    The displacements $[d\mathbf{x}]$ in Eq.~(\ref{eq:dxdyBott})
    comply with the displacement
    constraints of Eq.~(\ref{eq:homogeneousC}),
    and the force increments
    $[d\mathbf{y}]$ are the reaction forces of constraint
    (superposed on $[d\boldsymbol{\mathfrak{p}}]$) that are necessary
    to satisfy the constraints.
    Note that $[d\mathbf{x}]$ and $[d\mathbf{y}]$ belong to
    orthogonal spaces,
    as $[d\mathbf{x}]^{\text{T}}
        [d\mathbf{y}] =
       -[d\mathbf{z}]^{\text{T}}[\mathbf{P}_{\mathcal{L}}]^{\text{T}}
        (([\mathbf{H}][\mathbf{P}_{\mathcal{L}}]
         + [\mathbf{P}_{\mathcal{L}^{\perp}}])[d\mathbf{z}]
         -[\mathbf{H}][\mathbf{P}_{\mathcal{L}}][d\mathbf{z}])=
         [d\mathbf{z}]^{\text{T}}[\mathbf{P}_{\mathcal{L}}]^{\text{T}}
         [\mathbf{P}_{\mathcal{L}^{\perp}}][d\mathbf{z}]=0$,
    a result that is consistent with
    d'Alembert's principle of virtual work.
  \item[Type~III constraint:]
    A set of $r$ linear holonomic
    \emph{non-homogeneous} constraints is applied to
    the displacements and rotations $[d\mathbf{x}]$,
    \begin{equation}\label{eq:nonhomogeneousC}
      \left[\mathbf{C}\right]_{r\times 6N}
      \left[d\mathbf{x}\right]%_{6N\times 1}
      =
      \left[d\mathbf{c}\right]_{r\times 1}
    \end{equation}
    Types~I and~II are special cases of this constraint.
    Similar to Type~II constraint, we seek a solution of
    the equations
    \begin{equation}\label{eq:nonhomosystem}
      \begin{aligned}
      \left[\mathbf{H}\right]
      \left[d\mathbf{x}\right] -
      \left[d\mathbf{y}\right] &=
      \left[d\boldsymbol{\mathfrak{p}}\right]\\
      \left[\mathbf{C}\right]
      \left[d\mathbf{x}\right]
      &=
      \left[d\mathbf{c}\right]
      \end{aligned}
    \end{equation}
    We assume that the constraint matrix $[\mathbf{C}]$
    is full rank and adequate to preclude rigid movements
    of the assembly.
    We also assume that the constraints are consistent,
    with $[d\mathbf{c}]$ in the range of $[\mathbf{C}]$,
    $[d\mathbf{c}]\in\mathcal{R}([\mathbf{C}])$.
    If Eq.~(\ref{eq:nonhomosystem}\textsubscript{2})
    is consistent,
    then for any generalized $\{1\}$-inverse
    $[\mathbf{C}]^{(1)}$,
    $[\mathbf{C}][\mathbf{C}]^{(1)}[d\mathbf{c}]=[d\mathbf{c}]$,
    since a $\{1\}$-inverse satisfies the weak property
    $[\mathbf{C}][\mathbf{C}]^{(1)}[\mathbf{C}]=[\mathbf{C}]$.
    A consistent Eq.~(\ref{eq:nonhomosystem}\textsubscript{2})
    has the general solution
    \begin{equation}
      \left[d\mathbf{x}\right] =
      \left[\mathbf{C}\right]^{(1)}
      \left[d\mathbf{c}\right] +
      \left(
        \left[\,\mathbf{I}\,\right]
        -
        \left[\mathbf{C}\right]^{(1)}
        \left[\mathbf{C}\right]
      \right)
      [d\mathbf{v}],\quad\forall
      [d\mathbf{v}]\in\mathbb{R}^{6N}
    \end{equation}
    for arbitrary $[d\mathbf{v}]$
    (see \S 2.1, \cite{BenIsrael2003a}).
    That is, $[d\mathbf{x}]$ is the sum of a solution
    of the inhomogeneous Eq.~(\ref{eq:nonhomosystem}\textsubscript{2})
    and any solution of its homogeneous complement,
    $[\mathbf{C}][d\mathbf{x}]=[\mathbf{0}]$.
    Substituting into
    Eq.~(\ref{eq:nonhomosystem}\textsubscript{1}),
    \begin{equation}\label{eq:Hvypc}
      \left[\mathbf{H}\right]
      \left(
        \left[\,\mathbf{I}\,\right]
        -
        \left[\mathbf{C}\right]^{(1)}
        \left[\mathbf{C}\right]
      \right)
      [d\mathbf{v}]
      - \left[d\mathbf{y}\right] =
      \left[d\boldsymbol{\mathfrak{p}}\right] -
      \left[\mathbf{H}\right]
      \left[\mathbf{C}\right]^{(1)}
      \left[d\mathbf{c}\right]
    \end{equation}
    The matrix $([\mathbf{I}]-[\mathbf{C}]^{(1)}[\mathbf{C}])$
    is idempotent and
    projects an arbitrary vector
    $[d\mathbf{v}]\in\mathbb{R}^{6N}$
    onto the null space
    of $[\mathbf{C}]$.
    This projection might be non-orthogonal (skew), but
    if we choose the Moore--Penrose inverse
    $[\mathbf{C}]^{\dagger}=
     [\mathbf{C}]^{\text{T}}([\mathbf{C}][\mathbf{C}]^{\text{T}})^{-1}$
    as the $\{1\}$-inverse,
    then the projection is orthogonal,
    and the projection matrices
    $[\mathbf{P}_{\mathcal{L}}]$ and $[\mathbf{P}_{\mathcal{L}^{\perp}}]$ in
    Eqs.~(\ref{eq:PL}) and~(\ref{eq:PLperp})
    are used in place of
    $([\mathbf{I}]-[\mathbf{C}]^{(1)}[\mathbf{C}])$
    and its complement.
    As such,
    Eq.~(\ref{eq:Hvypc}) is equivalent to the
    system
    \begin{equation}\label{eq:TypeIIIc}
      \left(\rule{0ex}{2.2ex}
      \left[\mathbf{H}\right]
      \left[\mathbf{P}_{\mathcal{L}}\right] +
      \left[\mathbf{P}_{\mathcal{L}^{\perp}}\right]
      \right)
      \left[d\mathbf{z}\right] =
      \left[d\boldsymbol{\mathfrak{p}}\right] -
      \left[\mathbf{H}\right]
      \left[\mathbf{C}\right]^{\dagger}
      \left[d\mathbf{c}\right]
    \end{equation}
    If the sum
    $([\mathbf{H}][\mathbf{P}_{\mathcal{L}}]+[\mathbf{P}_{\mathcal{L}^{\perp}}])$
    is non-singular, then
    \begin{align}\label{eq:dxdyBottIII}
      &
      \begin{aligned}
      \left[d\mathbf{x}\right] &=
        \left[\mathbf{C}\right]^{\dagger}\left[d\mathbf{c}\right] +
        \left[\mathbf{P}_{\mathcal{L}}\right]\left[d\mathbf{z}\right]
        \\
      &=\left[\mathbf{C}\right]^{\dagger}\left[d\mathbf{c}\right] +
      \left[\mathbf{P}_{\mathcal{L}}\right]
      \left(\rule{0ex}{2ex}
        \left[\mathbf{H}\right]
        \left[\mathbf{P}_{\mathcal{L}}\right] +
        \left[\mathbf{P}_{\mathcal{L}^{\perp}}\right]
      \right)^{-1}
      \left(
        \left[d\boldsymbol{\mathfrak{p}}\right] -
        \left[\mathbf{H}\right]
        \left[\mathbf{C}\right]^{\dagger}
        \left[d\mathbf{c}\right]
      \right)
      \end{aligned}
      \\\label{eq:dyBottIII}
      &\left[d\mathbf{y}\right] =
        -\left[d\boldsymbol{\mathfrak{p}}\right]
%        \left[\mathbf{H}\right]
%        \left[\mathbf{C}\right]^{\dagger}
%        \left[d\mathbf{c}\right] -
      +\left[\mathbf{H}\right]\left[d\mathbf{x}\right]
    \end{align}
    where $[d\mathbf{x}]$ is the solution vector of incremental
    displacements and rotations, and $[d\mathbf{y}]$ contains the
    incremental
    reaction forces of constraint (superposed on $[d\boldsymbol{\mathfrak{p}}]$)
    that are necessary to satisfy the movement
    constraints.
  \item[Type~IV constraint:]
     %\subsection{\normalsize Isolated systems}
     %This section might not be needed!
     %            Depends on whether examples
     %            of isolated systems are used.
    \par
    We now consider the special case of \emph{no displacement constraints}
    on the particles' movements,
    which we call an \emph{isolated system}.
    Isolated systems
    are the discrete analog of full load-control
    and were the sole focus of \cite{Kuhn:2005b}.
    Type~IV constraint
    (or lack of constraint) can be used to examine the intrinsic
    \emph{material behavior} of an assembly, without the
    disrupting influence of platens or other displacement
    constraints.
    \par
    Because such isolated systems are not anchored to a Gallilean foundation
    (imagine a granular asteroid tumbling through space),
    the stiffness matrix $[\mathbf{H}]$ is singular of rank
    $6N-6$, and Eq.~(\ref{eq:Hdudb}) represents a
    bijective mapping from
    the quotient space of deformation-producing
    movements $[d\mathbf{x}]$ to the
    subspace of $[d\mathbf{p}]$ that is comprised only of
    equilibrium forces.
    \par
    With the absence of any displacement constraints that would
    otherwise prevent rigid-body motions,
    two discrepancies arise when applying
    %special treatment must be given to
    the stiffness relationship in Eqs.~(\ref{eq:Hdudb})
    and~(\ref{eq:Hdxfrakp})
    and when applying the second-order work product
    $[d\mathbf{b}/d\mathbf{w}]^{\text{T}}[d\mathbf{u}/d\boldsymbol{\theta}]$
    in Eq.~(\ref{eq:W2}).
    First, the use of the stiffness
    $[\mathbf{H}]$ in Eq.~(\ref{eq:Hdudb}) will yield incorrect
    external forces
    $[d\mathbf{p}]=[d\mathbf{b}/d\mathbf{w}]$
    when the displacements
    $[d\mathbf{x}]=[d\mathbf{u}/d\boldsymbol{\theta}]$
    correspond to a
    \emph{rigid rotation} of an entire isolated assembly,
    as the existing forces $[\mathbf{b}/\mathbf{w}]$
    will rotate, in the manner of follower forces, during
    a rotation of the isolated assembly.
    Secondly, a rigid rotation of an isolated assembly
    (or alternatively, an observer rotation)
    will produce apparent alterations
    $[d\mathbf{p}]$ and $[d\mathbf{x}]$
    %$[d\mathbf{b}/d\mathbf{w}]$ and
    %$[d\mathbf{u}/d\boldsymbol{\theta}]$
    that have a \emph{non-zero} inner product,
    %$[d\mathbf{b}/d\mathbf{w}]^{\text{T}}[d\mathbf{u}/d\boldsymbol{\theta}]$.
    $[d\mathbf{p}]^{\text{T}}[d\mathbf{x}]\neq 0$.
    That is, the second-order work $\mathcal{W}_{2}$ in
    Eq.~(\ref{eq:W2}) is not objective.
    \par
    A solution to the second discrepancy was proposed in \cite{Kuhn:2005b},
    which involved projecting forces and movements onto a special
    subspace.
    Here, we resolve \emph{both} discrepancies by
    using a generalized matrix inverse.
    Consider a $6N\times 6$ matrix $[\mathbf{R}]$
    that shifts and rotates the entire assembly as a
    rigid body:
\begin{equation}
  \def\arraystretch{1.40}
  \left[
    \begin{array}{@{\extracolsep{\fill}}c@{\extracolsep{\fill}}}
      d\mathbf{u}^{\text{rigid}} \\\hdashline[1pt/1pt]
      d\boldsymbol{\theta}^{\text{rigid}}
    \end{array}
  \right]_{6N \times 1}
  =
  \left[ \mathbf{R} \right]_{6N\times 6}
  \def\arraystretch{1.40}
  \left[
    \begin{array}{@{\extracolsep{\fill}}c@{\extracolsep{\fill}}}
      d\overline{\mathbf{u}} \\\hdashline[1pt/1pt]
      d\overline{\boldsymbol{\theta}}
    \end{array}
  \right]_{6 \times 1}
\end{equation}
    where $d\overline{\mathbf{u}}$ and $d\overline{\boldsymbol{\theta}}$
    are the shift and rotation vectors that are applied uniformly
    to the entire assembly,
    and $d\mathbf{u}^{\text{rigid}}$ and
    $d\boldsymbol{\theta}^{\text{rigid}}$ are the resulting
    individual particle motions and rotations.
    \par
    For an isolated system, we seek solutions of the
    stiffness Eq.~(\ref{eq:Hdudb}) that preclude
    rigid-body motions, such that
    $[d\mathbf{x}]=[d\mathbf{u}/d\boldsymbol{\theta}]$ belongs
    to the left null-space of $[\mathbf{R}]$:
    \begin{equation}\label{eq:homogeneousR}
      \left[\mathbf{R}\right]_{6\times 6N}^{\text{T}}
      \left[d\mathbf{x}\right]%_{6N\times 1}
      =
      \left[\mathbf{0}\right]_{6\times 1}
    \end{equation}
    This restriction reduces the analysis of an isolated
    system to that of a system with homogeneous Type~II constraints,
    as in Eq.~(\ref{eq:homogeneousC}), and we
    can follow a similar approach for finding solutions and
    for identifying stiffness pathologies.
    As with Eqs.~(\ref{eq:PL})--(\ref{eq:PLperp}),
    the matrices $[\mathbf{P}^{\text{r-r}}]$
    and $[\mathbf{P}^{\text{n-r-r}}]$ project a general
    set of particle
    motions onto the sub-space of rigid-body motions
    and the sub-space that precludes rigid-body motions:
    \begin{align}\label{eq:Prr}
      \left[\mathbf{P}^{\,\text{r-r}}\right]_{6N\times 6N}
      &=
      \left[ \mathbf{R} \right]
      \left( \left[ \mathbf{R} \right]^{\text{T}}
      \left[ \mathbf{R} \right]\right)^{-1}
      \left[ \mathbf{R} \right]^{\text{T}}
      \\
      \label{eq:Pnrr}
      \left[ \mathbf{P}^{\,\text{n-r-r}}\right]_{6N\times 6N}
      &=
      \left[ \,\mathbf{I}\,\right]_{6N\times 6N}
      - \left[\mathbf{P}^{\text{r-r}}\right]
    \end{align}
    The consistency of the stiffness problem in
    Eqs.~(\ref{eq:Hdudb}) and~(\ref{eq:Hdxfrakp}) along with
    the constraints of Eq.~(\ref{eq:homogeneousR}) is equivalent
    to consistency of the equation
    \begin{equation}\label{eq:TypeIVa}
      \left(\rule{0ex}{2.2ex}
      \left[\mathbf{H}\right]
      \left[\mathbf{P}^{\text{n-r-r}}\right] +
      \left[\mathbf{P}^{\text{r-r}}\right]
      \right)
      \left[d\mathbf{z}\right] =
      \left[d\boldsymbol{\mathfrak{p}}\right]
    \end{equation}
    If the summed matrix on the left is non-singular,
    then this equation has the unique solution
    \begin{align}
      [d\mathbf{x}]
      &=
      \left[ \mathbf{P}^{\text{n-r-r}}\right]
      \left(\rule{0ex}{2.2ex}
      \left[\mathbf{H}\right]
      \left[\mathbf{P}^{\text{n-r-r}}\right] +
      \left[\mathbf{P}^{\text{r-r}}\right]
      \right)^{-1}
      \left[d\boldsymbol{\mathfrak{p}}\right]
      \\
      [d\mathbf{y}]
      &=
      -\left[d\boldsymbol{\mathfrak{p}}\right]
      +
      \left[\mathbf{H}\right] [d\mathbf{x}]
    \end{align}
    where $[d\mathbf{y}]$ are the reaction forces of
    constraint, distributed among all particles in the system,
    that are
    required to prevent a rigid-body motion of the system.
    The matrix $([\mathbf{H}][\mathbf{P}^{\text{n-r-r}}] +[\mathbf{P}^{\text{r-r}}])$
    in Eq.~(\ref{eq:TypeIVa}) can be used for examining
    questions of non-uniqueness and instability
    of granular systems
    (Section~\ref{sec:Pathologies}).
    %
%
%  The search for possible displacement bifurcations in an isolated
%  system requires another transformed stiffness,
%
%  \begin{equation}
%    \LB \mathbb{H} \RB =
%    \LB \mathbf{H} \RB
%    \LB \mathbf{P}^{\text{n-r-r}}\RB
%  \end{equation}
%
%  When multiplied by the displacements
%  $[d\mathbf{u}/d\boldsymbol{\theta}]$, this
%  stiffness matrix filters any displacments that correspond to
%  rigid rotations of the system (see Eq.~\ref{eq:Pnnr}),
%  yielding the force increments
%  $[\mathbbm{d}\mathbf{b}/\mathbbm{d}\mathbf{w}]$ that result
%  in the absence of such rigid rotation:
%
%  \begin{equation}
%    \LB \mathbb{H} \RB_{6N \times 6N}
%    \left[\begin{MAT}(r){l}
%          d\mathbf{u}\\:d\boldsymbol{\theta}\\ \end{MAT}
%    \right]_{6N \times 1}
%    = \left[\begin{MAT}(r){l}
%            \mathbbm{d}\mathbf{b}\\:
%            \mathbbm{d}\mathbf{w}\\ \end{MAT}
%      \right]_{6N \times 1}
%  \end{equation}
%
\end{description}
\subsection{\normalsize Linear-frictional contact model}\label{sec:contacts}
Many constitutive models have been proposed for the contact
force--displacement relationship between two bodies in contact,
such as the general form of Eqs.~(\ref{eq:constitutive})--(\ref{eq:constitutiveM}).
These models include
%of Eqs.~(\ref{eq:constitutive})--(\ref{eq:constitutiveM}),
%including the complex,
the complex, history-dependent models of Cattaneo \cite{Cattaneo:1938a}
and Mindlin \cite{Mindlin:1953a}
and of Kalker \cite{Kalker:1967a};
models that include resistive moments \cite{Iwashita:1998a};
and models that include incipient, grazing contact.
Among the simplest models is a zero-tension
zero-moment contact with
linear stiffnesses in both normal and tangential directions
but with a frictional limit on the tangential force
(for example, \cite{Michalowski:1978a,Radi:1999a}).
This \emph{linear-frictional} contact model with
no contact moment is illustrated in
Fig.~\ref{fig:Springs} and is widely used in
DEM simulations (e.g., \cite{Cundall:1979a}), and it
and will be exclusively applied in the paper's examples.
With this model,
the stiffness of a contact $pq$ is incrementally nonlinear
with two branches:
an elastic (no-slip) branch that is characterized
with positive normal and tangential stiffnesses
$k^{pq}$ and $\alpha k^{pq}$,
and a sliding (slip) branch that is
characterized by the friction coefficient
$\mu^{pq}$.
Whenever the friction limit is reached,
the active branch is
determined by the direction of the contact
deformation $\delta\mathbf{u}^{pq,\text{def}}$.
Sliding occurs %at a firm contact
when two conditions are met:
\begin{enumerate}
\item
When the current contact force has reached
the friction limit, satisfying the
yield condition $G^{pq}=0$:
\begin{equation}\label{eq:yield}
G^{pq} = G(\mathbf{f}^{pq} ) =
\left |
\mathbf{f}^{pq} - (\mathbf{n}^{pq} \cdot \mathbf{f}^{pq}) \mathbf{n}^{pq}
\right |
+
\mu \mathbf{f}^{pq} \cdot \mathbf{n}^{pq}
= 0
\end{equation}
This yield condition depends upon the current
contact force $\mathbf{f}^{pq}$,
which must be known %\emph{a priori}.
before the contact stiffness can be constructed.
Sliding also requires the second condition.
%With the isotropic frictional behavior
%in Eq.~(\ref{eq:yield}), the yield condition is axisymmetric
%within the contact plane
%(see \cite{Michalowski:1978a} for alternative, non-symmetric forms).
%
\item
When the contact deformation $\delta\mathbf{u}^{pq,\text{def}}$
is directed outward
from the yield surface (in displacement space):
%the condition $S^{pq}>0$:
%
\begin{equation}\label{eq:flow}
S^{pq} = S(\mathbf{f}^{pq} ,
           \delta\mathbf{u}^{pq,\text{def}}) = \mathbf{g}^{pq}
         \cdot \delta\mathbf{u}^{pq,\text{def}} > 0
%  k \left( \alpha \mathbf{h}^{pq} + \mu \mathbf{n}^{pq} \right) \cdot \Dudef > 0 \;,
\end{equation}
where vector $\mathbf{g}^{pq}$ is
normal to the yield surface:  %has the normal direction
\begin{equation}\label{eq:G}
\mathbf{g}^{pq} =
  k \left( \alpha \mathbf{h}^{pq} + \mu \mathbf{n}^{pq} \right)
\end{equation}
and the unit sliding direction $\mathbf{h}^{pq}$
is tangent to the contact plane
and aligned with the tangential component of the
current force $\mathbf{f}^{pq}$:
\begin{equation}\label{eq:Hdirection}
\mathbf{h}^{pq} = \frac{  \mathbf{f}^{pq} - (\mathbf{n}^{pq} \cdot\mathbf{f}^{pq} )
                       \mathbf{n}^{pq}   }
            {| \mathbf{f}^{pq} - (\mathbf{n}^{pq} \cdot\mathbf{f}^{pq} )
            \mathbf{n}^{pq} |}
\end{equation}
\end{enumerate}
With this simple linear-frictional
model and its hardening modulus of zero,
the contact stiffness tensor $\mathbf{F}^{pq}$
in Eq.~(\ref{eq:constitutive}) has two branches,
elastic and sliding (no-slip and slip), given by
\begin{equation}\label{eq:ContactF}
\mathbf{F}^{pq} = \begin{cases}
\mathbf{F}^{pq\text{, elastic}} =
  k\left[ \alpha\mathbf{I}_3 + (1-\alpha)\mathbf{n}^{pq}\otimes\mathbf{n}^{pq} \right]
  & \text{if } G^{pq}<0 \text{ or } S^{pq}\leq 0 \\
  \mathbf{F}^{pq\text{, sliding}} =
  \mathbf{F}^{pq\text{, elastic}} -
  \mathbf{h}^{pq}\otimes\mathbf{g}^{pq}
  & \text{if } G^{pq}=0 \text{ and } S^{pq}>0
\end{cases}
\end{equation}
where $\mathbf{I}_3$ is the $3\times 3$ identity matrix.
%(or $2\times 2$ for two-dimensional assemblies).
Because the sliding and yield directions do not coincide
($\mathbf{h}^{pq} \neq \mathbf{g}^{pq}$), sliding is non-associative.
The presence of the dyad
$\mathbf{h}^{pq}\otimes\mathbf{g}^{pq}$ creates a non-symmetric
contact stiffness in Eq.~(\ref{eq:ContactF}$_2$),
%is non-symmetric,
leading to a non-symmetric global mechanical stiffness
$[\mathbf{H}^{\text{m}}]$
%That is, $[\mathbf{H}^{\text{m}}]$ is non-symmetric
whenever one or more contacts have reached the friction limit
(with $G=0$) and are sliding (with $S>0$).
%and may lead to negative second-order work at the contact.
The sliding behavior possesses deviatoric associativity, however,
since the sliding direction $\mathbf{h}^{pq}$ is aligned with
the tangential component of the yield surface normal
$\mathbf{g}^{pq}$ \cite{Bigoni:2000a}.
\par
Another characteristic embedded in Eq.~(\ref{eq:ContactF})
is the possibility of negative second-order work at a contact
(i.e., contact weakening),
in which the second-order quantity
$\mathfrak{d}\mathbf{f}^{pq}\cdot\delta\mathbf{u}^{pq,\text{def}}$
becomes negative.
Substituting Eqs.~(\ref{eq:G})--(\ref{eq:Hdirection})
and assuming $k>0$,
the second-order work is negative at sliding contacts,
$G^{pq}=0$, that meet
the following displacement condition:
%
%\begin{equation}
%  G^{pq}=0,\; %\text{ and }
%  \mu\mathbf{h}^{pq}\cdot\delta\mathbf{u}^{pq,\text{def}}
%  > \mathbf{n}^{pq}\cdot\delta\mathbf{u}^{pq,\text{def}}
%  \quad\Rightarrow\quad
%  \mathfrak{d}\mathbf{f}^{pq}\cdot\delta\mathbf{u}^{pq,\text{def}}
%  <0
%\end{equation}
%
\begin{equation}\label{eq:negsec}
 % \begin{aligned}
  \delta\mathbf{u}^{\text{def}}\cdot
  \left[
  \mathbf{n}\otimes\mathbf{n}
  +
  \alpha\left(
    \mathbf{I}_{3}
    -\mathbf{n}\otimes\mathbf{n}
    -\mathbf{h}\otimes\mathbf{h}
  \right)
  -\mu\,\mathbf{h}\otimes\mathbf{n}
  \right]\cdot
  \delta\mathbf{u}^{\text{def}} < 0%\\
  \;\iff\;
  \mathfrak{d}\mathbf{f}^{pq}\cdot\delta\mathbf{u}^{pq,\text{def}}<0
 % \end{aligned}
\end{equation}
an expression that we have
shortened by removing the $pq$ superscripts.
This condition, when expressed in sufficient abundance among
the sliding contacts, can
lead to a cumulative negative second-order work of the entire assembly
(as in \cite{Nicot:2007b}).
%\todo{Scaling of Hm and Hg. Symmetry of the two parts.}
Equation~(\ref{eq:negsec})
can be simplified by noting that
$\delta\mathbf{u}\cdot
 (\mathbf{n}\otimes\mathbf{n})
 \cdot\delta\mathbf{u}$
equals the squared magnitude of the normal movement,
$|\delta\mathbf{u}^{\text{n,def},pq}|^{2}$,
and that the middle term is
the squared magnitude of the component of
movement that is orthogonal to both the contact normal
and the current tangential force
(i.e., the squared magnitude of the tangential
component of movement that is orthogonal
to the current tangential force,
see Eq.~\ref{eq:Hdirection}):
\begin{equation}
  \left|\delta\mathbf{u}^{\text{n,def},pq}\right|^{2} +
  \alpha \left|\delta\mathbf{u}^{\text{h}_{\perp},\text{def},pq}\right|^{2}
  -
  \mu\,\delta\mathbf{u}^{\text{def},pq}\cdot
  \left(
    \mathbf{h}^{pq}\otimes\mathbf{n}^{pq}
  \right)\cdot\delta\mathbf{u}^{\text{def},pq} > 0
  \;\iff\;
  \mathfrak{d}\mathbf{f}^{pq}\cdot\delta\mathbf{u}^{pq,\text{def}}<0
\end{equation}
Figure~\ref{fig:slip} summarizes the conditions for yielding
(i.e., $G^{pq}=0$), for sliding ($S^{pq}>0$), and for
negative second-order work, within a two-dimensional setting.
\begin{figure}
  \centering
  \includegraphics{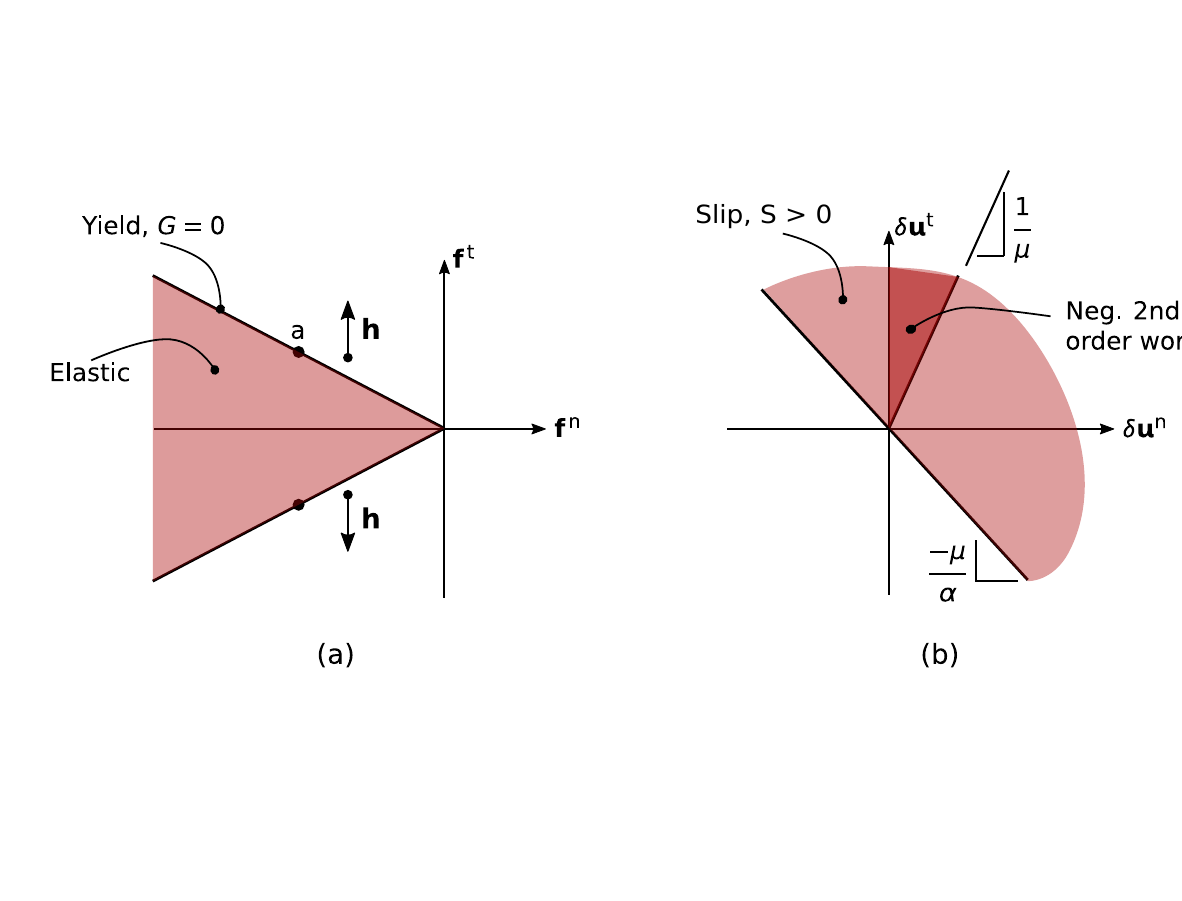}
  \caption{Conditions for yield, sliding, and negative
   second-order work in a two-dimensional setting:
   (a)~admissible force domain
   and the friction limit for normal and tangential
   contact forces, $f^{\,\text{n},pq}$
   and $f^{\,\text{t},pq}$.
   (b)~for the case ``a'' of forward sliding,
   the conditions for sliding and for negative second-order work
   with respect to the normal and tangential contact
   movements, $\delta u^{\text{n},pq}$ and
   $\delta u^{\text{t},pq}$.
   \label{fig:slip}}
\end{figure}
Negative second-order work can only occur when
the contact is being unloaded in the normal direction
($\mathbf{n}^{pq}\cdot\delta\mathbf{u}^{\text{def},pq}>0$) while
sliding continues in the direction of the current tangential
force ($\mathbf{h}^{pq}\cdot\delta\mathbf{u}^{\text{def},pq}>0$).
%\par
%We note that the contact stiffness
%$\mathbf{F}^{pq}$ in Eq.~(\ref{eq:constitutive})
%is not a continuous function of
%displacement $d\mathbf{u}^{pq,\text{def}}$, but rather
%changes abruptly when contact transitions between zones of elastic
%and sliding and plastic behavior (Eq.~{\ref{eq:ContactF}).
%Such discontinuous change will be shown to
%enable stable bifurcations of the global behavior.
\par
We had earlier assumed that the mapping between the contact
movement $\delta\mathbf{u}^{\text{def},pq}$ and the contact force
increment $\mathfrak{d}\mathbf{f}^{pq}$ is homogeneous
of degree~1 (see Eq.~\ref{eq:constitutive}),
and the stiffness in Eq.~(\ref{eq:ContactF})
is consistent with this assumption.
Although the mapping is incrementally non-linear and is not additive,
it is continuous, as the two stiffness branches
produce the same increment
increment $\mathfrak{d}\mathbf{f}^{pq}$ at the
transition hyper-plane
$S^{pq}=0$ \cite{Hashiguchi:1993a}.
\subsection{\normalsize Multiple stiffness branches}\label{sec:branches}
%
%Returning to the global
%stiffnesses in Eqs.~(\ref{eq:GeomMech}) and~(\ref{eq:GeomMech2}),
%the geometric stiffness $[\mathbf{H}^{\text{g}}]$
%Eqs.~(\ref{eq:GeomMech}) and~(\ref{eq:GeomMech2})
%is independent of the
%loading direction,
%provided that all particle pairs share, at most, a single contact
%and that the particles' surfaces have continuous
%curvatures at the contact point
%(thus preventing certain rocking modes).
The geometric stiffness
$[\mathbf{H}^{\text{g}}]$
in Eq.~(\ref{eq:GeomMech})
is independent of the loading direction.
However,
the incremental mechanical stiffness $[\mathbf{H}^{\text{m}}]$
can depend
upon the directions of movements at the contacts. 
For example,
with the simple linear--frictional contact model,
described in the previous section,
both elastic and sliding
contact stiffnesses are available at each
contact that has reached the friction limit,
%On the other hand,
%the slip condition for
%frictional contacts (Eq.~\ref{eq:yield})
%at a single contact or
%at the contacts, % within a granular assembly,
%which will lead to a combined
so that the mechanical stiffness of the entire assembly,
$[\mathbf{H}^{\text{m}}(\boldsymbol{\nu})]$,
is incrementally nonlinear with multiple stiffness branches,
whenever at least one contact has reached the friction limit.
As a second example, certain models of contact rolling-friction
involve \emph{four} branches of stiffness for a single
contact,
due to the coupling of translational and rotational
movements \cite{Ai:2011a}.
As another example, two particles can be in nascent
contact with zero force,
such that the particles' incremental approach mobilizes the
contact's stiffness, but incremental withdrawal maintains
the zero force.
Such nascent (grazing) contacts have three stiffness branches:
a non-slip approach, an approach that initiates slip,
and a zero-stiffness withdrawal.
In the following, we consider only the two-branch linear-frictional
model of Section~\ref{sec:contacts}.
%with the active branch dependent on the direction
%of the movement vector $[d\mathbf{u}/d\boldsymbol{\theta}]$.
\par
With linear-frictional contacts,
the sliding conditions of all $M$ contacts can be collected
in a single vector $[\boldsymbol{\mathfrak{s}}]$
by first gathering the vectors $\mathbf{g}^{pq}$ of
Eq.~(\ref{eq:G}) as the rows of a matrix $[\mathbf{G}]$,
\begin{equation}
\mathbf{g}^{pq}
\rightsquigarrow
\left[\mathbf{G}\right]_{2M\times 2(6M)}
\end{equation}
but placing rows of zeros into $[\mathbf{G}]$ for
those contacts that have not reached
the friction limit ($G^{pq}<0$ in Eq.~\ref{eq:yield}).
The $2M\times 1$ vector $[\boldsymbol{\mathfrak{s}}]$ is then computed as
\begin{equation}
  \left[\rule{0ex}{2.0ex}
  \boldsymbol{\mathfrak{s}}\left([d\mathbf{x}]\right)\right]_{2M\times 1}
  =
  \sign\left(\rule{0ex}{2.2ex}\left[\mathbf{G}\right]
  \left[d\mathbf{u}^{\text{def},pq}\right]\right)
  =
  \sign\left(\rule{0ex}{2.2ex}\left[\mathbf{G}\right]
  \left[\mathbf{B}\right]
  \left[d\mathbf{x}\right]\right)
\end{equation}
where the ``sign'' function
has a value $1$ (or $-1$) for a contact that is at
the friction limit and is sliding (or is non-sliding).
The function $\boldsymbol{\mathfrak{s}}(\cdot)$ has a
value of $0$ for each contact that has not reached the friction limit
($G^{pq}<0$).
Note that the $pq$ and $qp$ contact variants share the same value:
$\mathfrak{s}^{pq}=\mathfrak{s}^{qp}$.
\par
When at least one contact has reached the friction limit,
allowing the two branches in Eq.~(\ref{eq:ContactF}),
the global mechanical stiffness
$[\mathbf{H}^{\text{m}}(\boldsymbol{\nu})]$
in Eq.~(\ref{eq:GeomMech}) will have multiple branches.
The various branches of
$[\mathbf{H}^{\text{m}}(\boldsymbol{\nu})]$
comprise a set
$\{\mathbf{H}^{\text{m}}\}$ of
matrices
$[\mathbf{H}^{\text{m,1}}]$,
$[\mathbf{H}^{\text{m,2}}]$,
$[\mathbf{H}^{\text{m,3}}]$, $\ldots\,$,
$[\mathbf{H}^{\text{m,}B}]$,
with a total of $B$ such branches,
corresponding the $B$ branches of the total stiffness:
\begin{equation}\label{eq:Himg}
  \left[\mathbf{H}^{i}\right]=
         \left[\mathbf{H}^{\text{m},i}\right]
         + \left[\mathbf{H}^{\text{g}}\right],\quad
  i = 1,2,\ldots,B
\end{equation}
Because linear--frictional contact behavior
is homogeneous of degree one
(as in Eqs. \ref{eq:constitutive}--\ref{eq:constitutiveM}),
the active branch
$[\mathbf{H}^{\text{m}}(\boldsymbol{\nu})]
\in\{\mathbf{H}^{\text{m}}\}$
is determined by the unit
direction $[\boldsymbol{\nu}]_{6N\times 1}$
of the movement vector $[d\mathbf{x}]$,
with
$[\boldsymbol{\nu}]=[d\mathbf{x}]/\sqrt{[d\mathbf{x}]^{\text{T}}[d\mathbf{x}]}$.
Each pair of conditions of the type in Eqs.~(\ref{eq:flow})
and~(\ref{eq:ContactF}) separates
the $\mathbb{R}^{6N}$ space of displacements $[d\mathbf{x}]$
into pairs of half-spaces, so that
$\mathbb{R}^{6N}$
%of displacement vectors $[d\mathbf{x}]$
is partitioned into $B$ convex pointed polygonal cone regions,
$[d\mathbf{x}]\in \Omega^{i}$, $i = 1,2,\ldots,B$.
Each cone $\Omega^{i}$ is formed from the intersection
of half-planes defined by the inequality in Eq.~(\ref{eq:flow}).
Each cone can be characterized with
a vector $[\mathbf{s}^{i}]_{2M\times 1}$ that is filled with
0's and 1's, with
0's for contacts that have not reached the friction limit,
and 1's for contacts at the friction limit.
%or are corresponding to contacts that are non-sliding but at the friction limit,
%to sliding contacts,
%and to contacts not yet at the friction limit,
%as described in the previous paragraph.
As such, a region $\Omega_{i}$ is defined as
%each with its particular stiffnesses $[\mathbf{H}^{\text{m}}]$
%and $[\mathbf{H}]$.
%
\begin{equation}\label{eq:condition}
\Omega^{i}=
\left\{[d\mathbf{x}]\in\mathbb{R}^{6N}\!:\;
   \left[\rule{0ex}{2ex}\boldsymbol{\mathfrak{s}}\left([d\mathbf{x}]\right)=1\right]
    =\left[\mathbf{s}^{i}\right]\right\}
,\quad
\bigcup_{i=1}^{B}\Omega^{i}
=
\mathbb{R}^{6N}
\end{equation}
and the union of these disjoint regions is a covering
of the full displacement space $\mathbb{R}^{6N}$.
In this expression,
$\left[\rule{0ex}{2ex}\boldsymbol{\mathfrak{s}}\left([d\mathbf{x}]\right)=1\right]$
is a vector of 0's and 1's, with 1's placed wherever
$\boldsymbol{\mathfrak{s}}\left([d\mathbf{x}]\right)=1$
and 0's elsewhere.
\par
%
%$[d\mathbf{u}/d\boldsymbol{\theta}]/|[d\mathbf{u}/d\boldsymbol{\theta}]|$.
%\par
%Although incrementally nonlinear, we assume that the
%incremental mapping
%$[\mathbf{H}^{\text{m}}(\boldsymbol{\nu})]:\,
%[d\mathbf{u}/d\boldsymbol{\theta}]\rightarrow[d\mathbf{b}/ d\mathbf{w}]$
%is continuous and piece-wise linear, so that two adjacent branches share the
%same stiffness along their shared boundary, and the behavior is linear within
%each branch
%(for example, see \cite{DarveOrLoret} for a discussion of
%such constitutive cones in
%a continuum setting).
%The simple linear-frictional
%model of a single contact,
%described in Section~\ref{sec:contacts},
%leads to an incrementally nonlinear mapping
%$[\mathbf{H}^{\text{m}}(\boldsymbol{\nu})]$
%having this piece-wise linear characteristic.
We now replace Eqs.~(\ref{eq:Hdxfrakp})--(\ref{eq:GeomMech})
with the following incrementally non-linear stiffness relation,
which specifically applies to the linear--frictional contact model
of the previous section:
\begin{equation}\label{eq:Hidxdp}
        \left[\mathbf{H}^{i}\right]
         \left[d\mathbf{x}\right] =
         \left[d\boldsymbol{\mathfrak{p}}\right]\\
         \quad
       \forall
       \left[d\mathbf{x}\right]
       \in\Omega^{i}
\end{equation}
where $[\mathbf{H}^{i}]$ is the sum of the mechanical
stiffness $[\mathbf{H}^{\text{m},i}]$ for the particular
loading direction and the generic geometric stiffness
$[\mathbf{H}^{\text{g}}]$
that applies to all loading directions (Eq.~\ref{eq:Himg}).
%
%\begin{equation}
%\left[\mathbf{H}^{i}\right]=
%  \left[\mathbf{H}^{\text{m},i}\right]
%  + \left[\mathbf{H}^{\text{g}}\right]
%\end{equation}
%
%where $[\mathbf{H}^{\text{m},i}]=[\mathbf{H}^{\text{m}}(\boldsymbol{\nu}^{i})]$.
%
%and ``$\forall$'' means ``for every.''
%Because the regions $\Omega^{i}$ are convex,
%Eq.~(\ref{eq:Hidxdp}\textsubscript{1})
%has the additive and homogeneous properties
%\begin{equation}
% \left.\begin{aligned}
%        &\left[\mathbf{H}^{i}\right]
%         \left[d\mathbf{x}^{a}\right] +
%         \left[\mathbf{H}^{i}\right]
%         \left[d\mathbf{x}^{b}\right] =
%         \left[\mathbf{H}^{a}\right]
%         \left(\left[d\mathbf{x}^{a}\right]
%         + \left[d\mathbf{x}^{b}\right]\right)\\
%        &\left[\mathbf{H}^{i}\right]=
%         \left[\mathbf{H}^{\text{m},i}\right]
%         + \left[\mathbf{H}^{\text{g}}\right]
%       \end{aligned}
% \right\}
%       \forall
%       \left[d\mathbf{x}^{a}\right],
%       \left[d\mathbf{x}^{b}\right]
%       \in\Omega^{i},\;
%       \lambda\in \left[0,\infty \right)
%\end{equation}
\par
With the linear--frictional contact model,
a contact has a single stiffness if it is elastic
($G<0$ in Eq.~\ref{eq:yield}), but it has two stiffness branches
when the friction limit (yield surface) has been reached.
If $M^{\text{limit}}$ represents the number of contacts that
are known to be at the friction limit
(and are potentially sliding, having a $G^{pq}=0$),
then the combined stiffness
$[\mathbf{H}^{\text{m}}(\boldsymbol{\nu})]$
has $B=2^{M^{\text{limit}}}$ branches.
For example,
a granular assembly with three potentially sliding contacts has
$2^{3}=8$ branches.
Each of the stiffness branches
$[\mathbf{H}^{i}]$ will, in general, be non-symmetric:
the geometric contributions
$[\mathbf{H}^{\text{g-1}}],\ldots,[\mathbf{H}^{\text{g-4}}]$
are non-symmetric, and the mechanical stiffness
$[\mathbf{H}^{\text{m},i}]$ will be non-symmetric
for a branch that involves any contact slip.
%The active branch is determined by applying the $M^{\text{s}}$
%independent sliding conditions in various combinations,
%each in the form of Eq.~(\ref{eq:flow}), noting that
%the vector of all contact movements,
%$[d\mathbf{u}^{pq,\text{def}}]$, is determined by applying the kinematics
%matrix $[\mathbf{B}]$ to the particle movements $[d\mathbf{x}]$,
%as in Eq.~(\ref{eq:Bmatrix}).
%\par
%We had noted that the stiffness
%$\mathbf{F}^{pq}$ of a single contact is a discontinuous function
%of contact movement $d\mathbf{u}^{pq,\text{def}}$
%and changes abruptly at the transition from elastic to sliding
%behaviors.
%As a result, stiffness
%$[\mathbf{H}(\boldsymbol{\nu})]$ is a discontinuous function
%of the radial vector $[\boldsymbol{\nu}]$.
%
%
%
%
\section{\large Stiffness pathologies}\label{sec:Pathologies}
%
%\todo[inline]{Rewrite this section.  Include other forms of instability.
%              Incoroporate the 4 types of constraint, using generalized
%              inverses.}
%
We now define three pathologies of the stiffness matrix
$[\mathbf{H}(\boldsymbol{\nu})]$.
Because
each pathology is associated with particular
directions of movement $[d\mathbf{x}]$, their
definitions are necessarily restricted in two ways:
\begin{enumerate}
\item
The displacements $[d\mathbf{x}]$ must be compatible with any
external constraints,
and each pathology will be placed in the
context of the constraints developed
in Section~\ref{sec:constraints}.
\item
The deformation direction
$[\boldsymbol{\nu}]=[d\mathbf{x}]/\sqrt{[d\mathbf{x}]^{\text{T}}[d\mathbf{x}]}$
that is associated with the movements $[d\mathbf{x}]$
of a pathology must be compatible
with the stiffness matrix $[\mathbf{H}(\boldsymbol{\nu})]$
of the particular pathology.
For systems with linear--frictional contacts,
the restricted domain $\Omega^{i}$
in Eq.~(\ref{eq:Hidxdp}) must be enforced, and
this restriction leads to results that differ from
those of elastic systems.
\end{enumerate}
The three pathologies are neutral equilibrium,
bifurcation (and path instability),
and instability of equilibrium.
%
%\begin{itemize}
  \subsection{\normalsize Neutral equilibrium}\label{sec:NE}
  Neutral equilibrium is a condition of non-uniqueness
  and is the existence of
  adjacent displaced equilibrium configurations
  that can be reached with neutral (zero) loadings
  that are infinitely close to the current state.
  Also called a \emph{loss of control} or \emph{divergence instability},
  neutral equilibrium can be considered
  a form of incipient bifurcation, but one in which
  the adjacent equilibrium
  states correspond to
  neutral loading, $[d\boldsymbol{\mathfrak{p}}]=[\mathbf{0}]$, and
  lie within
  the same domain $\Omega^{i}$ of a
  stiffness branch  $[\mathbf{H}^{i}]$.
  That is, the null space of the stiffness matrix
  $[\mathbf{H}^{i}]$ includes non-zero
  displacement vectors that lie within
  the branch's domain $\Omega^{i}$.
  Setting aside
  (for the moment) the matter of displacement constraints,
  neutral equilibrium is defined in a generic manner as
  \begin{equation}\label{eq:NE1}
    \exists
    \left[d\mathbf{x}\right]\in\Omega^{i}\setminus [\mathbf{0}]:\;\;
    \left[\mathbf{H}^{i}\right]
    \left[d\mathbf{x}\right]
    =
    \left[\mathbf{0}\right]
  \end{equation}
  When this criterion is met,
  Eq.~(\ref{eq:Hidxdp}) has multiple solutions
  $[d\mathbf{x}]$ for those vectors $[d\boldsymbol{\mathfrak{p}}]$
  within the range of $[\mathbf{H}^{i}]$, that is, for
  vectors $[d\boldsymbol{\mathfrak{p}}]\in\mathcal{R}([\mathbf{H}^{i}])$.
  On the other hand, for loading vectors
  $[d\boldsymbol{\mathfrak{p}}]$ that lie
  outside of
  this range (i.e., vectors within the left null space
  of $[\mathbf{H}^{i}]$), Eq.~(\ref{eq:Hidxdp})
  has no available solution
  $[d\mathbf{x}]\in\Omega^{i}$,
  even for displacements that are infinitely large.
  Again, without yet considering displacement constraints,
  neutral equilibrium implies singularity
  (i.e. zero determinant) of a
  stiffness branch $[\mathbf{H}^{i}]$,
  for which the null space includes vectors within its domain
  $\Omega^{i}$:
  \begin{equation}\label{eq:NE2}
    \exists
    \left[\mathbf{H}^{i}\right]:\;\;
    \detr\left([\mathbf{H}^{i}]\right)=0,\;
    \mathcal{N}\left([\mathbf{H}^{i}]\right)\cap\Omega^{i}\neq[\mathbf{0}]
  \end{equation}
  An alternative definition of neutral equilibrium is
  the existence of a stiffness branch
  $[\mathbf{H}^{i}]$
  with a zero eigenvalue and for which
  the corresponding eigenvector
  lies within $\Omega^{i}$.
  If $\{\lambda([\mathbf{Z}])\}$ is the set of
  eigenvalues of a square matrix $[\mathbf{Z}]$,
  and
  $\{\boldsymbol{\eta}_{0}([\mathbf{Z}])\}$
  is the subspace spanned by those eigenvectors whose
  eigenvalues are zero, then
  we can write an alternative (generic) definition as
  \begin{equation}\label{eq:NE3}
    \exists
    \left[\mathbf{H}^{i}\right]:\;\;
    0\in \{\lambda([\mathbf{H}^{i}])\},\;
    \{\boldsymbol{\eta}_{0}([\mathbf{H}^{i}])\}
    \cap
    \Omega^{i}
    \neq
    [\mathbf{0}]
  \end{equation}
  The three Eqs.~(\ref{eq:NE1}), (\ref{eq:NE2}), and~(\ref{eq:NE3})
  are generic definitions of neutral equilibrium that do
  not account for any constraints on the displacements.
  \par
  In Table~\ref{table:NE}, these three definitions are placed
  in the context of
  the four types of constraints of Section~\ref{sec:constraints}.
  For Type~I constraint, the stiffness problem can
  be written in the form of Eqs.~(\ref{eq:IHH})--(\ref{eq:IHH}),
  and the coefficient matrix on the left of these equations
  is singular if and only if $[\mathbf{H}^{\text{f{}f},i}]$
  is singular.
  For constraints of Types~II, III, and~IV,
  neutral equilibrium is associated with singularity of the
  matrices on the left of Eqs.~(\ref{eq:TypeIIc}), (\ref{eq:TypeIIIc}),
  and~(\ref{eq:TypeIVa}).
  \begin{table}
    \renewcommand{\arraystretch}{1.3}
    \centering
    \caption{Neutral equilibrium with four types of displacement
             constraints.\rule[-2ex]{0ex}{2ex}\label{table:NE}}
    \begin{tabular}{@{}rl}
    \toprule
    \multicolumn{2}{@{}l}{Generic neutral equilibrium
                       (without considering constraints,
                       Eqs.~\ref{eq:NE1}--\ref{eq:NE3})}\\
    \quad a) &
      $\exists\, [d\mathbf{x}]\in\Omega^{i}\setminus[\mathbf{0}]:\;\;
       [\mathbf{H}^{i}]
       [d\mathbf{x}] = [\mathbf{0}]$\\
    b) &
      $\exists\, [\mathbf{H}^{i}]:\;\;
      \det ([\mathbf{H}^{i}])=0,\;
      \mathcal{N}([\mathbf{H}^{i}]) \cap \Omega^{i} \neq [\mathbf{0}]$\\
    c) &
      $\exists\,[\mathbf{H}^{i}]:\;\;
      0\in\{\lambda([\mathbf{H}^{i}])\},\;
      \{\boldsymbol{\eta}_{0}([\mathbf{H}^{i}])\}\cap\Omega^{i}\neq[\mathbf{0}]$\\
    \midrule
    \multicolumn{2}{@{}l}{Type~I constraint$^{\ast}$}\\
    a) &
      $\exists\,[d\mathbf{x}]=
%     [d\mathbf{x}^{\text{c}}/d\mathbf{x}^{\text{f}\,}]
      [ \mathbf{0}_{r\times 1} \times d\mathbf{x}^{\text{f}\,}]
      \in\Omega^{i}\setminus [\mathbf{0}]:\;\;
       [\mathbf{H}^{\text{f{}f},i}][d\mathbf{x}^{\text{f}\,}]=[\mathbf{0}]$\\
    b) &
      $\exists\,[\mathbf{H}^{i}]:\;\;
       \det ([\mathbf{H}^{\text{f{}f},i}])=0,\;
       [ \mathbf{0}_{r\times 1} \times
       \mathcal{N}([\mathbf{H}^{\text{f{}f},i}])]
       \cap \Omega^{i}\neq[\mathbf{0}]$\\
     c) &
       $\exists\,[\mathbf{H}^{i}] :\;\;
        0\in\{\lambda([\mathbf{H}^{\text{f{}f},i}])\},\;
        [ \mathbf{0}_{r\times 1} \times
         \{\boldsymbol{\eta}_{0}([\mathbf{H}^{\text{f{}f},i}])\}]
         \cap\Omega^{i}\neq[\mathbf{0}]$\\
    \midrule
    \multicolumn{2}{@{}l}{Types~II and~III constraints,
                          $[\mathbf{X}^{i}]=
                           [\mathbf{H}^{i}]
                           [\mathbf{P}_{L}]
                           +[\mathbf{P}_{L^{\perp}}]$}\\
    a) &
      $\exists\,[d\mathbf{z}],\;[\mathbf{P}_{\mathcal{L}}][d\mathbf{z}]
                                \in\Omega^{i}\setminus [\mathbf{0}]:\;\;
       [\mathbf{X}^{i}][d\mathbf{z}]=[\mathbf{0}]$\\
    b) &
      $\exists\,[\mathbf{H}^{i}]:\;\;
       \det ([\mathbf{X}^{i}])=0,\;
       [\mathbf{P}_{\mathcal{L}}][\mathcal{N}([\mathbf{X}^{i}])]
       \cap \Omega^{i}\neq[\mathbf{0}]$\\
     c) &
       $\exists\,[\mathbf{H}^{i}]:\;\;
        0\in\{\lambda([\mathbf{X}^{i}])\},\;
        [\mathbf{P}_{\mathcal{L}}]
        [\{\boldsymbol{\eta}_{0}([\mathbf{X}^{i}])\}]
         \cap\Omega^{i}\neq[\mathbf{0}]$\\
    \midrule
    \multicolumn{2}{@{}l}{Type~IV constraint,
         $[\mathbf{X}^{i}]=[\mathbf{H}^{i}][\mathbf{P}^{\text{n-r-r}}] +
          [\mathbf{P}^{\text{r-r}}]$}\\
    a) &
      $\exists\,[d\mathbf{z}],\;[\mathbf{P}^{\text{n-r-r}}][d\mathbf{z}]
                                \in\Omega^{i}\setminus [\mathbf{0}]:\;\;
       [\mathbf{X}^{i}][d\mathbf{z}]=[\mathbf{0}]$\\
    b) &
      $\exists\,[\mathbf{H}^{i}]:\;\;
       \det ([\mathbf{X}^{i}])=0,\;
       [\mathbf{P}^{\text{n-r-r}}][\mathcal{N}([\mathbf{X}^{i}])]
       \cap \Omega^{i}\neq[\mathbf{0}]$\\
     c) &
       $\exists\,[\mathbf{H}^{i}]:\;\;
        0\in\{\lambda([\mathbf{X}^{i}])\},\;
        [\mathbf{P}^{\text{n-r-r}}]
        [\{\boldsymbol{\eta}_{0}([\mathbf{X}^{i}])\}]
         \cap\Omega^{i}\neq[\mathbf{0}]$\\
    \midrule
      \multicolumn{2}{@{}p{10.5cm}}{%
      {}$^{\ast}$ notation ``$\mathbf{0}_{r\times 1} \times$''
      means that $r$ zeros are placed in the
      constrained ``c'' locations of $[d\mathbf{x}]$.}\\
    \bottomrule
    \end{tabular}
  \end{table}
  \par
  Although, in principle, Eqs.~(\ref{eq:NE1})--(\ref{eq:NE3})
  provide criteria for
  determining neutral equilibrium,
  their implementation can present vexing problems,
  particularly when the dimension of the null-space
  $\mathcal{N}([\mathbf{H}^{i}])$ of a stiffness branch
  $[\mathbf{H}^{i}]$ is greater than~1.
  When the dimension is simply~1, the problem is more straightforward.
  One simply determines whether the single basis vector $[d\mathbf{x}]$
  of the null space, or its reversal $[-d\mathbf{x}]$, belong
  to the region $\Omega^{i}$, by testing the
  condition in Eq.~(\ref{eq:condition}),
  with both
  $\left[\rule{0ex}{2ex}\boldsymbol{\mathfrak{s}}\left([d\mathbf{x}]\right)=1\right]$
  and
  $\left[\rule{0ex}{2ex}\boldsymbol{\mathfrak{s}}\left([-d\mathbf{x}]\right)=1\right]$.
  Note that $\Omega^{i}$ is a pointed cone,
  so both $[d\mathbf{x}]$ and $[-d\mathbf{x}]$ must be tested.
  The situation is more complex when the null space is of dimension
  greater than 1.
  In this case, all of the basis vectors could lie outside of
  $\Omega^{i}$, even though a linear combination of the vectors
  can lie within $\Omega^{i}$ (see Fig.~\ref{fig:Nonconsistent}).
  \begin{figure}
    \centering
    \includegraphics{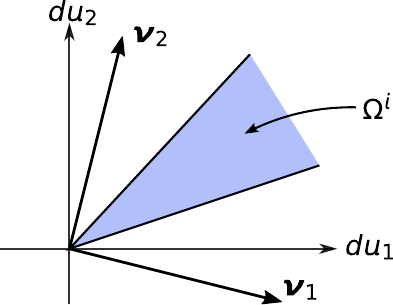}
    \caption{Example in which neither of the basis vectors of
             the null space (vectors $\boldsymbol{\nu}_1$
             and $\boldsymbol{\nu}_{2}$) lies within the
             convex region $\Omega^{i}$,
             but a linear combination of the vectors would
             lie within $\Omega^{i}$.
             \label{fig:Nonconsistent}}
  \end{figure}
  One must, therefore, check the full sub-space spanned by the
  basis vectors.
  Although Farkas' lemma might provide an elegant means
  of determining whether the sub-space intersects $\Omega^{i}$
  (thus indicating the existence of a legitimate null-vector),
  in the examples of Section~\ref{sec:Examples} we used
  a brute-force approach:  we maximized the smallest
  element-wise product $\mathfrak{s}_{i}s_{i}$ in
  Eq.~(\ref{eq:condition}) and tested whether the largest product
  was positive
  (the search was made on the unit cube $[-1,1]^{m}$, where $m$
  is the dimension of the null-space).
  \par
  Another difficulty can arise
  when the criteria of Eqs.~(\ref{eq:NE1})--(\ref{eq:NE3})
  are applied to realistic irregular assemblies of particles
  (as with the simulation data of Section~\ref{sec:example3}).
  When a time-stepping algorithm is used to adjust the
  particle positions,
  a granular system can momentarily pass through the condition of
  neutral equilibrium during a single time step,
  so that the determinant of Eq.~(\ref{eq:NE2}) is non-zero
  at the beginning and end of the step,
  although the neutral equilibrium
  condition (and the possibility of uncontrollable loading)
  would occur within the step.
  In this situation, the condition of neutral equilibrium
  would not be recognized.
  In analyzing such numerical data, we tested the
  \emph{condition number} of matrix $\mathbf{H}^{i}$ (i.e., the ratio
  of the largest and smallest singular values found
  in the decomposition
  $[\mathbf{H^{i}}]=[\mathbf{U}][\boldsymbol{\Sigma}][\mathbf{V}]$)
  to determine the near-singularity of $\mathbf{H}^{i}$.
  When a large condition number was detected,
  we applied the Eckart--Young--Mirsky principle to find
  the near-null space of $[\mathbf{H}^{i}]$,
  isolating those vectors within $[\mathbf{V}]$ that corresponded to
  the smallest singular values.
  These basis vectors were then tested for consistency with the
  domain $\Omega^{i}$.
  \par
  We finish this section by noting differences between neutral
  equilibrium and the \emph{loss of rigidity}
  \cite{Agnolin:2007a,Agnolin:2007c}.
  The analysis of rigidity (and the related conditions
  of force staticity or indeterminacy)
  are familiar to physicists and structural engineers, and the analysis
  treats a system as if it is composed of rigid components:
  for example, rigid particles with rolling but
  rigid, non-compliant contacts. 
  A rigid system is one that precludes articulated displacement
  modes in which no resistive forces are mobilized among the
  components; similarly, a hyperstatic or indeterminate
  system is one
  that admits a space of non-unique contact forces to support
  a given set of external forces.
  These conditions are established by considering only the
  kinematics (rigidity) matrix $[\mathbf{B}]$ of Eq.~(\ref{eq:Bmatrix})
  (or its transpose, the statics
  matrix $[\mathbf{A}]$ of Eq.~\ref{eq:Equilbrium0a})
  and apply when the null space
  of $[\mathbf{B}]$ exceeds 6 (the null space can be no smaller,
  since six modes of rigid translation and rotation produce
  no relative movements among the components).
  Loss of rigidity occurs when the null space of
  $[\mathbf{B}]$ exceeds 6, thus admitting articulated displacement
  modes.
  Rigidity depends entirely upon $[\mathbf{B}]$,
  which is computed for the undisplaced condition;
  whereas, neutral equilibrium (or bifurcation and instability,
  as described in the next sections) depends upon the full
  stiffness $[\mathbf{H}]$, which incorporates both contact and
  geometric stiffnesses.
  The mechanical stiffness $[\mathbf{H}^{\text{m}}]$
  involves a product with the matrix $[\mathbf{B}]$
  (see Eq.~\ref{eq:Hmprod}), so that the
  rank of $[\mathbf{H}^{\text{m}}]$
  can not exceed the rank of $[\mathbf{B}]$.
  The geometric stiffnesses, however,
  are not directly computed as products with
  $[\mathbf{B}]$, so that
  loss of rigidity does not imply neutral equilibrium, since articulated
  modes can be resisted by alterations of the system's geometry, 
  even though these modes
  do not mobilize any of the contacts' stiffnesses.
  \subsection{\normalsize Bifurcation and path instability}\label{sec:BPI}
  Bifurcation, whether stable or unstable, is the
  existence of multiple displaced equilibrium paths that
  emanate from the current equilibrium state for the particular
  loading increment $[d\boldsymbol{\mathfrak{p}}]$.
  Unlike neutral equilibrium (Section~\ref{sec:NE})
  and instability of an
  equilibrium state (Section~\ref{sec:IofE}),
  which are associated with a state,
  bifurcation involves a choice of \emph{equilibrium paths}
  and is associated with a deformation
  process (path):
  bifurcation is a condition of path rather than state
  \cite{Petryk:1991a}.
  Another difference from neutral equilibrium
  is that the stiffness matrix at incipient bifurcation
  retains full rank, and
  the multiple solutions of
  Eq.~(\ref{eq:Hidxdp}) are for different stiffness
  branches $[\mathbf{H}^{i}]$,
  each with a direction $[d\mathbf{x}]$
  that lies within its separate
  domain $\Omega^{i}$.
  The canonical example is the Shanley
  column, for which increased loading~--- rather
  than neutral loading~---
  can occur along multiple bifurcation branches
  and for which states
  along each branch exhibit stability of equilibrium
  (see \cite{Hill:1960a} and
  \cite{Shanley:1947a}, which includes the succinct discussion
  of von~K\'{a}rm\'{a}n).
  %Whether bifurcation occurs depends upon the
  %loading increment $[d\mathbf{p}]$, and
  \par
  Ignoring, for the moment, any
  constraints on displacement,
  bifurcation along a loading path $[d\boldsymbol{\mathfrak{p}}]$
  is defined as the existence of multiple solutions,
  say ``a'' and ``b'' solutions, each lying
  within a separate stiffness domain:
  \begin{equation}\label{eq:defBif}
    \exists
    \left[d\mathbf{x}^{a}\right]
    \in\Omega^{a},\;
    \left[d\mathbf{x}^{b}\right]
    \in\Omega^{b},\;
    \Omega^{a}\neq\Omega^{b}
    :\quad
    \left[\mathbf{H}^{a}\right]
    \left[d\mathbf{x}^{a}\right]
    =
    \left[\mathbf{H}^{b}\right]
    \left[d\mathbf{x}^{b}\right]
    =
    \left[d\boldsymbol{\mathfrak{p}}\right]
  \end{equation}
  Contrarily, a sufficient (but not necessary) condition
  for the uniqueness of a solution $[d\mathbf{x}^{0}]\in\Omega^{0}$
  is the Hill condition
  \begin{equation}
    \left(
      \left[\mathbf{H}\right]
      \left[d\mathbf{x}\right] -
      \left[\mathbf{H}^{0}\right]
      \left[d\mathbf{x}^{0}\right]
    \right)^{\text{T}}
    %\cdot
    \left(\left[d\mathbf{x}\right] - \left[d\mathbf{x}^{0}\right]\right)
    > 0
    ,\quad
    \forall\, \Omega^{i},\;\forall\,
    \left[d\mathbf{x}\right]\in\Omega^{i},
    \left[d\mathbf{x}\right]\neq\left[d\mathbf{x}^{0}\right]
  \end{equation}
  that is, a positive value is found
  for all $[d\mathbf{x}]$ that are consistent with the
  prescribed loading increments $[d\boldsymbol{\mathfrak{p}}]$,
  such that $[\mathbf{H}][d\mathbf{x}]=[d\boldsymbol{\mathfrak{p}}]$
  (see \cite{Hill:1959a,Hill:1961a,Petryk:1991a}).
%  \todo{Rewrite, since interior body forces are zero, only
%        boundary forces and displacements are involved.}
   \par
   Table~\ref{table:bif} presents the bifurcation criterion
   of Eq.~(\ref{eq:defBif}) in the contexts of
   the four types of displacement constraints
   in Section~\ref{sec:constraints}.
   With Type~I constraint, we must check consistency with
   a domain $\Omega^{i}$ by reshuffling the constrained
   displacements $[d\mathbf{x}^{\text{c}}]$ into the
   solved displacements $[d\mathbf{x}^{\text{f}}]$.
   This difficulty is avoided with the other constraint types.
   \par
   When multiple equilibrium
   solutions exist, one should then determine which of the
   alternative solutions
   is followed as the more \emph{stable path}.
   Ba\v{z}ant \cite{Bazant:1988b,Bazant:1989b}
   addressed this issue by deriving the
   increments of internal entropy
   along each solution path of the external force
   $[d\mathbf{p}]$ and its
   associated movement $[d\mathbf{x}]$~--- an increase
   $\Delta S_{\text{in}}$ that resembles the incremental
   form $[d\mathbf{p}]^{\text{T}}[d\mathbf{x}]$
   of the second-order external
   work $\mathcal{B}_{2}$ in Eq.~(\ref{eq:W2}).
   However, Ba\v{z}ant noted that
   components of the vectors $[d\mathbf{x}]$ and $[d\mathbf{p}]$
   must be treated differently for the separate cases of controlled
   movements and of controlled forces:
   (1)~for those movements that are controlled,
   the subset $[d\mathbf{x}^{\text{c}}]$
   is paired with the
   multiple sets of complementary restraining
   forces $[d\mathbf{p}^{\text{c},\alpha}]$
   along the alternative $\alpha$-branches; and
   (2)~for those external forces that are controlled,
   the subset $[d\mathbf{p}^{\text{f}\,}]$ is paired
   with the complementary solution
   movements $[d\mathbf{x}^{\text{f},\alpha}]$ along the
   alternative $\alpha$-branches.
   With this understanding, the second-order quantity
   \begin{equation}\label{eq:I2}
     \mathcal{I}_{2}(d\mathbf{x}^{\alpha})
     =
     \frac{1}{2}\left(
     \left[d\mathbf{p}^{\text{c},\alpha}\right]^{\text{T}}
     \left[d\mathbf{x}^{\text{c}}\right]
     -
     \left[d\mathbf{p}^{\text{f}}\right]^{\text{T}}
     \left[d\mathbf{x}^{\text{f},\alpha}\right]
     \right)
   \end{equation}
   can be computed for each $\alpha$ branch
   (a similar quantity was introduced
   by Hill \cite{Hill:1961a} in developing
   stability extremum principles in a \emph{continuum} setting).
   Ba\v{z}ant hypothesized that the system approaches
   equilibrium along the branch that maximizes $\mathcal{I}_{2}$.
   Petryk \cite{Petryk:1991a}, also using energy arguments,
   defined a stable path as a branch $[d\mathbf{x}^{0}]\in A$
   that satisfies the condition
   \begin{equation}
     \mathcal{I}_{2}(d\mathbf{x}^{0}) \ge
     \mathcal{I}_{2}(d\mathbf{x}^{\alpha}),\quad
     \forall
     \left[d\mathbf{x}^{\alpha}\right]\in A
   \end{equation}
   for the set $A$ of bifurcation paths,
   $A=\{[d\mathbf{x}^{a}],[d\mathbf{x}^{b}],[d\mathbf{x}^{0}],\ldots\}$,
   that satisfy Eq.~(\ref{eq:defBif}).
   Contrarily, an unstable path $[d\mathbf{x}^{0}]\in A$,
   exhibiting \emph{path instability},
   is one for which
   \begin{equation}\label{eq:PI2}
     \exists\,[d\mathbf{x}^{\alpha}]\in A,\quad
     \mathcal{I}_{2}(d\mathbf{x}^{\alpha}) >
     \mathcal{I}_{2}(d\mathbf{x}^{0})
   \end{equation}
   An example is the Shanley column with an elasto-plastic hinge,
   for which three branches apply.
   In this case, an elasto-plastic column that is loaded
   to the tangent limit can remain straight with
   continued loading $d\mathbf{p}$
   (i.e., the fundamental deformation $d\mathbf{x}^{0}$),
   or the column can buckle to the left or to the right.
   The fundamental deformation exhibits path instability,
   as the column is inclined to buckle by the criterion
   of Eq.~(\ref{eq:PI2}).
   \par
   In Table~\ref{table:bif},
   the generic definition of path instability in Eq.~(\ref{eq:PI2})
   is given in the context of the four types
   of displacement constraints that were described
   in Section~\ref{sec:constraints}.
  \begin{table}
    \renewcommand{\arraystretch}{1.3}
    \centering
    \caption{Bifurcation with four types of displacement
             constraints and the corresponding
             definitions of path stability.%
             \rule[-2ex]{0ex}{2ex}\label{table:bif}}
    \begin{tabular}{@{}l}
    \toprule
    Generic bifurcation (without considering constraints,
                         Eq.~\ref{eq:defBif})\\\quad\quad
      $\exists\, [d\mathbf{x}^{a}]\in\Omega^{a},\,
                 [d\mathbf{x}^{a}]\in\Omega^{a},\,
                 \Omega^{a}\neq\Omega^{b}:\;\;
       [\mathbf{H}^{a}] [d\mathbf{x}^{a}] =
       [\mathbf{H}^{b}] [d\mathbf{x}^{b}] =
       [d\mathbf{p}]$\\\quad
       Path stability,
         $\mathcal{I}_{2}(d\mathbf{x}^{\alpha})=
          \tfrac{1}{2}\left([d\mathbf{p}^{\text{c},\alpha}]^{\text{T}}
                       [d\mathbf{x}^{\text{c}}]
                      -[d\mathbf{p}^{\text{f}}]^{\text{T}}
                      [d\mathbf{x}^{\text{f},\alpha}]\right)$%
          :\textsuperscript{\S}\\\quad\quad
       $\mathcal{I}_{2}(d\mathbf{x}^{0})\ge
        \mathcal{I}_{2}(d\mathbf{x}^{\alpha}),
        \forall [d\mathbf{x}^{\alpha}]\in A$\\
    \midrule
    Type~I constraint\\\quad
      Bifurcation\\\quad\quad
      $\exists\,
      [d\mathbf{x}^{\text{c}} \times d\mathbf{x}^{\text{f},a}]
      \in\Omega^{a},
      [d\mathbf{x}^{\text{c}} \times d\mathbf{x}^{\text{f},b}]
      \in\Omega^{b},
      \Omega^{a}\neq\Omega^{b}$:\\
      \multicolumn{1}{r}{%
         $[\mathbf{U}^{a}]
         [d\boldsymbol{\mathfrak{p}}^{\text{c}}/d\mathbf{x}^{\text{f},a}]
         =
         [\mathbf{V}^{a}]
         [d\mathbf{x}^{\text{c}}/d\boldsymbol{\mathfrak{p}}^{\text{f}\,}],\;
         [\mathbf{U}^{b}]
         [d\boldsymbol{\mathfrak{p}}^{\text{c}}/d\mathbf{x}^{\text{f},b}]
         =
         [\mathbf{V}^{b}]
         [d\mathbf{x}^{\text{c}}/d\boldsymbol{\mathfrak{p}}^{\text{f}\,}]$
      }\\
      \quad
      Path stability of $[d\mathbf{x}^{0}]$,
      $\mathcal{I}_{2}(d\mathbf{x}^{a})=
       \frac{1}{2}[d\mathbf{x}^{\text{c}}/
       (-d\boldsymbol{\mathfrak{p}}^{\text{f}})]^{\text{T}}
       [\mathbf{U}^{a}]^{-1}
       [\mathbf{V}^{a}][d\mathbf{x}^{\text{c}}/
       d\boldsymbol{\mathfrak{p}}^{\text{f}\,}]$:%
       \textsuperscript{\S}
      \\\quad\quad
       $\mathcal{I}_{2}(d\mathbf{x}^{0}) \ge
        \mathcal{I}_{2}(d\mathbf{x}^{\alpha}),\;
        \forall
        [d\mathbf{x}^{\alpha}]\in A_{\mathbf{x}}$
      \\
    \midrule
    Type~II constraint,
                          $[\mathbf{X}^{i}]=
                           [\mathbf{H}^{i}]
                           [\mathbf{P}_{\mathcal{L}}]
                           +[\mathbf{P}_{\mathcal{L}^{\perp}}]$\\\quad
      Bifurcation:\\\quad\quad
      $\exists\,[d\mathbf{z}^{a}],\;
                [d\mathbf{z}^{b}],\;
                [\mathbf{P}_{\mathcal{L}}][d\mathbf{z}^{a}]\in\Omega^{a},\;
                [\mathbf{P}_{\mathcal{L}}][d\mathbf{z}^{b}]\in\Omega^{b},\;
                \Omega^{a}\neq\Omega^{b}:\;\;$\\
      \multicolumn{1}{r}{%
       $[\mathbf{X}^{a}][d\mathbf{z}^{a}]=
       [\mathbf{X}^{b}][d\mathbf{z}^{b}]=[d\boldsymbol{\mathfrak{p}}]$}
       \\\quad
       Path stability of $[d\mathbf{z}^{0}]$,
       $\mathcal{I}_{2}(d\mathbf{z}^{\text{a}})=
        \tfrac{1}{2}\left(
        [d\mathbf{y}^{\text{a}}]^{\text{T}}
        [d\mathbf{x}^{\text{a}}]-
        [d\boldsymbol{\mathfrak{p}}]^{\text{T}}
        [d\mathbf{x}^{\text{a}}]\right)$:%
        $^{\ast\ast,\S}$\\ \quad\quad
       $\mathcal{I}_{2}(d\mathbf{z}^{0}) \ge
        \mathcal{I}_{2}(d\mathbf{z}^{\alpha}),\;
        \forall
        [d\mathbf{z}^{\alpha}]\in A_{\mathbf{x}}$\\
    \midrule
    Type~III constraint,
                          $[\mathbf{X}^{i}]=
                           [\mathbf{H}^{i}]
                           [\mathbf{P}_{\mathcal{L}}]
                           +[\mathbf{P}_{\mathcal{L}^{\perp}}]$,
                          $[d\boldsymbol{\mathfrak{y}}^{i}]=
                           [\mathbf{H}^{i}][\mathbf{C}]^{\dagger}[d\mathbf{c}]$\\\quad
      Bifurcation:$^{\ast}$\\\quad\quad
      $\exists\,[d\mathbf{z}^{a}],\;
                [d\mathbf{z}^{b}],\;
                [\mathbf{C}]^{\dagger} [d\mathbf{c}] +
                [\mathbf{P}_{\mathcal{L}}][d\mathbf{z}^{a}]\in\Omega^{a},\;
                [\mathbf{C}]^{\dagger} [d\mathbf{c}] +
                [\mathbf{P}_{\mathcal{L}}][d\mathbf{z}^{b}]\in\Omega^{b},\;
                \Omega^{a}\neq\Omega^{b}:\;\;$\\
       \multicolumn{1}{r}
       {
       $[\mathbf{X}^{a}][d\mathbf{z}^{a}]=[d\boldsymbol{\mathfrak{p}}]
       -[d\boldsymbol{\mathfrak{y}}^{a}],\;
       [\mathbf{X}^{b}][d\mathbf{z}^{b}]=[d\boldsymbol{\mathfrak{p}}]
       -[d\boldsymbol{\mathfrak{y}}^{b}]$}
       \\\quad
       Path stability of $[d\mathbf{z}^{0}]$,
       $\mathcal{I}_{2}(d\mathbf{z}^{\text{a}})=
        \frac{1}{2}\left([d\mathbf{y}^{\text{a}}]^{\text{T}}
        [d\mathbf{x}^{\text{a}}]-
        [d\boldsymbol{\mathfrak{p}}]^{\text{T}}
        [d\mathbf{x}^{\text{a}}]\right)$:%
        $^{\rule{0.3ex}{0ex}\ast\ast,\S}$\\ \quad\quad
       $\mathcal{I}_{2}(d\mathbf{z}^{0}) \ge
        \mathcal{I}_{2}(d\mathbf{z}^{\alpha}),\;
        \forall
        [d\mathbf{z}^{\alpha}]\in A_{\mathbf{x}}$\\
    \midrule
    Type~IV constraint,
         $[\mathbf{X}^{i}]=[\mathbf{H}^{i}][\mathbf{P}^{\text{n-r-r}}] +
          [\mathbf{P}^{\text{r-r}}]$\\\quad
      Bifurcation:\\\quad\quad
      $\exists\,[d\mathbf{z}^{a}],\;
                [d\mathbf{z}^{b}],\;
                [\mathbf{P}^{\text{n-r-r}}][d\mathbf{z}^{a}]\in\Omega^{a},\;
                [\mathbf{P}^{\text{n-r-r}}][d\mathbf{z}^{b}]\in\Omega^{b},\;
                \Omega^{a}\neq\Omega^{b}:\;\;$\\
      \multicolumn{1}{r}{%
       $[\mathbf{X}^{a}][d\mathbf{z}^{a}]=
       [\mathbf{X}^{b}][d\mathbf{z}^{b}]=[d\boldsymbol{\mathfrak{p}}]$}
       \\\quad
       Path stability of $[d\mathbf{z}^{0}]$,
       $\mathcal{I}_{2}(d\mathbf{z}^{\text{a}})=
        \tfrac{1}{2}\left(
        [d\mathbf{y}^{\text{a}}]^{\text{T}}
        [d\mathbf{x}^{\text{a}}]-
        [d\boldsymbol{\mathfrak{p}}]^{\text{T}}
        [d\mathbf{x}^{\text{a}}]\right)$:%
        \textsuperscript{\ddag,\S}\\ \quad\quad
       $\mathcal{I}_{2}(d\mathbf{z}^{0}) \ge
        \mathcal{I}_{2}(d\mathbf{z}^{\alpha}),\;
        \forall
        [d\mathbf{z}^{\alpha}]\in A_{\mathbf{x}}$\\
    \midrule
    {}$^{\ast}$
    $[d\boldsymbol{\mathfrak{y}}^{a}]=
     [\mathbf{H}^{a}][\mathbf{C}]^{\dagger}[d\mathbf{c}]$\\
    $^{\ast\ast}$
    $[d\mathbf{y}^{\alpha}]=
    -[d\boldsymbol{\mathfrak{p}}] +
     [\mathbf{H}^{\alpha}][d\mathbf{x}^{\alpha}]$,
     $[d\mathbf{x}^{\alpha}]=[\mathbf{P}_{\mathcal{L}}][d\mathbf{z}^{\alpha}]$\\
    \textsuperscript{\ddag}
    $[d\mathbf{y}^{\alpha}]=
    -[d\boldsymbol{\mathfrak{p}}] +
     [\mathbf{H}^{\alpha}][d\mathbf{x}^{\alpha}]$,
     $[d\mathbf{x}^{\alpha}]=[\mathbf{P}^{\text{n-r-r}}][d\mathbf{z}^{\alpha}]$\\
    %{}$^{\s}$
    \textsuperscript{\S}
    $A_{\mathbf{x}}=
    \{[d\mathbf{x}^{a}],[d\mathbf{x}^{b}],[d\mathbf{x}^{0}],\ldots\}$,
    the set of all solutions\\
    \bottomrule
    \end{tabular}
  \end{table}
   With simple Type~I constraint,
   the $\alpha$-branch is the mixed solution of Eq.~(\ref{eq:IHH}),
   $[d\boldsymbol{\mathfrak{p}}^{\text{c}}/
     d\mathbf{x}^{\text{f}}]=
    [\mathbf{U}^{\alpha}]^{-1}
    [\mathbf{V}^{\alpha}][d\mathbf{x}^{\text{c}}/
    d\boldsymbol{\mathfrak{p}}^{\text{f}}]$,
  and the second-order quantity $\mathcal{I}_{2}(d\mathbf{x}^{\alpha})$
  is the product
  $[d\mathbf{x}^{\text{c}}/
    (-d\boldsymbol{\mathfrak{p}}^{\text{f}})]
    [\mathbf{U}^{\alpha}]^{-1}
   [\mathbf{V}^{\alpha}][d\mathbf{x}^{\text{c}}/
    d\boldsymbol{\mathfrak{p}}^{\text{f}}]$,
    noting the negative sub-vector
    $[-d\boldsymbol{\mathfrak{p}}^{\text{f}}]$,
    as in Eq.~(\ref{eq:I2}).
    With Type~II and~III constraints,
    the term
    $[d\mathbf{p}^{\text{f}}]^{\text{T}}
     [d\mathbf{x}^{\text{f},\alpha}]$
    in Eq.~(\ref{eq:I2})
    is the product of the vectors $[d\boldsymbol{\mathfrak{p}}]$
    and $[d\mathbf{x}]$ that appear in
    Eqs.~(\ref{eq:dxdyBott}) and~(\ref{eq:dxdyBottIII});
    whereas the term
    $[d\mathbf{p}^{\text{c},\alpha}]^{\text{T}}
     [d\mathbf{x}^{\text{c}}]$
    is the product of the reaction forces
    $[d\mathbf{y}]$
    and the movements $[d\mathbf{x}]$,
    as in Eqs.~(\ref{eq:dxdyBott}), (\ref{eq:dxdyBottIII}),
    and~(\ref{eq:dyBottIII}).
    Type~IV constraint is analyzed in a similar manner,
    with the projection matrix $[\mathbf{P}^{\text{n-r-r}}]$
    taking the place of matrix $[\mathbf{P}_{\mathcal{L}}]$
    in computing vectors $[d\boldsymbol{\mathfrak{p}}]$ and $[d\mathbf{x}]$.
  \subsection{\normalsize Instability of equilibrium}\label{sec:IofE}
  With instability of equilibrium,
  an impulsive departure from an equilibrium
  state is energetically available without a change
  in the loading parameters.
  %infinitely small disturbances in forces or imperfections in
  %configuration can produce finite deviations in
  %system displacements.
  Eq.~(\ref{eq:Eddot3})
  of Section~\ref{sec:stability} gives
  the rate of kinetic
  energy $\ddot{E}$ (or rate of internal entropy)
  for an equilibrium system perturbed across a time span of
  $\Delta t$ by the rates of
  movement and loading, $[\dot{\mathbf{x}}]$
  and $[\dot{\mathbf{q}}]$.
  If the loading rate $[\dot{\mathbf{q}}]$ is suspended
  at a particular state (i.e., is momentarily zero),
  so that $\mathcal{B}_{2}=0$,
  a spontaneous increase in $E$ is energetically
  favored when the second-order work $\mathcal{W}_{2}$
  is negative in a particular movement direction
  of $[\dot{\mathbf{x}}]$.
  Because this movement must be consistent with
  the directional stiffness $[\mathbf{H}^{i}]$,
  instability of equilibrium is defined as
  \begin{equation}\label{eq:Ins1}
    \exists
    \left[d\mathbf{x}\right]\in\Omega^{i}:\;\;
    \left[d\mathbf{x}\right]^{\text{T}}
    \left[\mathbf{H}^{i}\right]
    \left[d\mathbf{x}\right]
    < 0
  \end{equation}
  (note that this definition is generic,
  as we have not yet considered any movement constraints).
  Even with a non-vanishing loading rate,
  $[\dot{\mathbf{q}}]\neq 0$, such that
  $\mathcal{B}_{2}(\dot{\mathbf{x}})\neq 0$,
  the rate $\ddot{E}$ in
  Eq.~(\ref{eq:Eddot3}) is dominated by $\mathcal{W}_{2}$,
  which is quadratic in $[d\mathbf{x}]$, so that $[d\mathbf{q}]$,
  in principle,
  can be treated as zero.
  \par
  Because Eq.~(\ref{eq:Ins1}) involves a scalar quadratic form,
  an alternative definition of instability of equilibrium is
  the existence of a stiffness matrix
  $[\mathbf{H}^{i}]$
  with a negative eigenvalue of its symmetric part,
  $[\widehat{\mathbf{H}}^{i}]=
    \tfrac{1}{2}([\mathbf{H}^{i}]+[\mathbf{H}^{i}]^{\text{T}})$,
  for which
  the corresponding eigenvector
  lies within $\Omega^{i}$.
  Similar to Eq.~(\ref{eq:NE3}),
  we designate $\{\lambda([\widehat{\mathbf{Z}}])\}$
  as the set of
  eigenvalues of the symmetric matrix $[\widehat{\mathbf{Z}}]$,
  and
  $\{\boldsymbol{\eta}_{-}([\widehat{\mathbf{Z}}])\}$
  as the subspace spanned by those eigenvectors corresponding
  to the negative eigenvalues of the symmetric part.
  An alternative (generic) definition of instability of
  equilibrium is
  \begin{equation}\label{eq:Ins2}
    \exists
    \left[\mathbf{H}^{i}\right]:\;\;
    \{\lambda([\widehat{\mathbf{H}^{i}}]): \lambda<0\}\neq\varnothing,\;
    \{\boldsymbol{\eta}_{-}([\widehat{\mathbf{H}^{i}}])\}
    \cap
    \Omega^{i}
    \neq
    [\mathbf{0}]
  \end{equation}
  \par
  The above two definitions are
  closely related to the concept of unsustainability
  proposed by Nicot and Darve \cite{Nicot:2007b}.
  In a continuum setting, unsustainablity arises
  when a change in the stress and strain of a region
  can be reached through a dynamic process without any
  change in control parameters applied to the region's boundary
  and interior.
  In a discrete setting, this condition is equivalent to
  the generic criterion $\mathcal{W}_{2}(\dot{\mathbf{x}})<0$
  of Eqs.~(\ref{eq:Ins1})--(\ref{eq:Ins2}), noting that notions of
  ``boundary'' and ``interior'' are to be avoided with discrete
  systems. 
  We also note that stability criteria in 
  Eqs.~(\ref{eq:Ins1}) and (\ref{eq:Ins2})
  are restricted
  to a quasi-static setting, in which the system is initially
  in stationary equilibrium (Section~\ref{sec:stability}).
  It is possible to show, for instance, that in a full
  dynamic setting, flutter can arise
  in non-conservative systems before
  a mode of negative second-order work is available
  \cite{Challamel:2010a}.
  \par
  Much attention has been given to the hierarchy of
  the different pathologies,
  in particular to the question of whether instability of equilibrium
  precedes neutral equilibrium.
  Such hierarchy is
  based upon the following matrix property:
  the real parts of the eigenvalues of a matrix $[\mathbf{H}]$
  are bounded by the smallest and largest eigenvalues of
  its symmetric counterpart,
  \begin{equation}
    \text{min}\,\{\lambda([\widehat{\mathbf{H}}])\}
    \le
    \{\text{Re}\left(\lambda([\mathbf{H}])\right)\}
    \le
    \text{max}\,\{\lambda([\widehat{\mathbf{H}}])\}
  \end{equation}
  As such, for elastic systems
  that undergo a smooth, continuous transition of
  stiffness during a loading program,
  instability of equilibrium is encountered
  before neutral equilibrium:
  a negative eigenvalue of the symmetric matrix
  $[\widehat{\mathbf{H}}]$ is encountered before
  a zero eigenvalue appears
  (along with a zero determinant) with the full matrix $[\mathbf{H}]$.
  For a smooth stiffness operator, Challamel et al. \cite{Challamel:2010a}
  have shown
  that the second-order work criterion for a non-conservative
  system coincides with neutral equilibrium with one homogeneous
  (i.e., Type~II) constraint, and Lerbet et al. \cite{Lerbet:2012b}
  have generalized this result to systems with $n$ constraints.
  This hierarchy of second-order work and 
  neutral equilibrium does not necessarily apply to
  non-smooth inelastic systems.
  The eigenvectors of the full stiffness
  matrix and of its symmetric counterpart
  are not equal,
  and it is possible that the negative eigenvalues
  of the symmetric stiffness $[\widehat{\mathbf{H}}^{i}]$
  correspond to eigenvectors that
  lie \emph{outside} the stiffness's domain $\Omega^{i}$;
  whereas, the full matrix $[\mathbf{H}^{i}]$ can have
  a zero eigenvalue with a different eigenvector, but one
  that lies within $\Omega^{i}$.
  \par
  Table~\ref{table:Int} places these generic definitions
  of instability of equilibrium
  in the context of the four types of movement constraints
  described in Section~\ref{sec:constraints}.
  \begin{table}
    \renewcommand{\arraystretch}{1.3}
    \centering
    \caption{Instability of equilibrium
             with four types of displacement
             constraints.\rule[-2ex]{0ex}{2ex}\label{table:Int}}
    \begin{tabular}{@{}rl}
    \toprule
    \multicolumn{2}{@{}l}{Generic instability of equilibrium
                       (without considering constraints,
                       Eqs.~\ref{eq:Ins1}--\ref{eq:Ins2})}\\
    \quad a) &
      $\exists\, [d\mathbf{x}]\in\Omega^{i}:\;\;
       [d\mathbf{x}]^{\text{T}}
       [\mathbf{H}^{i}]
       [d\mathbf{x}] < 0$\\
    b) &
      $\exists\,[\mathbf{H}^{i}]:\;\;
    \{\lambda([\widehat{\mathbf{H}^{i}}])\!: \lambda<0\}\neq\varnothing,\;
    \{\boldsymbol{\eta}_{-}([\widehat{\mathbf{H}^{i}}])\}
    \cap
    \Omega^{i}
    \neq
    [\mathbf{0}]$\\
    \midrule
    \multicolumn{2}{@{}l}{Type~I constraint}\\
    a) &
      $\exists\,[d\mathbf{x}]=
      [d\mathbf{x}^{\text{c}}/d\mathbf{x}^{\text{f}\,}]
      \in\Omega^{i}:\;\;
       [d\mathbf{x}^{\text{f}\,}]^{\text{T}}
       [\mathbf{H}^{\text{f{}f},i}][d\mathbf{x}^{\text{f}\,}]<0$\\
     b) &
       $\exists\,[\mathbf{H}^{i}] :\;\;
        \{\lambda([\widehat{\mathbf{H}^{\text{f{}f},i}}])\!: \lambda<0\}\neq\varnothing,\;
    \left(
    [d\mathbf{x}^{\text{c}}]\times
    \{\boldsymbol{\eta}_{-}([\widehat{\mathbf{H}^{\text{f{}f},i}}])\}\right)
    \cap
    \Omega^{i}
    \neq
    [\mathbf{0}]$\\
    \midrule
    \multicolumn{2}{@{}l}{Types~II and III constraints}\\
    a) &
      $\exists\,[d\mathbf{z}],\;[\mathbf{P}_{\mathcal{L}}][d\mathbf{z}]\in\Omega^{i}:\;\;
       [d\mathbf{z}]^{\text{T}}
       \left(
       [\mathbf{P}_{\mathcal{L}}]^{\text{T}}
       [\mathbf{H}^{i}]
       [\mathbf{P}_{\mathcal{L}}]\right)
       [d\mathbf{z}]
       <0$\\
     b) &
       $\exists\,[\mathbf{H}^{i}]:\;\;
        \left\{\lambda\!\left([\mathbf{P}_{\mathcal{L}}]^{\text{T}}
                            [\widehat{\mathbf{H}^{i}}]
                            [\mathbf{P}_{\mathcal{L}}]\right)\!: \lambda<0\right\}
        \neq\varnothing,\;
        \left\{\boldsymbol{\eta}_{-}\!
          \left([\mathbf{P}_{\mathcal{L}}]^{\text{T}}[\widehat{\mathbf{H}^{i}}]
                [\mathbf{P}_{\mathcal{L}}]\right)\right\}
        \cap
        \Omega^{i}
        \neq
        [\mathbf{0}]$\\
    \midrule
    \multicolumn{2}{@{}l}{Type~IV constraint}\\
    a) &
      $\exists\,[d\mathbf{z}],\;[\mathbf{P}_{\mathcal{L}}][d\mathbf{z}]\in\Omega^{i}:\;\;
       [d\mathbf{z}]^{\text{T}}
       \left(
       [\mathbf{P}^{\text{n-r-r}}]^{\text{T}}
       [\mathbf{H}^{i}]
       [\mathbf{P}^{\text{n-r-r}}]\right)
       [d\mathbf{z}]
       <0$\\
     b) &
       $\exists\,[\mathbf{H}^{i}]:\;\;
        \left\{\lambda\!\left([\mathbf{P}^{\text{n-r-r}}]^{\text{T}}
                            [\widehat{\mathbf{H}^{i}}]
                            [\mathbf{P}^{\text{n-r-r}}]\right)\!: \lambda<0\right\}
        \neq\varnothing,\;
        \left\{\boldsymbol{\eta}_{-}\!
          \left([\mathbf{P}^{\text{n-r-r}}]^{\text{T}}[\widehat{\mathbf{H}^{i}}]
                [\mathbf{P}^{\text{n-r-r}}]\right)\right\}
        \cap
        \Omega^{i}
        \neq
        [\mathbf{0}]$\\
    \bottomrule
    \end{tabular}
  \end{table}
%\end{itemize}
%
  For example, with Type~I constraint, the known, constrained
  displacements $[d\mathbf{x}^{\text{c}}]$ are treated as zero,
  since the quadratic form of Eq.~(\ref{eq:Ins1}) is dominated
  by the product
  $[d\mathbf{x}^{\text{f}\,}]^{\text{T}}
   [\mathbf{H}^{\text{f{}f},i}][d\mathbf{x}^{\text{f}\,}]$.
%   with arbitrarily large $[d\mathbf{x}^{\text{f}}]$
 %  rather than by products involving the fixed $[d\mathbf{x}^{\text{c}}]$.
   Restricting the question of instability to a test of whether
   the submatrix $[\mathbf{H}^{\text{f{}f},i}]$ is semi-definite or indefinite
   reduces the displacement-space that is available for instability
   modes from $\mathbb{R}^{6N}$ to $\mathbb{R}^{6N-r}$
   (recall that $r$ displacements are constrained).
   Similarly, with
   Types~II and ~III constraints,
   one need not query
   the full space $[d\mathbf{z}]\in\mathbb{R}^{6N}$;
   rather, one must only consider the projection
   $[\mathbf{P}_{\mathcal{L}}][d\mathbf{z}]$ of $\mathbb{R}^{6N}$ onto
   the subspace $\mathcal{L}$ that satisfies the displacement
   constraints (Eqs.~\ref{eq:homogeneousC}
   and~\ref{eq:nonhomogeneousC}).
   With Type~IV constraint,
   we are only concerned with instability associated with
   particle motions
   that are not rigid-body movements of the entire assembly,
   and by investigating the product
   $[\mathbf{P}^{\text{n-r-r}}]^{\text{T}}
    [\widehat{\mathbf{H}^{i}}]
    [\mathbf{P}^{\text{n-r-r}}]$
   we eliminate any second-order work that is merely an
   artifact of the rigid rotation of the assembly and its
   external forces.
   \par
   Finally, we note that a numerical search for modes
   of unstable equilibrium can encounter the same
   difficulties faced in determining neutral equilibrium,
   as were discussed and resolved in the final two
   paragraphs of Section~\ref{sec:NE}.
\section{\large Examples}\label{sec:Examples}\label{sec:threedisks}
We now consider three examples of increasing complexity,
involving three, twelve, and sixty-for disks.
In each example, we identify multiple
(sometimes thousands of) stiffness pathologies.
Of the several criteria for distinguishing
neutral equilibrium and instability of equilibrium,
we applied the particular criteria of
Eqs.~(\ref{eq:NE2}) and~(\ref{eq:Ins2}).
For these two-dimensional situations,
the curvature tensors $[\mathbf{K}^{pq,p}]$
and $[\mathbf{K}^{pq,q}]$ are simplified
as $-(1/\rho^{p})[\mathbf{n}^{pq}]^{\text{T}}[\mathbf{n}^{pq}]$
and $-(1/\rho^{q})[\mathbf{n}^{pq}]^{\text{T}}[\mathbf{n}^{pq}]$,
where $\rho^{p}$ and $\rho^{q}$ are the particles' radii
of curvature at the contact point $pq$.
Each example was numerically analyzed with Matlab code.
%
%\todo[inline]{Revise. Add other examples.  Ziegler column?
%              Friction columns?  Squirted disk?
%              Add various types of constraints.
%              Add example with flexible membrane or follower forces?}
%
\subsection{Example~1: three disks}
In this simplest example,
we examine a system of three disks (Fig.~\ref{fig:Squirt2}).
\begin{figure}
  \centering
  \includegraphics{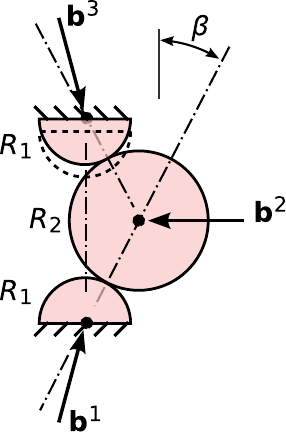}
  \caption{Example~1: three disks, in which the two contacts
           are at the friction limit.
           Rotation and horizontal movement is prevented with
           the top and bottom disks, while the top disk
           is displaced downward.
           \label{fig:Squirt2}}
\end{figure}
The top and bottom disks have the same radius $R_{1}$
and are centered on a vertical line;
whereas, the middle disk has radius $R_{2}$
and is offset by angle $\beta$ from the top and
bottom disks.
The two contacts share the same contacts stiffness
and friction coefficient
(values $k=1$, $\alpha=1$, and $\mu=0.5$).
The top and bottom disks are not allowed to rotate or to
move horizontally,
and the bottom disk is constrained from vertical movement.
The three disks touch at two contacts with an equal
normal contact force of 0.001$k$,
and both contacts are assumed at the friction limit,
with the directions of their tangential force being consistent
with a vertical compression of the assembly.
The system is loaded by displacing the top disk downward toward
the bottom disk (i.e., $\dot{u}_{2}^{3}< 0$).
Briefly, the non-rotating top and bottom disks are pressed
vertically upon the middle disk and a constant force
$\mathbf{b}^{2}$ places the two contacts at the friction limit.
The system has only four kinematic degrees of freedom
($u_{2}^{1}$, $u_{2}^{2}$, $\theta_{3}^{2}$, and $u_{2}^{3}$), 
with five forces of constraint
(the reactions $b_{1}^{1}$, $b_{2}^{1}$, $w_{3}^{1}$,
$b_{1}^{3}$, and $w_{3}^{3}$).
\par
%We also considered the case of Type~IV constraint in which
%all motions could freely occur.
We consider three relative radii of the disks
($R_2 < R_1$, $R_2 = R_1$, and $R_2 > R_1$),
and three relative offset angles $\beta$.
When $\beta=\cot(\mu)$, the top and bottom external forces,
$\mathbf{b}^{1}$ and $\mathbf{b}^{3}$, are vertical,
and no external force acts upon the middle particle
($\mathbf{b}^{2}=\mathbf{0}$);
however, when $\beta$ is less than (greater than) $\cot(\mu)$,
equilibrium requires that
the external forces $\mathbf{b}^{1}$ and $\mathbf{b}^{3}$
must have a leftward (rightward) component,
and the external force $\mathbf{b}^{2}$ must have
a rightward (leftward) component,
so that the two contacts are held at the friction limit.
Figure~\ref{fig:Squirt2} depicts the
case of $R_2 > R_1$ and $\beta<\cot(\mu)$,
with $\mathbf{b}^{2}$ acting toward the left.
\par
Although this three-disk assembly seems fairly simple,
it displays several pathologies,
which are summarized in Table~\ref{table:threedisks}.
\begin{table}
  \centering
  \caption{Stiffness pathologies of the three-disk
           system in Fig.~\ref{fig:Squirt2},
           in which the contacts' radii of curvature
           are the same as the disks' physical radii
           ($\rho=R$).
           Each $\Box$ indicates a solution of the
           stiffness problem, with multiple $\Box$'s
           representing a bifurcation.
           Each $\triangle$ indicates an instability
           of equilibrium, with multiple
           $\triangle$'s representing multiple
           unstable modes.
           A $\boxtimes$ means that no solution
           exists.
           \label{table:threedisks}}
  \begin{tabular}{lccc}
    \toprule
    & $R_{2}<R_{1}$ & $R_{2}=R_{1}$ & $R_{2}>R_{1}$ \\
    \midrule
    $\beta<\cot(\mu)$
                      & $\square\square$ & $\square\square$ & $\square\square$ \\
%                     &  & & \\
%                     & $\triangle$ & $\triangle$ & $\triangle$ \\
    \midrule
    \multirow{2}{*}{$\beta=\cot(\mu)$}
                      & $\boxtimes$ & $\boxtimes$ & $\boxtimes$ \\
                      & $\triangle$ & $\triangle$ & $\triangle\triangle\triangle$ \\
%                     & $\triangle$ & $\triangle$ & $\triangle\triangle\triangle$ \\
    \midrule
    \multirow{2}{*}{$\beta>\cot(\mu)$}
              & $\square\square\square$ & $\square\square\square$ & $\square\square\square$\\
              & $\triangle$ & $\triangle$ & $\triangle\triangle\triangle$ \\
    \bottomrule
  \end{tabular}
\end{table}
Because both contacts are at the friction limit,
we must investigate $2^2=4$ possible slip (or no-slip) combinations
and check whether the displacement of the solutions
(or pathologies) of each combination
are consistent with the assumed slip (or no-slip) combination.
As an example, when $\beta=\cot(\mu)$ (i.e., when the
equilibrating side force
$\mathbf{b}^{2}$ is zero), no solution exists that will maintain
the constant, zero side force.
Indeed,this arrangement places the
system in an untenable, unstable condition,
and the middle particle will squirt to the
right without being prompted by any movement of the top and
bottom particles: a negative second-order work is
associated with both contacts during a
horizontal movement of the middle disk.
When the middle disk is larger than the top and bottom
disks and
$\beta=\cot(\mu)$, three instability modes are available:
a mode whereby horizontal
movement of the middle disk toward the right
causing frictional slip of both contacts,
and two modes in which only one contact slips while the
middle disk both rotates and obliquely moves as it
``escapes'' from the other disks.
\par
The case of $\beta>\cot(\mu)$ is also unusual,
as three solutions are available, creating a possible trifurcation
with path instability (i.e., the triplets of squares in the bottom
of Table~\ref{table:threedisks}).
Note that when $\beta>\cot(\mu)$, the external force $\mathbf{b}^{2}$
is pulling the middle disk toward the right to bring the two contacts
to the friction limit.
Each solution is along a different stiffness branch
(i.e., with a different combination of slip and no-slip
contacts).
The first solution is a slight rightward movement of the middle disk,
causing the two contacts to elastically unload with no slip.
The other two solutions involve one slipping contact and one
elastic contact and a rotation
and rightward movement of the middle disk.
Of the three solutions, the first solution is the more stable,
with the largest $\mathcal{I}_{2}(d\mathbf{x}^{\alpha})$ value.
But even this solution branch is unstable,
since a rightward movement of the middle disk
\emph{in the absence of movements of the top and bottom disks}
will exhibit negative second-order work with frictional slip
in both contacts (the triangles in the bottom row of the table).
\par
We now consider the extent to which geometric effects are manifested
in this simple system, which we illustrate in two ways.
Equation~(\ref{eq:GeomMech}) gives the stiffness as a sum
of mechanical and geometric parts, $[\mathbf{H}^{\text{m}}]$
and $[\mathbf{H}^{\text{g}}]$.
The results of our analysis are quite different
when the geometric stiffness is ignored and only the mechanical
stiffness is considered.
For example, \emph{neutral equilibrium} would be predicted
in each of the $\beta=\cot(\mu)$
cases reported in Table~\ref{table:threedisks}.
Neglecting the geometric stiffness, no rotation of the
contact forces would occur during a vertical movement of
both the top and bottom particles, and no change in
the $\mathbf{b}^{2}$ force would be predicted:  that is,
the vector $[\mathbf{p}]$ in Eq.~(\ref{eq:NE1}) would be zero,
indicating neutral equilibrium.
This result is corrected with the inclusion of the
$[\mathbf{H}^{\text{g}}]$ stiffness.
\par
Another illustration of geometric effects
is presented in Table~\ref{table:threedisks2}.
\begin{table}
  \centering
  \caption{Stiffness pathologies of a three-disk
           system. Unlike Fig.~\ref{fig:Squirt2}
           and Table~\ref{table:threedisks},
           the radii of curvature are increased by
           a factor of~2 at the contact points,
           $\rho=2R$.
           Each $\Box$ indicates a solution of the
           stiffness problem, with multiple $\Box$'s
           representing a bifurcation.
           Each $\triangle$ indicates an instability
           of equilibrium, with multiple
           $\triangle$'s representing multiple
           unstable modes.
           A $\boxtimes$ means that no solution
           exists.
           Solutions with symbol ``$\boxminus$''
           differ from those in Table~\ref{table:threedisks}.
           \label{table:threedisks2}}
  \begin{tabular}{lccc}
    \toprule
    & $R_{2}<R_{1}$ & $R_{2}=R_{1}$ & $R_{2}>R_{1}$ \\
    \midrule
    $\beta<\cot(\mu)$
                      & $\boxminus$ & $\boxminus$ & $\boxminus$ \\
%                     &  & & \\
%                     & $\triangle$ & $\triangle$ & $\triangle$ \\
    \midrule
    \multirow{2}{*}{$\beta=\cot(\mu)$}
                      & $\boxtimes$ & $\boxtimes$ & $\boxtimes$ \\
                      & $\triangle$ & $\triangle$ & $\triangle\triangle\triangle$ \\
%                     & $\triangle$ & $\triangle$ & $\triangle\triangle\triangle$ \\
    \midrule
    \multirow{2}{*}{$\beta>\cot(\mu)$}
                      & $\square\,\boxminus$ & $\square\,\boxminus$ & $\square\,\boxminus$\\
                      & $\triangle$ & $\triangle$ & $\triangle\triangle\triangle$ \\
    \bottomrule
  \end{tabular}
\end{table}
This table is for a three-disk system in which the shapes
of the particles have been altered at the contact points.
Here, the particles have the radii $R_{1}$ and $R_{2}$,
but at their points of contact, we have slightly
``flattened'' the particles' contours by artificially
increasing their radii of curvature by a factor of~2.
The amended results in Table~\ref{table:threedisks2} can be
compared with those of Table~\ref{table:threedisks}, in which the
particles' radii of curvature are equal the disks' bulk curvature.
The numbers of solutions, designated with squares in the tables,
are different for the cases of $\rho=R$ and $\rho=2R$,
and solutions in Table~\ref{table:threedisks2} that are marked
as ``$\boxminus$'' occur with a different combination of slip/no-slip
contacts
than those in Table~\ref{table:threedisks}.
The subtle geometry of the particles' contacts clearly
affect the assembly's stiffness and the solutions of
a displacement problem.
\subsection{Example~2: regular array of twelve disks}\label{sec:twelvedisks}
The second example involves a regular arrangement
of twelve equal-size disks (Fig.~\ref{fig:Regular_Array_1}).
\begin{figure}
  \centering
  \includegraphics{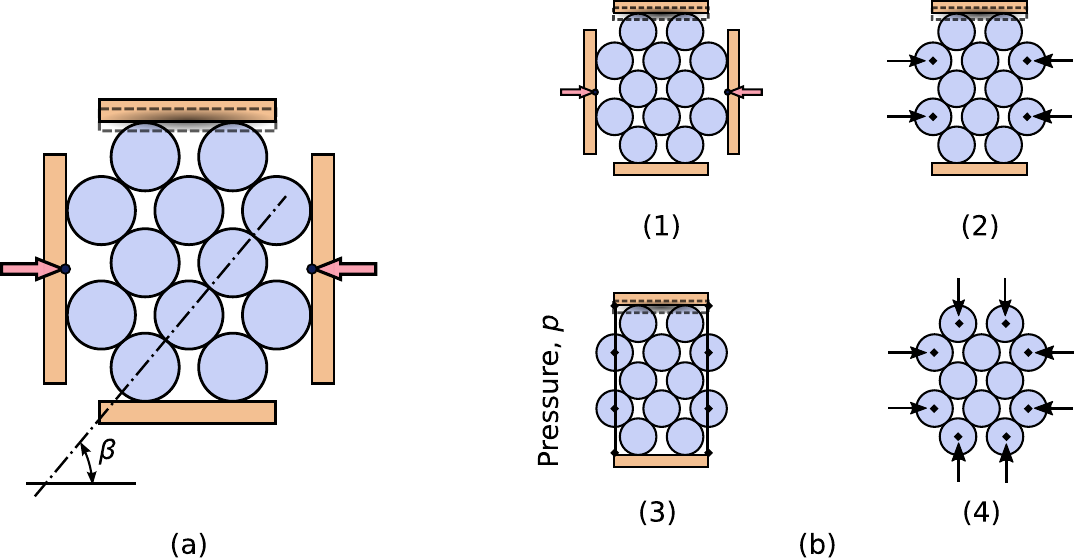}
  \caption{(a) regular array of twelve disks examined
           in Example~2, with
           each disk-disk contact at the friction limit.
           (b) Four different boundary types are considered.
           With the first three boundary types, the top and
           bottom platens are displaced vertically toward each
           other.
           \label{fig:Regular_Array_1}}
\end{figure}
Four different problems are investigated,
each with its own types of boundaries:
(1)~four platens that restrict the movements of the outer disks;
(2)~only top and bottom platens, with constant external forces applied to the
side disks;
(3)~only top and bottom platens, with a flexible boundary along the
side disks, applying a constant pressure;
and (4)~Type~IV constraint with no displacement constraints applied to the
particles.
All disk-disk and disk-platen contacts share the same stiffness and friction
coefficient
(values $k=1$, $\alpha=1$, and $\mu=0.5$),
and each contacting disk-disk pair
has the same contact orientation $\beta$.
All disk-disk contacts have an equal normal
force of 0.001$k$, and all disk-disk contacts
are placed at the friction limit,
with the directions of their tangential force being consistent
with a vertical compression of the assembly.
Because of the symmetric disk arrangement,
the disk-platen forces have no tangential component,
so that equal vertical forces are applied at all top and
bottom disk-platen contacts, and
equal horizontal forces are applied at all side disk-platen
contacts.
\par
The first three problems include top and bottom platens,
and these platens are constrained from
moving horizontally and are vertically pressed toward (approaching) each
other with a vertical displacement.
These displacement conditions can be modeled as either
Type~I or Type~III constraints.
With the four-platen problem (problem~1), we include
side platens but disallow their vertical movement,
and a constant horizontal
external forces $\mathbf{b}$ are applied to these side platens.
The side platens are removed for problem~2,
and their horizontal disk-platen contact forces are replaced with
constant horizontal external forces $\mathbf{b}$,
applied at the centers of the
side particles.
With problem~3, the horizontal external forces are produced
by imaginary membranes
(one membrane on the left, another membrane on the right)
that are draped between the top
and bottom platens and pass through the centers of the
side particles.
A constant pressure $p$ is applied to the membranes,
sufficient to equilibrate the contact forces that act upon the
side disks.
The external forces applied by the membranes
to the disks are position-dependent:
the forces are altered by changes in the disk-disk and
disk-platen distances,
and their directions are altered by any relative shifting
of the disks.
That is, the membrane forces are position-dependent
and involve introducing a $[\mathbf{H}^{\text{g-4}}]$
geometric stiffness.
With problem~4, we remove all platens and membranes
and treat the twelve disks as an isolated system.
\par
There are 16 contacts among the twelve disks,
and because each disk-disk contact is at the friction limit,
we must search for stiffness pathologies among
the $2^{16}=$\,65,536 possible combinations of slip and no-slip
at the contacts.
Results for the four problems are given in Table~\ref{table:regular}
for the case of $\beta=50^{\circ}$, for which the side particles
(or platens)
must press inward on the disks,
to place each disk-disk contact at the friction limit.
\begin{table}
  \centering
  \caption{Stiffness pathologies of the twelve-disk systems in
           Fig.~\ref{fig:Regular_Array_1}.
           \label{table:regular}}
  \begin{tabular}{lccc}
  \toprule
       & Number of      & Number of & Number of \\
       & neutral modes  & solutions & unstable modes\\
  \midrule
  Problem~1: four platens & 0 & 231 & 7,272 \\
  Problem~2: two platens, constant side forces & 0 & 251 & 13,320 \\
  Problem~3: two platens, constant side pressure & 0 & 70 & 13,247 \\
  Problem~4: isolated system & 1 & --- & 52,033 \\
  \bottomrule
  \end{tabular}
\end{table}
The assembly with four platens (problem~1) does not have a
unique solution: there are 231 separate solutions, offering
a multitude of bifurcation opportunities~--- a multi-furcation.
Each of the solutions would produce slip at a subset of the contacts
while producing elastic (no-slip) movements at the remaining contacts,
and with each solution, the combination of contacts that are in a
slip or no-slip condition is consistent with the particles' movements
(Section~\ref{sec:branches}).
Of the 231 solutions, 34 exhibit path stability,
as the other solutions yield smaller values
of $\mathcal{I}_{2}(d\mathbf{x}^{\alpha})$.
The 34 stable solutions
involve symmetric systems of
rotations of the particles and platens and produce
slip displacements at 8 of the 16 disk-disk contacts,
while the other 8 contacts are displaced elastically.
These 34 solutions produce the same displacements of
the four platens.
The other $231-34=197$ solutions are unstable bifurcation paths:
even though they produce
the same downward movement of the top platen, these modes
have different
movements of the side platens (i.e., different dilation rates).
\par
However,
even when all platens are stationary,
the system is unstable,
as the constrained stiffness $[\mathbf{H}]$
has 7272 branches (modes) that are unstable.
This situation is not surprising,
since each of the disk-disk contacts is initially
loaded to the friction limit,
and there are numerous opportunities
for the spontaneous generation
of negative second-order work
by the shifting of particles.
We also investigated a variation of problem~1,
in which we disallowed rotations of the four platens.
The numbers of solutions and unstable branches for problem~1 are
greatly reduced when rotations of the platens are prevented.
This additional constraint reduces the number of solutions to 118
and the number of unstable branches to 2460.
\par
The other three problems in Fig.~\ref{fig:Regular_Array_1}b
have similar characteristics
(see Table~\ref{table:regular}).
The only example of neutral equilibrium, however, occurs
with the isolated assembly of problem~4.
In this problem, we applied no change of the external
forces, such that $[d\mathbf{p}]=[\mathbf{0}]$,
and the neutral, uncontrollable mode is
simply the synchronized gear-like rotations of all 12 disks,
in which each disk rolls against its neighbors but none of
the gears are displaced.
This rolling mode produces no changes in the contact forces
and no change in the external forces.
As such, any non-zero set of equilibrated force increments
$[d\mathbf{p}]$ can be produced by a superposition of
a set of disk movements and the gear-like mode.
\subsection{Example~3: irregular assembly of 64 disks}\label{sec:example3}
An assembly of 64 particles was created and then loaded with
the discrete element (DEM) finite-difference algorithm,
and at particular stages of loading,
the assembly's condition was analyzed
using the stiffness methods derived in the paper
(Fig.~\ref{fig:irregular}).
\begin{figure}
  \centering
  \includegraphics{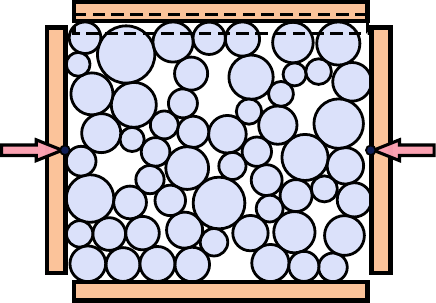}
  \caption{Assembly of 64 disks that was created with
           a DEM simulation.
           \label{fig:irregular}}
\end{figure}
The DEM method applies
Newton's equations in a time-stepping algorithm to
advance the particles' positions,
while maintaining near-equilibrium of the forces among particles
and approximating the particles' rearrangements.
The particles are never in equilibrium, however,
as the algorithm relies upon dis-equilibrium to impel
particles to new locations.
Therefore, questions of bifurcation and instability are not directly
addressed with the DEM, as the particles' inertias sweep them
along a solution path, possibly passing through the various
pathologies that were described in Section~\ref{sec:Pathologies}
and that would otherwise be
confronted under true quasi-static loading conditions.
\par
The irregular arrangement of the 64 disks
was created by placing them sparsely
within four flat platens, assigning the disks random velocities,
and isotropically reducing the assembly dimensions.
The average disk radius $D$ was about 0.5, and the largest disk
was about 3 times larger than the smallest.
All disk-disk and disk-platen contacts were assigned
the same properties:  $k=1$, $\alpha=1$, and $\mu=0.5$.
Once compacted, the average contact force was about 0.0034$kD$.
The assembly was then compressed in the vertical direction
at a constant rate of strain
while maintaining constant horizontal stress
by adjusting the separation of the side platens.
The deviator stress $q$ and vertical
strain $\varepsilon_{22}$ are shown in
Fig.~\ref{fig:stress} for this stage of biaxial compression.
\begin{figure}
  \centering
  \includegraphics{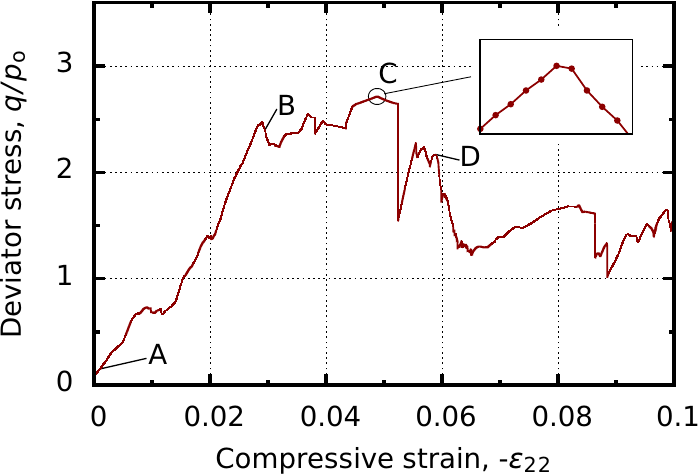}
  \caption{Stress and strain from DEM simulation of biaxial compression
           of an irregular arrangement of 64 disks.
           \label{fig:stress}}
\end{figure}
The ragged, serrated nature of the evolving stress is typical
of DEM simulations of small assemblies and is also observed in
laboratory tests of large specimens of glass beads
\cite{Kuhn:2009b,KuhnDaouadji:drops,Michlmayr:2013a}.
We focus on a brief episode of strain ``C'',
located at the peak deviator stress which is
circled in Fig.~\ref{fig:stress} and detailed in the inset figure.
Other strains (''A'', ''B'', and ''D'') are considered
near the end of this section.
\par
At the peak stress ``C'', we determined which stiffness pathologies
were present with two types of boundaries.
First, we analyzed the same conditions that were used
in the DEM simulation:
zero movement of the bottom platen,
zero rotation of all four platens,
constant horizontal stress on the two side platens,
and a relative downward vertical movement of the top
platen.
To maintain constant horizontal stress, we introduced
a geometric stiffness $[\mathbf{H}^{\text{g-4}}]$ that would
reduce the horizontal forces on the side platens with a reduction
in the assembly's height.
This first type of boundaries was analyzed using Type~III constraint,
although the same results would be found by
applying Type~I constraint.
The second type of boundaries was simply Type~IV constraint,
in which we removed the platens and treated the disk
assembly as
an isolated system without any displacement constraints.
This Type~IV constraint is intended to capture the inherent
material behavior in the absence of any restrictive boundaries
that might suppress the onset of bifurcations
or other stiffness pathologies.
\par
A snapshot of the DEM data was taken at the peak deviator
stress ``C'' (circled in Fig.~\ref{fig:stress}),
which included all particle positions
and all contact forces and orientations.
Of the original 68 particles (64 disks and 4 platens),
seven disks were ``rattler'' particles having
zero or one contacts with other disks,
and these rattlers were ignored in the analysis and are
not shown in Fig.~\ref{fig:irregular}.
Because a DEM simulation merely approximates equilibrium,
the contact forces in the snapshots were slightly out-of-equilibrium,
with small but non-zero net forces and moments,
$d\mathbf{b}^{p}$ and $d\mathbf{w}^{p}$, on the particles.
The data was pre-conditioned to restore equilibrium
by slightly adjusting the contact
forces $\mathbf{f}^{pq}$ with small adjustments $d\mathbf{f}^{pq}$
by projecting the out-of-equilibrium forces
$[d\mathbf{p}]=[d\mathbf{b}/d\mathbf{w}]$
onto the equilibrium sub-space given by the columns
of the statics matrix $[\mathbf{A}]$ (see Eq.~\ref{eq:Equilbrium0a}):
\begin{equation}
  \left[\boldsymbol{\mathfrak{f}}\right] \leftarrow
  \left[\boldsymbol{\mathfrak{f}}\right] - [d\boldsymbol{\mathfrak{f}}],
  \quad\quad
  \left[d\boldsymbol{\mathfrak{f}}\right] =
  \left[\mathbf{A}]^{\dagger} [d\mathbf{p}\right]
\end{equation}
where $[d\boldsymbol{\mathfrak{f}}]=[d\mathbf{f}/d\mathbf{m}]$,
and $[\mathbf{A}]^{\dagger}$ is the Moore--Penrose
inverse $[\mathbf{A}]^{\text{T}}([\mathbf{A}][\mathbf{A}]^{\text{T}})^{-1}$.
\par
There were 97 contacts among the disks and platens at ``C'',
and because 14 of these contacts were at the friction limit,
we searched for stiffness pathologies among
the $2^{14}=$\,16,384 possible combinations of slip and no-slip
at the 14 contacts.
Our results show that only a
single solution exists for the problem
(i.e., no bifurcation of solutions),
and this solution yielded platen movements that
were similar to those
found in the DEM simulation.
No modes of neutral
equilibrium were found, but the system exhibited 306 unstable modes.
That is, if the platens' movements were frozen,
particle movements could spontaneously occur in 306 possible patterns.
We also considered an assembly with the same arrangement
of particles but without the four platens:
a Type~IV isolated system.
By removing the platens and replacing them with
external forces on the peripheral particles, the disks have more freedom,
and the 306 unstable modes increased to 807.
\par
The fact that this granular system can exhibit so many instabilities
may seem curious,
as one might think that DEM results, such as those
in Fig.~\ref{fig:stress}
are the unique result of an initial particle arrangement
and the imposed boundary movements.
One must remember, however, that the DEM simulates
a dynamic system, and the particles' movements are produced by
a continual condition of dis-equilibrium \cite{Nguyen:2016a}.
Internal instability can, in fact, be present, and should be expected.
DEM modelers have noted that simulation results
are sensitive to the DEM loading rate and to the DEM damping constant:
if one changes the rate at which the boundaries are moved or the amount
of damping,
the DEM results will be altered.
In our example,
the 306 instabilities will impel particles to new positions,
even while the assembly is being compressed.
As seen with Eqs.~(\ref{eq:quadratic})--(\ref{eq:W2})
and~(\ref{eq:Ins1}),
an unstable stiffness branch $[\mathbf{H}^{i}]$ produces
an increase in kinetic energy that is quadratic in time.
Each of the 306 unstable branches (modes) can produce quadratically
increasing movements that
occur concurrently %with each other and
with the movements imposed by the moving platens.
The resulting particle positions
will depend, therefore, upon the various unstable modes and,
in the context of either DEM or physical experiments,
upon the platens' velocities and upon the numerical
(or physical) damping within the system.
However,
the quadratic increases in the particles' kinetic energies
that are produced by each particular instability mode
would not progress unabated.
New particle positions will cause some contacts to disengage
and other contacts to be newly established
and will cause some contacts to reach newly reach the friction
even as other contacts withdraw from the friction limit.
These rearrangements will occur while the assembly is being compressed,
and the new arrangements will deactivate and nullify past
instability modes while creating fresh, new ones.
Moreover, all 306 modes would not occur concurrently,
as each unstable mode can only occur within its region $\Omega^{i}$,
which might be incompatible with other modes.
As a final note on DEM simulations and physical experiments,
these realizations typically exhibit stress
relaxation, an observation that is entirely consistent with the
existence of unstable modes: these modes can proceed in the
absence of any boundary movements, since
$\mathcal{W}_{2}$ is dominated by the unconstrained
movements (see the discussions of Eq.~\ref{eq:quadratic}
and of Table~\ref{table:Int}).
\par
Although we have focused on the 64 particles at their
moment of peak stress (Fig.~\ref{fig:stress}),
we also examined the assembly near the start of
loading ``A'',
after a smaller peak ``B'',
and at a post-peak instant of softening ''D''.
With two exceptions, the results were similar to the peak ``C'':
a single unique solution with multiple modes of
unstable equilibrium, even near the start of loading ``A''.
A mode of \emph{neutral equilibrium} that occurred
immediately after the smaller peak ``B''.
That is, controllability was lost after this peak,
such that the singular stiffness matrix would allow
multiple solutions.
A bifurcation occurred at ``D'', with two possible equilibrium solutions.
As was previously noted,
in regard to modes of instability of equilibrium,
a DEM algorithm or a physical process
will yield its own solution,
although this solution
will depend upon the platens' and the particles' velocities
when the singularity is encountered.
\section{Discussion}
We have developed the incremental stiffness relationship
between particle movements and external forces for
a system of durable particles that interact at their
points of contact.
The curvatures of the particles' surfaces at their contacts
were shown to affect an assembly's stiffness,
as the contact forces can behave as internal follower forces
in three different ways, leading to three types of geometric
stiffness.
A fourth geometric stiffness applies to systems
with \emph{external} follower forces.
The stiffness relation of granular systems exhibits
an inherent incremental non-linearity, as each contact
at the friction limit brings its own
incrementally non-linear stiffness.
\par
The examples in Section~\ref{sec:threedisks} reveal
several aspects of granular behavior:
\begin{itemize}
\item
When examining potential stiffness pathologies,
one must include geometric effects.
Ignoring these effects,
by considering only the mechanical stiffnesses of the contacts,
leads to incorrect solutions and to an incorrect assessment
of neutral equilibrium and instability.
Moreover, potential instabilities either
can be arrested or provoked by simply altering
the particles' curvatures (flatness or pointiness) at their contacts.
\item
Multiple solutions (bifurcations) were exposed in two of the
examples,
and the Hill--Ba\v{z}ant--Petryk
second-order $\mathcal{I}_{2}$ can be used for determining which
solutions are stable and unstable.
\item
When even a few contacts within an assembly reach the friction limit,
multiple unstable branches are typically available,
and each instability can produce a sudden burst
in kinetic energy.
Multiple unstable modes were present in each of the three
small examples,
but such instability is surely a pervasive presence during
the deformation of granular materials.
Extrapolating our results to large specimens,
instabilities occur as localized pathologies and are likely
the cause of certain observed phenomena,
such as the sudden, irregular alterations of stress and volume,
the sudden generation of kinetic energy,
and the emission of acoustic energy \cite{KuhnDaouadji:drops}. 
For the most part, these localized events are self-arresting:
a sudden shifting of particles is eventually met with
a more stable local configuration
that withdraws contacts from the brink
of the friction limit or creates fresh supporting contacts.
Although each unstable, dynamic event is limited in scope,
when considered together,
these instabilities place a limit on bulk strength,
as they provide limitations on the local particle arrangements
and distributions of contact forces.
\end{itemize}
Of course, the paper's methods are currently
impractical for analyzing
potential pathologies of entire assemblies with more than a
few hundreds of particles, as the $2^{16}$ possible branches in
the second example required several hours of compute time.
The methods could be readily used, however, for examining
small sub-regions within an entire assembly, by forgoing
an exact representation of a sub-region's surroundings and treating
the sub-region as an isolated, Type~IV system.
Finally,
we note that our use of Type~III constraint and of generalized
inverses proved superior to the conventional Type~I approach
with static condensation.
Using generalized inverses obviates the need to rearrange rows
and columns (and then later un-rearranging them);
allows a more direct analysis of
neutral equilibrium, the second-order $\mathcal{I}_{2}$,
and unstable equilibrium;
and simplifies the process of checking whether
a solution $[d\mathbf{x}]$ lies within its proper region
$\Omega^{i}$.
\newpage
\bibliographystyle{unsrt}
%\bibliography{Kuhn}
%

\end{document}